\documentclass[
pra,
aps,
twocolumn,
reprint,
showpacs,
groupedaddress,
floatfix, 
superscriptaddress,
longbibliography]{revtex4-1}

\pdfoutput=1 

\usepackage{general_latex}
\usepackage{dsfont}

\usepackage{amsmath}
\usepackage{amssymb}
\usepackage{amsthm}

\newtheoremstyle{mystyle}%                % Name
  {}                                      % Space above
  {}                                      % Space below
  {\itshape}                              % Body font
  {}                                      % Indent amount 
  {\bfseries}                             % Theorem head font
  {.}                                     % Punctuation after theorem head
  { }                                     % Space after theorem head, ' ', or \newline
  {\thmname{#1}\thmnumber{ #2}\thmnote{ (#3)}}%    
\theoremstyle{mystyle}

\usepackage{array, makecell}
\usepackage{boldline}

\begin{document}
\title{Encoding an oscillator into many oscillators} 
\author{Kyungjoo Noh}\email{noh827@gmail.com}
\affiliation{Departments of Applied Physics and Physics, Yale University, New Haven, Connecticut 06520, USA}
\affiliation{Yale Quantum Institute, Yale University, New Haven, Connecticut 06520, USA}
%\author{TBD}
%\affiliation{TBD}
\author{S. M. Girvin}
\affiliation{Departments of Applied Physics and Physics, Yale University, New Haven, Connecticut 06520, USA}
\affiliation{Yale Quantum Institute, Yale University, New Haven, Connecticut 06520, USA}
\author{Liang Jiang}\email{liang.jiang@uchicago.edu}
\affiliation{Departments of Applied Physics and Physics, Yale University, New Haven, Connecticut 06520, USA}
\affiliation{Yale Quantum Institute, Yale University, New Haven, Connecticut 06520, USA}
\affiliation{Pritzker School of Molecular Engineering, University of Chicago, 5640 South Ellis Avenue, Chicago, Illinois 60637, USA}
%s\date{July 19, 2017}
%%%%%%%%%%%%%%%%%%%%%%%%%%%%%%%%%%%%%%%%%%%%%%%%%%%%%%%%%%%%%%%%%%
\begin{abstract}
An outstanding challenge for quantum information processing using bosonic systems is Gaussian errors such as excitation loss and added thermal noise errors. Thus, bosonic quantum error correction (QEC) is essential. Most bosonic QEC schemes encode a finite-dimensional logical qubit or qudit into noisy bosonic oscillator modes. In this case, however, the infinite-dimensional bosonic nature of the physical system is lost at the error-corrected logical level. On the other hand, there are several proposals for encoding an oscillator mode into many noisy oscillator modes. However, these oscillator-into-oscillators encoding schemes are in the class of Gaussian quantum error correction. Therefore, these codes cannot correct practically relevant Gaussian errors due to the established no-go theorems which state that Gaussian errors cannot be corrected by using only Gaussian resources. Here, we circumvent these no-go results and show that it is possible to correct Gaussian errors by using Gottesman-Kitaev-Preskill (GKP) states as non-Gaussian resources. In particular, we propose a non-Gaussian oscillator-into-oscillators code, the two-mode GKP-repetition code, and demonstrate that it can correct additive Gaussian noise errors. In addition, we generalize the two-mode GKP-repetition code to an even broader class of non-Gaussian oscillator codes, namely, GKP-stabilizer codes. Specifically, we show that there exists a highly hardware-efficient GKP-stabilizer code, the GKP-two-mode-squeezing code, that can quadratically suppress additive Gaussian noise errors in both the position and momentum quadratures up to a small logarithmic correction. Moreover, for any GKP-stabilizer code, we show that logical Gaussian operations can be readily implemented by using only physical Gaussian operations. Furthermore, we show that our non-Gaussian oscillator encoding scheme can also be used to correct excitation loss and thermal noise errors, which are dominant error sources in many realistic bosonic systems.  
\end{abstract}
\maketitle
%%%%%%%%%%%%%%%%%%%%%%%%%%%%%%%%%%%%%%%%%%%%%%%%%%%%%%%%%%%%%%%%%%
%%%%%%%%%%%%%%%%%%%%%%%%%%%%%%%%%%%%%%%%%%%%%%%%%%%%%%%%%%%%%%%%%%

\section{Introduction}
\label{section:Introduction}

Continuous-variable (CV) bosonic quantum systems are ubiquitous in various quantum computing and communication architectures \cite{Braunstein2005,Weedbrook2012} and provide unique advantages to, e.g., quantum simulation of bosonic systems such as boson sampling \cite{Aaronson2011,Hamilton2017,GarciaPatron2019,Wang2019b} and simulation of vibrational quantum dynamics of molecules \cite{Huh2015,Sparrow2018,Clements2018,Wang2019}. Since CV quantum information processing tasks are typically implemented by using harmonic oscillator modes in photonic or phononic systems, realistic imperfections such as excitation loss and added thermal noise errors are major challenges for realizing large-scale and fault-tolerant CV quantum information processing. 

Quantum error correction (QEC) is essential for scalable and fault-tolerant quantum information processing \cite{Gottesman2009}. Recently, there have been significant advances in using bosonic systems to realize QEC in a more hardware-efficient manner \cite{Albert2018}. In many bosonic QEC schemes proposed so far, a finite-dimensional quantum system (e.g., a qubit) is encoded into an oscillator \cite{Cochrane1999,Gottesman2001,Leghtas2013,Mirrahimi2014,Michael2016,Li2017} or into many oscillators \cite{Chuang1997,Harrington2001,Bergmann2016a,Fukui2017,
Niu2018,Fukui2018a,Fukui2018b,Vuillot2019,Fukui2019,Noh2019b}. For example, the four-component cat code \cite{Leghtas2013} encodes a logical qubit into an oscillator using cat states with even number of excitations, i.e., $|0_{L}\rangle \propto |\alpha\rangle+|-\alpha\rangle$ and $|1_{L}\rangle \propto |i\alpha\rangle+|-i\alpha\rangle$, which can correct single-excitation loss errors by monitoring the parity of the excitation number. Thanks to the inherent hardware efficiency, various qubit-into-an-oscillator bosonic QEC schemes have been realized experimentally \cite{Leghtas2015,Ofek2016,Touzard2017,Hu2019,Fluhmann2018,Fluhmann2019,
Fluhmann2019b,Grimm2019,Campagne2019}. However, in such qubit-into-an-oscillator encoding schemes, the infinite-dimensional bosonic nature of the physical oscillator modes is lost at the logical level because the error-corrected logical system is described by discrete variables (DV) such as Pauli operators. Therefore, the error-corrected logical DV system is not itself tailored to CV quantum information processing tasks.  

On the other hand, if an infinite-dimensional oscillator mode is encoded into many noisy oscillator modes (i.e., oscillator-into-oscillators encoding), such an error-corrected oscillator mode will still be tailored to various continuous-variable quantum information processing tasks. 
%, e.g., boson sampling \cite{Aaronson2011,Hamilton2017} and simulation of vibrational quantum dynamics of molecules \cite{Huh2015,Sparrow2018,Clements2018,Wang2019}. 
So far, there have been several proposals for encoding an oscillator into many oscillators \cite{Lloyd1998,Braunstein1998,Braunstein1998b,Aoki2009,
Hayden2016,Hayden2017,Faist2019,Woods2019}. For example, in the case of the three-mode Gaussian-repetition code \cite{Lloyd1998,Braunstein1998}, an infinite-dimensional oscillator mode is encoded into three oscillators by repeatedly appending the position eigenstates: $|\hat{q}_{L}=q\rangle \equiv |\hat{q}_{1}=q\rangle|\hat{q}_{2}=q\rangle|\hat{q}_{3}=q\rangle$. Note that in this case, the logical Hilbert space is infinite-dimensional because $q$ can be any real number.  

\begin{figure*}[t!]
\centering
\includegraphics[width=6.4in]{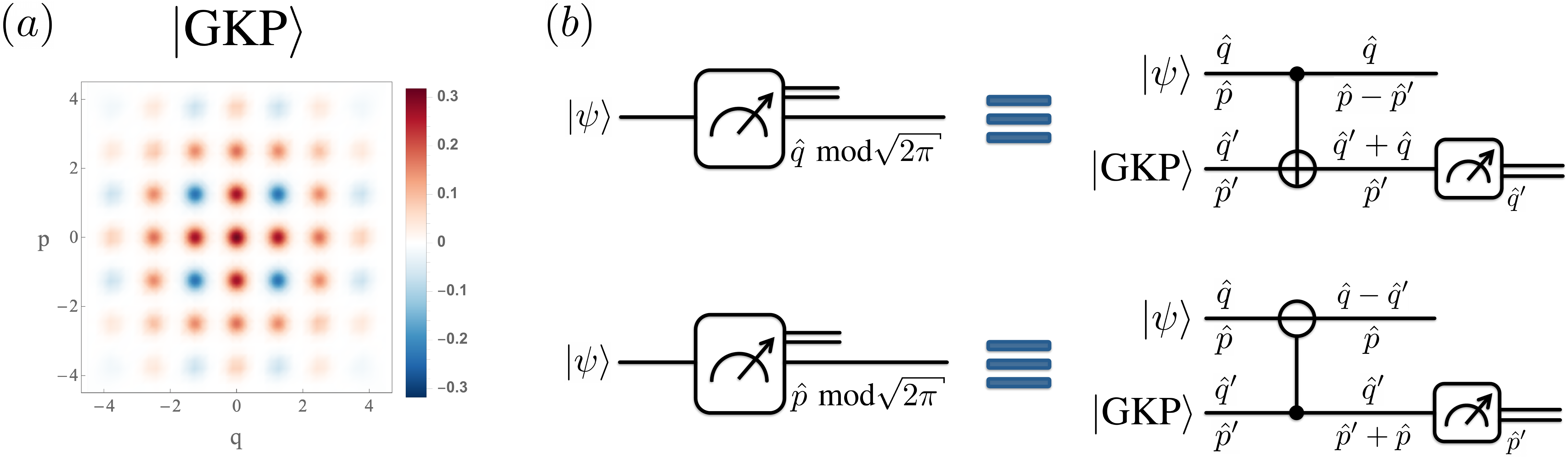}
\caption{(a) An approximate GKP state with an average photon number $\bar{n}=5$. (b) Measurement of the position or momentum operator modulo $\sqrt{2\pi}$. The controlled-$\oplus$ and $\ominus$ symbols respectively represent the SUM and the inverse-SUM gates.}
\label{fig:GKP state and measurements}
\end{figure*}

In general, oscillator-into-oscillators bosonic QEC is more challenging than qubit-into-oscillators bosonic QEC because in the former we aim to protect an infinite-dimensional bosonic Hilbert space against relevant errors, whereas in the latter we only aim to protect a finite-dimensional Hilbert space embedded in infinite-dimensional bosonic modes. Indeed, while there exist many qubit-into-oscillators codes that can correct experimentally relevant Gaussian errors, none of the previously proposed oscillator-into-oscillators codes can correct Gaussian errors. This is because they are \textit{Gaussian quantum error correction} schemes and the established no-go theorems state that Gaussian errors cannot be corrected by using only Gaussian resources \cite{Eisert2002,Niset2009,Vuillot2019}. Since Gaussian errors include excitation losses, thermal noise and additive Gaussian noise errors which are ubiquitous in many realistic CV quantum systems, these no-go results set a hard limit on the practical utility of the proposed Gaussian QEC schemes. 

%Therefore, it is highly desirable to have \textit{non-Gaussian quantum error correction} schemes (of the CV-into-CV type) that can correct Gaussian errors while still maintaining the bosonic nature of the encoded system. 

We circumvent the established no-go results and provide \textit{non-Gaussian} oscillator-into-oscillators codes that can correct Gaussian errors using Gottesman-Kitaev-Preskill (GKP) states \cite{Gottesman2001} as non-Gaussian resources. Our paper is organized as follows: In Section \ref{section:GKP states as non-Gaussian resources}, we briefly review common non-Gaussian resources and summarize known properties of the GKP state that will be used in later sections. In Section \ref{section:GKP-repetition codes}, we introduce a non-Gaussian oscillator-into-oscillators code, namely, the two-mode GKP-repetition code and demonstrate that it is indeed possible to correct Gaussian errors using the two-mode GKP-repetition code. In Section \ref{section:Generalization of GKP-repetition codes}, we generalize the two-mode GKP-repetition code and propose an even broader class of non-Gaussian oscillator codes, called GKP-stabilizer codes. In particular, in Subsection \ref{subsection:The GKP-two-mode-squeezing code}, we show that there exists a highly hardware-efficient GKP-stabilizer code, the GKP-two-mode-squeezing code, that can quadratically suppress additive Gaussian noise errors in both the position and momentum quadrature. In Subsection \ref{subsection:GKP-stabilizer codes}, we show that, for any GKP-stabilizer codes, logical Gaussian operations can be readily implemented by using only physical Gaussian operations which are available in many experimental systems. In Section \ref{section:Discussion}, we discuss experimental realization of our schemes and analyze the adverse effects of realistic imperfections. We also discuss potential applications of our schemes.

\section{GKP states as non-Gaussian resources}
\label{section:GKP states as non-Gaussian resources}

%\textit{GKP states as non-Gaussian resources--}
The established no-go theorems on Gaussian QEC schemes \cite{Eisert2002,Niset2009,Vuillot2019} make it clear that non-Gaussian resources \cite{Zhuang2018,Takagi2018} are necessary for correcting Gaussian errors while preserving the bosonic nature at the error-corrected logical level. Examples of non-Gaussian resources include the single-photon Fock state and photon-number-resolving measurements \cite{Knill2001,Aaronson2011}, Kerr nonlinearities \cite{Lloyd1999}, cubic phase state and gate \cite{Gottesman2001}, SNAP gate \cite{Krastanov2015}, Schr\"odinger cat states \cite{Cochrane1999}, and GKP states \cite{Gottesman2001,Baragiola2019}. Among these non-Gaussian resources, we demonstrate that GKP states are particularly useful for encoding an oscillator into many oscillators in a robust way. Specifically, we cast the GKP state as a tool to work around the Heisenberg uncertainty principle.

%\textit{The canonical GKP state--}
The Heisenberg uncertainty principle states that the position and momentum operators $\hat{q} \equiv (\hat{a}^{\dagger}+\hat{a})/\sqrt{2}$ and $\hat{p}\equiv i(\hat{a}^{\dagger}-\hat{a})/\sqrt{2}$ cannot be measured simultaneously because they do not commute with each other (i.e., $[\hat{q},\hat{p}]=i\neq 0$). Despite the Heisenberg uncertainty principle, the following displacement operators
\begin{align}
\hat{S}_{q}\equiv e^{i\sqrt{2\pi}\hat{q}} \,\,\,\textrm{and}\,\,\,\hat{S}_{p} \equiv e^{-i\sqrt{2\pi}\hat{p}}
\end{align}
do commute with each other and therefore can be measured simultaneously \cite{Gottesman2001}. Note that measuring $\hat{S}_{q} = \exp[i\sqrt{2\pi}\hat{q}]$ and $\hat{S}_{p} \equiv \exp[-i\sqrt{2\pi}\hat{p}]$ is equivalent to measuring their exponents (or phase angles) $i\sqrt{2\pi}\hat{q}$ and $-i\sqrt{2\pi}\hat{p}$ modulo $2\pi i$. Thus, the commutativity of $\hat{S}_{q}$ and $\hat{S}_{p}$ implies that the position and momentum operators can indeed be measured simultaneously if they are measured modulo $\sqrt{2\pi}$. The canonical GKP state (or the grid state) \cite{Gottesman2001,Duivenvoorden2017} is defined as the unique (up to an overall phase) simultaneous eigenstate of the two commuting displacement operators $\hat{S}_{q}$ and $\hat{S}_{p}$ with unit eigenvalues. Explicitly, the canonical GKP state is given by
\begin{align}
|\textrm{GKP}\rangle \propto \sum_{n\in\mathbb{Z}}|\hat{q}=\sqrt{2\pi}n\rangle \propto \sum_{n\in\mathbb{Z}}|\hat{p}=\sqrt{2\pi}n\rangle, 
\end{align}     
and thus clearly has definite values for both the position and momentum operators modulo $\sqrt{2\pi}$, i.e., $\hat{q}=\hat{p}=0$ mod $\sqrt{2\pi}$.

Ideally, the canonical GKP state has an infinite average photon number because it is superpositions of infinitely many ($\sum_{n\in\mathbb{Z}}$) infinitely squeezed states ($|\hat{q}=\sqrt{2\pi}n\rangle$ or $|\hat{p}=\sqrt{2\pi}n\rangle$). However, one can define an approximate GKP state with a finite average photon number by applying a non-unitary operator $\exp[-\Delta \hat{n}]$ to the canonical GKP state and then normalizing the output state: $|\textrm{GKP}_{\Delta}\rangle \propto \exp[-\Delta \hat{n}]|\textrm{GKP}\rangle$ \cite{Gottesman2001}. In Fig.\ \ref{fig:GKP state and measurements} (a), we plot the Wigner function of the canonical GKP state with an average photon number $\bar{n}=5$. Negative peaks in the Wigner function clearly indicate that the canonical GKP state is a non-Gaussian state.

There have been many proposals for preparing an approximate GKP state in various experimental platforms \cite{Gottesman2001,Travaglione2002,Pirandola2004,Pirandola2006,
Vasconcelos2010,Terhal2016,Motes2017,Weigand2017,Arrazola2019,
Su2019,Eaton2019,Shi2019,Weigand2019,Hastrup2019}. Notably, the proposal in Ref.\ \cite{Travaglione2002} has recently been realized in a trapped ion system \cite{Fluhmann2018,Fluhmann2019,Fluhmann2019b} and a variation of the scheme in Ref.\ \cite{Terhal2016} has recently been realized in a circuit QED system \cite{Campagne2019}. In the interest of clarity, we only consider the ideal canonical GKP state when we present our main results. Issues related to the use of the use of more realistic approximate GKP states will be addressed in Section \ref{section:Discussion}. 

Clearly, the ability to measure the position and momentum operators modulo $\sqrt{2\pi}$ allows us to prepare the canonical GKP state. Remarkably, the converse is also true. That is, we can measure the quadrature operators modulo $\sqrt{2\pi}$ given GKP states and Gaussian operations as resources: As shown in Fig.\ \ref{fig:GKP state and measurements} (b), one can measure the position (momentum) operator modulo $\sqrt{2\pi}$ by using a canonical GKP state, the SUM (inverse-SUM) gate and the homodyne measurement of the position (momentum) operator. The SUM gate is a Gaussian operation and is defined as $\textrm{SUM}_{j\rightarrow k} \equiv \exp[ -i\hat{q}_{j}\hat{p}_{k}]$, which maps $\hat{q}_{k}$ to $\hat{q}_{k}+\hat{q}_{j}$. The inverse-SUM gate is defined as the inverse of the SUM gate. The canonical GKP state and the modulo simultaneous quadrature measurement are the key non-Gaussian resources of our oscillator-into-oscillators encoding schemes which we introduce below.

In the following section, we construct a non-Gaussian oscillator-into-oscillators code, namely, the two-mode GKP-repetition code and demonstrate that it can correct additive Gaussian noise errors. That is, we circumvent the established no-go results on Gaussian QEC schemes \cite{Eisert2002,Niset2009,Vuillot2019} by using the canonical GKP state as a non-Gaussian resource. 

%In Section \ref{section:Discussion}, we will address issues related to the use of more realistic approximate GKP states.  

\section{The two-mode GKP-repetition code}
\label{section:GKP-repetition codes}

\begin{figure*}[t!]
\centering
\includegraphics[width=5.8in]{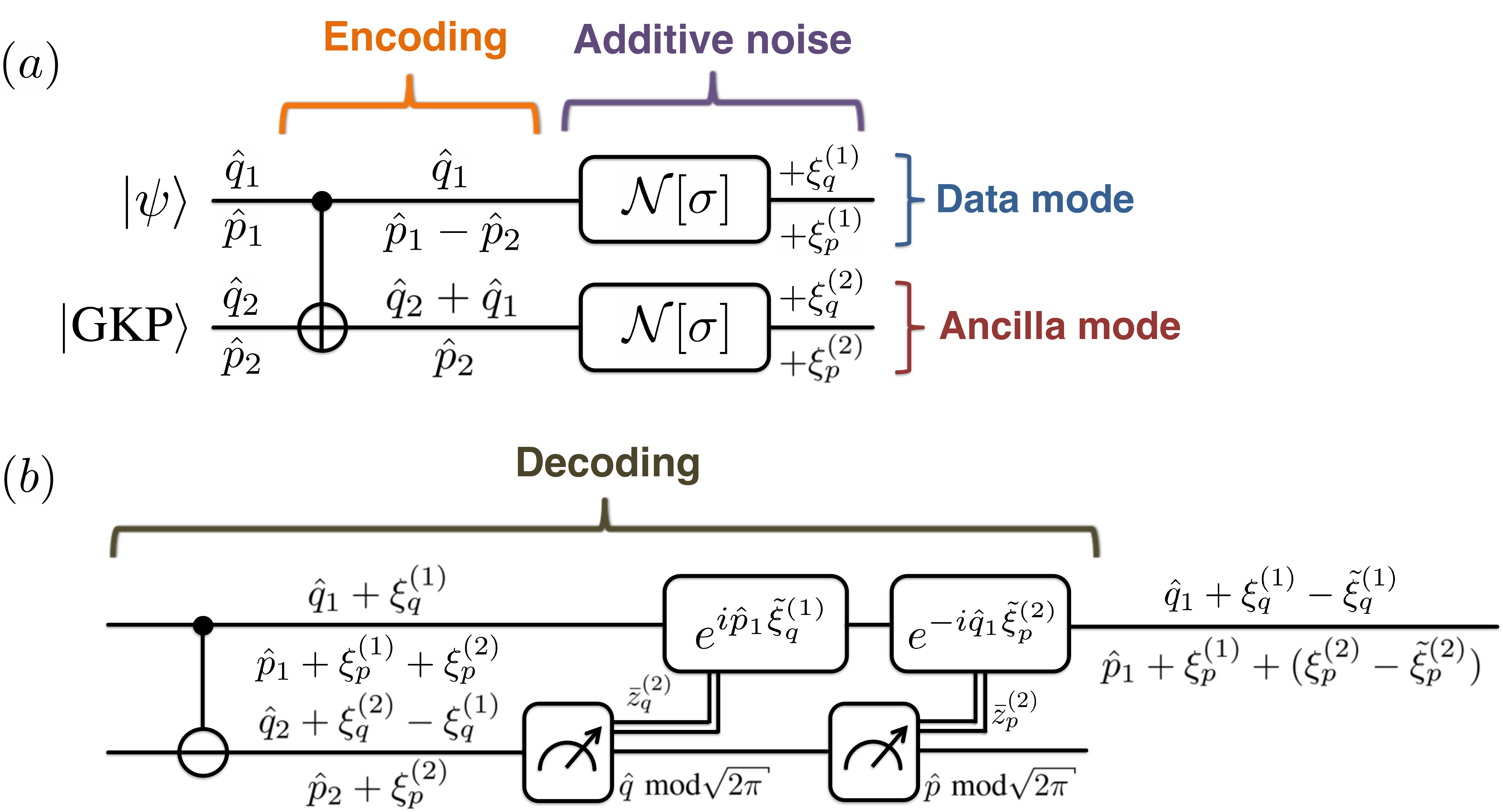}
\caption{(a) Encoding circuit of the two-mode GKP-repetition code subject to independent and identically distributed additive Gaussian noise errors. (b) Decoding circuit of the two-mode GKP-repetition code. The controlled-$\oplus$ and $\ominus$ symbols respectively represent the SUM and the inverse-SUM gates. Note that the circuits for the measurements of the position and the momentum operators modulo $\sqrt{2\pi}$ (at the end of the decoding) are defined in Fig.\ \ref{fig:GKP state and measurements} (b). }
\label{fig:GKP repetition codes}
\end{figure*}

Quantum error-correcting codes work by hiding the quantum information from the environment by storing the logical state in a non-local entangled state of the physical components. In the case of the scheme we present below, a data oscillator mode is entangled via a Gaussian operation with an ancillary oscillator mode, which is initially in the canonical GKP state, in a manner that prevents the environment from learning about the logically encoded state. Like qubit-into-an-oscillator GKP codes \cite{Gottesman2001}, our oscillator-into-oscillators codes are specifically designed to protect against random displacement errors. Also, it succeeds because we assume access to non-Gaussian resources (GKP states and modular quadrature measurements) that are unavailable to the environment.   

%makes a generic data state appear to be similar to a random thermal state and information lost to the local environment provides no information on the encoded state.  The special initial GKP states in the ancillae become randomly displaced GKP states in a manner that also prevents local measurements from learning about the logically encoded state. 

%\textit{GKP-repetition codes--}

%GKP states have been mostly viewed as encoded logical states of a family of DV-into-CV codes, namely, GKP codes \cite{Gottesman2001}. Here, we take a different perspective and view the canonical GKP state as a non-Gaussian resource that can be useful for many other CV information processing tasks besides encoding a DV system into CV systems. Specifically, we construct a family of non-Gaussian CV-into-CV QEC codes that can correct various Gaussian errors. 

More explicitly, we propose the following encoding of an arbitrary bosonic state $|\psi\rangle = \int dq \psi(q)|\hat{q}_{1} = q\rangle$ into two oscillator modes, 
\begin{align}
&|\psi_{L}\rangle = \textrm{SUM}_{1\rightarrow 2} |\psi\rangle \otimes |\textrm{GKP}\rangle, \label{eq:GKP-repetition code states}
\end{align}  
where $|\textrm{GKP}\rangle$ is the canonical GKP state in the second mode. We call this encoding the two-mode GKP-repetition code (see Fig.\ \ref{fig:GKP repetition codes} (a) for the encoding circuit). Also, we refer to the first mode as the data mode and the second mode as the ancilla mode because the information was stored only in the first mode before the application of the encoding circuit. 
%$|\textrm{GKP}_{k}\rangle\propto \sum_{n\in\mathbb{Z}}|\hat{q}_{k}=\sqrt{2\pi}n\rangle$ is the canonical GKP state in the $k^{\textrm{th}}$ bosonic mode where $k\in\lbrace 2,\cdots,N\rbrace$. 
The $\textrm{SUM}$ gate $\textrm{SUM}_{j\rightarrow k}\equiv \exp[-i\hat{q}_{j}\hat{p}_{k}]$ ($j\neq k$) is a CV analog of the $\textrm{CNOT}$ gate which, in the Heisenberg picture, transforms $\hat{q}_{k}$ and $\hat{p}_{j}$ into $\hat{q}_{k}+\hat{q}_{j}$ and $\hat{p}_{j}-\hat{p}_{k}$ and leaves all the other quadrature operators unchanged. The quadrature operators are transformed by the encoding circuit into 
\begin{alignat}{2}
&\hat{q}_{1}  \rightarrow \hat{q}'_{1}\equiv \hat{q}_{1},\quad && \hat{p}_{1}\rightarrow \hat{p}'_{1}\equiv  \hat{p}_{1} - \hat{p}_{2} , 
\nonumber\\
&\hat{q}_{2} \rightarrow \hat{q}'_{2} \equiv \hat{q}_{2}+\hat{q}_{1}, \quad && \hat{p}_{2} \rightarrow \hat{p}'_{2}\equiv \hat{p}_{2}. \label{eq:quadrature transformation due to encoding circuit}
\end{alignat}
Note that the $\textrm{SUM}$ gate in the encoding circuit is analogous to the CNOT gate in the encoding circuit of the two-bit repetition code for qubit bit-flip errors: This is why we refer to the encoding in Eq. \eqref{eq:GKP-repetition code states} as the two-mode GKP-repetition code.

Before moving on to the detailed analysis of the two-mode GKP-repetition code, we point out that our two-mode GKP-repetition code can be regarded as a non-Gaussian modification of the two-mode Gaussian-repetition code (see Refs. \cite{Lloyd1998,Braunstein1998} for the three-mode Gaussian-repetition code and Appendix \ref{appendix:Gaussian-repetition codes} for its generalization to $N$ modes). We also remark that the two-mode GKP-repetition code is not very efficient as it is significantly outperformed by the GKP-two-mode-squeezing code introduced in Subsection \ref{subsection:The GKP-two-mode-squeezing code}. Nevertheless, we introduce the two-mode GKP-repetition code to explain the key elements of our non-Gaussian oscillator encoding schemes and to contrast our schemes with the previous Gaussian encoding schemes.  

%In particular, we explain in Appendix \ref{appendix:Gaussian-repetition codes} (also later in this section) how the Gaussian-repetition codes fail to correct additive Gaussian noise errors. Before analyzing the performance of the GKP-repetition codes, 

%%%%%%%and $\textrm{Sq}_{1}(1/\lambda)$ is the single-mode squeezing operation acting on the data mode which, in the Heisenberg picture, transforms $\hat{q}_{1}$ and $\hat{p}_{1}$ into $\hat{q}_{1}/\lambda$ and $\lambda\hat{p}_{1}$. The squeezing parameter $\lambda$ is a free parameter that can be chosen by our will. 

Let us now analyze the performance of the two-mode GKP-repetition code. We assume that the oscillator modes undergo independent and identically distributed (iid) additive Gaussian noise errors (or Gaussian random displacement errors) $\mathcal{N}^{(1)}[\sigma] \otimes \mathcal{N}^{(2)}[\sigma]$. Here, $\mathcal{N}^{(k)}[\sigma]$ is an additive Gaussian noise error acting on the $k^{\textrm{th}}$ mode which, in the Heisenberg picture, adds Gaussian random noise $\xi_{q}^{(k)}$ and $\xi_{p}^{(k)}$ to the position and momentum operators of the $k^{\textrm{th}}$ mode, i.e., 
\begin{align}
\hat{q}'_{k}\rightarrow \hat{q}''_{k}\equiv  \hat{q}'_{k}+\xi_{q}^{(k)} \,\,\,\textrm{and}\,\,\, \hat{p}'_{k}\rightarrow \hat{p}''_{k}\equiv  \hat{p}'_{k}+\xi_{p}^{(k)},  \label{eq:quadrature transformation due to added noise}
\end{align}
where $k\in \lbrace 1,2\rbrace$. Also, $\xi_{q}^{(k)}$ and $\xi_{p}^{(k)}$ are independent Gaussian random variables with zero mean and variance $\sigma^{2}$. That is, $(\xi_{q}^{(1)},\xi_{p}^{(1)},\xi_{q}^{(2)},\xi_{p}^{(2)})\sim_{\textrm{iid}}\mathcal{N}(0,\sigma^{2})$. 

%The notation $B_{2}$ is based on the classification of one-mode Gaussian channels carried out in Ref. \cite{Holevo2007}. 

We emphasize that additive Gaussian noise errors are generic in the sense that any excitation loss and thermal noise errors can be converted into an additive Gaussian noise error by applying a suitable quantum-limited amplification channel \cite{Albert2018,Noh2019}. For example, a pure excitation loss error with loss probability $\gamma$ can be converted into an additive Gaussian noise error $\mathcal{N}[\sigma]$ with $\sigma = \sqrt{ \gamma} $ (see Lemma 6 and Table 1 in Ref. \cite{Noh2019}). 

The decoding procedure (shown in Fig.\ \ref{fig:GKP repetition codes} (b)) begins with the inverse of the encoding circuit, i.e., with $\textrm{SUM}^{\dagger}_{1\rightarrow 2} = \exp[i\hat{q}_{1}\hat{p}_{2}]$. Upon the inverse of the encoding circuit, the transformed quadrature operators in Eq. \eqref{eq:quadrature transformation due to encoding circuit} are transformed back to the original quadrature operators but the added quadrature noise $\xi_{q/p}^{(1)}$ and $\xi_{q/p}^{(2)}$ in Eq. \eqref{eq:quadrature transformation due to added noise} are reshaped, i.e., 
\begin{align}
\hat{q}''_{k}  &\rightarrow \hat{q}_{k}+z_{q}^{(k)},\quad  \hat{p}''_{k}\rightarrow  \hat{p}_{k}+z_{p}^{(k)}  
\end{align}
for $k\in\lbrace 1,2 \rbrace$ where the reshaped quadrature noise $z_{q}^{(k)}$ and $z_{p}^{(k)}$ are given by 
\begin{alignat}{2}
z_{q}^{(1)}&\equiv  \xi_{q}^{(1)},\qquad  && z_{p}^{(1)}\equiv \xi_{p}^{(1)} + \xi_{p}^{(2)},
\nonumber\\
z_{q}^{(2)}&\equiv \xi_{q}^{(2)}-\xi_{q}^{(1)},\qquad  &&  z_{p}^{(2)}\equiv \xi_{p}^{(2)}.   \label{eq:noise quadrature transformation}
\end{alignat}
Note that the position quadrature noise of the data mode $\xi_{q}^{(1)}$ is transferred to the position quadrature of the ancilla mode (see $-\xi_{q}^{(1)}$ in $z_{q}^{(2)}$), whereas the momentum quadrature noise of the ancilla mode $\xi_{p}^{(2)}$ is transferred to the momentum quadrature of the data mode (see $+\xi_{p}^{(2)}$ in $z_{p}^{(1)}$).

In the remainder of the decoding procedure, both the position and momentum quadrature noise of the ancilla mode are measured simultaneously modulo $\sqrt{2\pi}$ (using the measurement circuits shown in Fig.\ \ref{fig:GKP state and measurements} (b)). By doing so, we measure both $\hat{q}'''_{2}\equiv \hat{q}_{2}+z_{q}^{(2)}$ and $\hat{p}'''_{2}\equiv \hat{p}_{2}+z_{p}^{(2)}$ modulo $\sqrt{2\pi}$. Note that such measurements of $\hat{q}'''_{2}$ and $\hat{p}'''_{2}$ modulo $\sqrt{2\pi}$ are equivalent to measurements of only the reshaped ancilla quadrature noise $z_{q}^{(2)}$ and $z_{p}^{(2)}$ modulo $\sqrt{2\pi}$. This is because the ancilla modes were initially in the canonical GKP state and thus $\hat{q}_{2}= \hat{p}_{2}=0$ mod $\sqrt{2\pi}$ holds. The extracted information about $z_{q}^{(2)} = \xi_{q}^{(2)}-\xi_{q}^{(1)}$ and $z_{p}^{(2)} = \xi_{p}^{(2)}$ will then be used to estimate the data position quadrature noise $\xi_{q}^{(1)}$ and the ancilla momentum quadrature noise $\xi_{p}^{(2)}$ such that the uncertainty of the data position quadrature noise is reduced while the ancilla momentum quadrature noise (transferred to the data momentum quadrature) does not degrade the momentum quadrature of the data mode. Below, we provide a detailed description of this estimation procedure.

From the outcomes of the measurements of $z_{q}^{(2)}$ and $z_{p}^{(2)}$ modulo $\sqrt{2\pi}$, we assume that the true values of $z_{q}^{(2)}$ and $z_{p}^{(2)}$ are the ones with the smallest length among the candidates that are compatible with the modular measurement outcomes. That is, 
\begin{align}
\bar{z}_{q}^{(2)} &= R_{\sqrt{2\pi}}(z_{q}^{(2)})\,\,\,\textrm{and}\,\,\,\bar{z}_{p}^{(2)} = R_{\sqrt{2\pi}}(z_{p}^{(2)})  \label{eq:estimation of the reshaped ancilla noise GKP repetition}
\end{align}    
where $R_{s}(z)\equiv z-n^{\star}(z)s$ and $n^{\star}(z)\equiv \textrm{argmin}_{n\in\mathbb{Z}}|z-ns|$. More concretely, $R_{s}(z)$ equals a displaced sawtooth function with an amplitude and period $s$ and is given by $R_{s}(z) = z$ if $z\in[-s/2,s/2]$. Then, based on these estimates, we further estimate that the position quadrature noise of the data mode $\xi_{q}^{(1)}$ and the momentum quadrature noise of the ancilla mode $\xi_{p}^{(2)}$ are 
\begin{align}
\tilde{\xi}_{q}^{(1)} &= -\frac{\bar{z}_{q}^{(1)}+\bar{z}_{q}^{(2)}}{2} \,\,\, \textrm{and}\,\,\,  \tilde{\xi}_{p}^{(2)} = \bar{z}_{p}^{(2)}  \label{eq:GKP-rep estimates for quadrature noise}
\end{align}
The latter estimate was chosen simply because $\xi_{p}^{(2)} = z_{p}^{(2)}$ and the former choice is based on a maximum likelihood estimation explained in detail in Appendix \ref{appendix:Gaussian-repetition codes}. 

%Intuitively, the former estimate can be roughly understood as averaging the estimated values of the reshaped ancilla momentum quadrature noise $z_{q}^{(k)} = \xi_{q}^{(k)} - \xi_{q}^{(1)}$ such that the ancilla position quadrature noise $\xi_{q}^{(k)}$ are averaged out in the limit of large $N$. 

\begin{figure}[t!]
\centering
\includegraphics[width=3.3in]{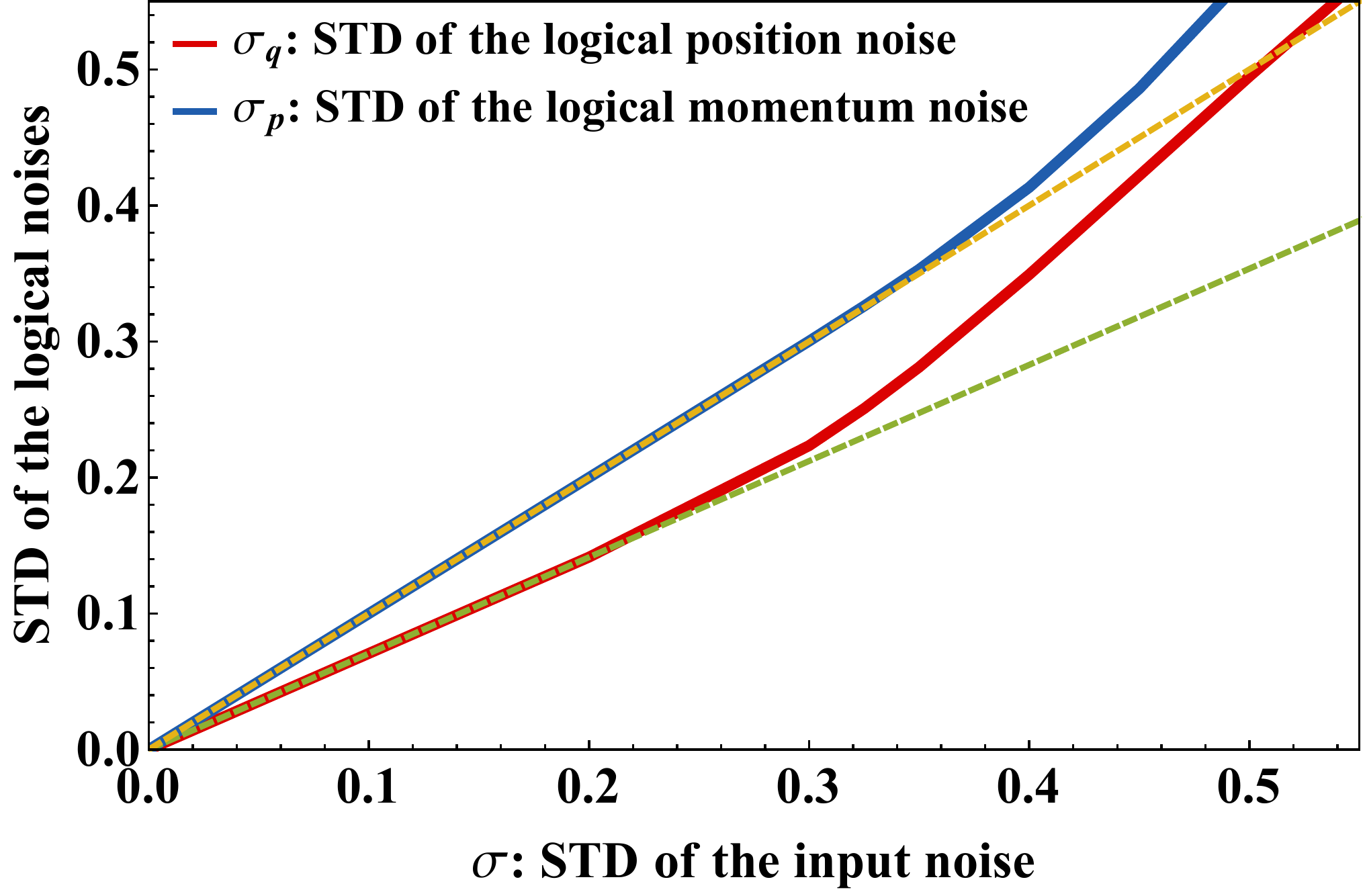}
\caption{Standard deviations of the logical quadrature noise $\sigma_{q}$ and $\sigma_{p}$ as a function of the input standard deviation $\sigma$ for the two-mode GKP-repetition code. The green and yellow dashed lines represent $\sigma_{q} = \sigma/\sqrt{2}$ and $\sigma_{p}=\sigma$. }
\label{fig:performance of GKP repetition codes}
\end{figure}

Finally, based on the estimates $\tilde{\xi}_{q}^{(1)}$ and $\tilde{\xi}_{p}^{(2)}$, we apply counter displacement operations $\exp[i\hat{p}_{1} \tilde{\xi}_{q}^{(1)}]$ and $\exp[-i\hat{p}_{1}\tilde{\xi}_{p}^{(2)}]$ to the data mode and end up with the following logical position and momentum quadrature noise
\begin{align}
\xi_{q} &\equiv z_{q}^{(1)} - \tilde{\xi}_{q}^{(1)} = \xi_{q}^{(1)} + \frac{1}{2}R_{\sqrt{2\pi}}(\xi_{q}^{(2)}-\xi_{q}^{(1)})  , 
\nonumber\\
\xi_{p} &\equiv z_{p}^{(1)} -  \tilde{\xi}_{p}^{(2)} =  \xi_{p}^{(1)} + \xi_{p}^{(2)}-R_{\sqrt{2\pi}}(\xi_{p}^{(2)}). \label{eq:logical quadrature noise}
\end{align}
In Appendix \ref{appendix:Probability density of the logical quadrature noise: GKP-repetition codes}, we provide explicit expressions for the probability density functions of $\xi_{q}$ and $\xi_{p}$ (which are used to obtain Fig.\ \ref{fig:performance of GKP repetition codes}) in the most general case. Here, we instead focus on a simple (but important) case where $\sigma$ is much smaller than $\sqrt{2\pi}$. In this case, the reshaped ancilla quadrature noise $z_{q}^{(2)}=\xi_{q}^{(2)}-\xi_{q}^{(1)}$ and $z_{p}^{(2)}=\xi_{p}^{(2)}$ lie in the unambiguously distinguishable range $[-\sqrt{\pi/2},\sqrt{\pi/2}]$ with a very high probability and thus we have $R_{\sqrt{2\pi}}(\xi_{q}^{(2)}-\xi_{q}^{(1)}) = \xi_{q}^{(2)}-\xi_{q}^{(1)}$ and $R_{\sqrt{2\pi}}(\xi_{p}^{(2)}) =\xi_{p}^{(2)}$. Then, the logical position and momentum quadrature noise are given by
\begin{align}
\xi_{q} & \xrightarrow{\sigma\ll \sqrt{2\pi}} \frac{\xi_{q}^{(1)}+\xi_{q}^{(2)}}{2}  \sim \mathcal{N}\Big{(}0,\sigma_{q}^{2} = \frac{\sigma^{2}}{2}\Big{)},
\nonumber\\
\xi_{p} &\xrightarrow{\sigma\ll \sqrt{2\pi}} \xi_{p}^{(1)} \sim \mathcal{N}\Big{(}0,\sigma_{p}^{2} = \sigma^{2} \Big{)}.  \label{eq:logical quadrature noise small sigma}
\end{align} 
That is, the variance of the logical position quadrature noise is reduced by a factor of $2$. This is due to the syndrome measurement of the reshaped ancilla position quadrature noise $z_{q}^{(2)} = \xi_{q}^{(2)}-\xi_{q}^{(1)}$ modulo $\sqrt{2\pi}$ which is then used to reduce the uncertainty of the data position quadrature noise $\xi_{q}^{(1)}$. Moreover, the variance of the logical momentum quadrature noise remains unchanged despite the temporary increase ($\xi_{p}^{(1)}\rightarrow z_{p}^{(1)} = \xi_{p}^{(1)} +\xi_{p}^{(2)}$) during the decoding procedure. Again, this is due to the syndrome measurement of the reshaped ancilla momentum quadrature noise $z_{p}^{(2)} = \xi_{p}^{(2)}$ modulo $\sqrt{2\pi}$ which fully captures the transferred ancilla momentum quadrature noise $\xi_{p}^{(2)}$ if $\sigma \ll \sqrt{2\pi}$.

In Fig.\ \ref{fig:performance of GKP repetition codes}, we plot the standard deviations of the output logical quadrature noise for the two-mode GKP-repetition code. The standard deviation of the output logical position quadrature noise is indeed reduced by a factor of $\sqrt{2}$ (i.e., $\sigma_{q} = \sigma/\sqrt{2}$), while the standard deviation of the logical momentum quadrature noise remains unchanged (i.e., $\sigma_{p}=\sigma$) for $\sigma \lesssim 0.3$. Note that the condition $\sigma \lesssim 0.3$ is translated to $\gamma = \sigma^{2} \lesssim 0.1$ in the case of pure excitation losses, where $\gamma$ is the pure-loss probability \cite{Albert2018,Noh2019}. Thus, if the standard deviation of an additive Gaussian noise error is sufficiently small, our GKP-repetition coding scheme can successfully reduce the noise of the position quadrature, while keeping the momentum quadrature noise unchanged. That is, our (non-Gaussian) GKP-repetition codes can \textit{correct} additive Gaussian noise errors.

%under the above choices ($N=2$ and $c=0.1$), the variance of the \textcolor{red}{momentum} quadrature noise is indeed suppressed quadratically while the variance of the \textcolor{red}{position} quadrature noise remains unchanged for $\sigma \lesssim 0.3$ (which corresponds to $\gamma = \sigma^{2} \lesssim 0.1$, where $\gamma$ is the pure-loss probability \cite{Albert2018,Noh2019}). Such a quadratic suppression of the \textcolor{red}{momentum} noise variance is analogous to the quadratic suppression of the bit-flip error probability with multi-qubit repetition codes. Note, however, that only $2$ oscillator modes (one data mode and one ancilla mode) are needed to achieve the quadratic error suppression, whereas at least $3$ qubits are required in the multi-qubit repetition code case to achieve a quadratic error suppression.  

We remark that in an analogous two-mode Gaussian-repetition coding scheme (presented in detail in Appendix \ref{appendix:Gaussian-repetition codes}), the variance of the position quadrature noise is reduced by a factor of $2$ (i.e., $\sigma_{q}^{2} = \sigma^{2}/2$) similarly as in Eq. \eqref{eq:logical quadrature noise small sigma} but the variance of the momentum quadrature noise is increased by the same factor (i.e., $\sigma_{p}^{2} = 2\sigma^{2}$). This implies that, in the case of Gaussian-repetition codes, the position and momentum quadrature noise are only squeezed ($\sigma_{q}\sigma_{p}=\sigma^{2}$) instead of being corrected ($\sigma_{q}\sigma_{p}<\sigma^{2}$): This reaffirms the previous no-go results on Gaussian QEC schemes \cite{Eisert2002,Niset2009,Vuillot2019}.

The key difference between our GKP-repetition code and the previous Gaussian-repetition code is that in the latter case the ancilla momentum quadrature noise that is transferred to the data mode (see $+\xi_{p}^{(2)}$ in $z_{p}^{(1)}$) is left completely undetected, whereas in our case it is captured by measuring the ancilla momentum quadrature operator modulo $\sqrt{2\pi}$. Such a limitation of the Gaussian-repetition code is in fact a general feature of any Gaussian QEC scheme which relies on homodyne measurements of ancilla quadrature operators. With homodyne measurements, one can only monitor the noise in one quadrature of a bosonic mode, while the noise in the other conjugate quadrature is left completely undetected.  

On the other hand, we have worked around this barrier by using the canonical GKP state as a non-Gaussian resource which allows simultaneous measurements of both the position and momentum quadrature operators modulo $\sqrt{2\pi}$ (see Fig.\ \ref{fig:GKP state and measurements} (b)). In this regard, we emphasize that such non-Gaussian modular simultaneous measurements of both position and momentum quadrature operators (using canonical GKP states) are fundamentally different from Gaussian heterodyne measurements \cite{Shapiro1979} where both quadrature operators are measured simultaneously but in a necessarily noisy manner.    

%%%%%%%\textcolor{red}{Finally, we remark that the single-mode squeezing operation $\textrm{Sq}_{1}(1/\lambda)$ with a proper choice of $\lambda$ was essential for achieving the quadratic noise suppression demonstrated in Eq. \eqref{eq:position noise variance quadratic suppression} and Fig. \ref{fig:performance of GKP repetition codes}. The use of squeezing operations is a unique feature of GKP-repetition codes that is not present in its DV counterpart, i.e., multi-qubit repetition codes.     }

Finally, we compare our two-mode GKP-repetition code with the conventional three-bit repetition code (i.e., $|0_{L}\rangle = |0\rangle^{\otimes 3}$ and $|1_{L}\rangle = |1\rangle^{\otimes 3}$) which can correct single bit-flip errors \cite{Nielsen2000}. First, we observe that only two bosonic modes are sufficient to reduce the variance of the position quadrature noise in the case of the GKP-repetition code, whereas at least three qubits are needed to suppress qubit bit-flip errors in the case of mutli-qubit repetition code. However, while the GKP-repetition code can be implemented in a more hardware-efficient way, it does not reduce the variance of the position quadrature noise quadratically but instead only reduce the variance by a constant factor, i.e., $\sigma^{2}\rightarrow \sigma_{q}^{2} = \sigma^{2}/2$ if $\sigma \ll \sqrt{2\pi}$. On the other hand, the three-qubit repetition code can reduce the bit-flip error probability quadratically from $p$ to $p_{L} \simeq 3p^{2}$ if $p\ll 1$.

Note that such a quadratic (or even higher order) suppression of Pauli errors is a key step towards qubit-based fault-tolerant universal quantum computation \cite{Gottesman2009}. Thus, it is also highly desirable in the oscillator encoding case to have such a quadratic suppression of additive Gaussian noise errors, going way beyond the reduction by a constant factor which was shown above. 

Below, we introduce the GKP-two-mode-squeezing code and show that it can suppress both the position and momentum quadrature noise quadratically up to a small logarithmic correction. In particular, the proposed scheme can achieve such a quadratic noise suppression by using only two bosonic modes (one data mode and one ancilla mode) and therefore is hardware-efficient. In the case of qubit-based QEC, on the contrary, at least five qubits and high-weight multi-qubit gate operations are needed to suppress both the bit-flip and phase-flip errors quadratically \cite{Laflamme1996,Bennett1996}.

\begin{figure*}
\centering
\includegraphics[width=5.8in]{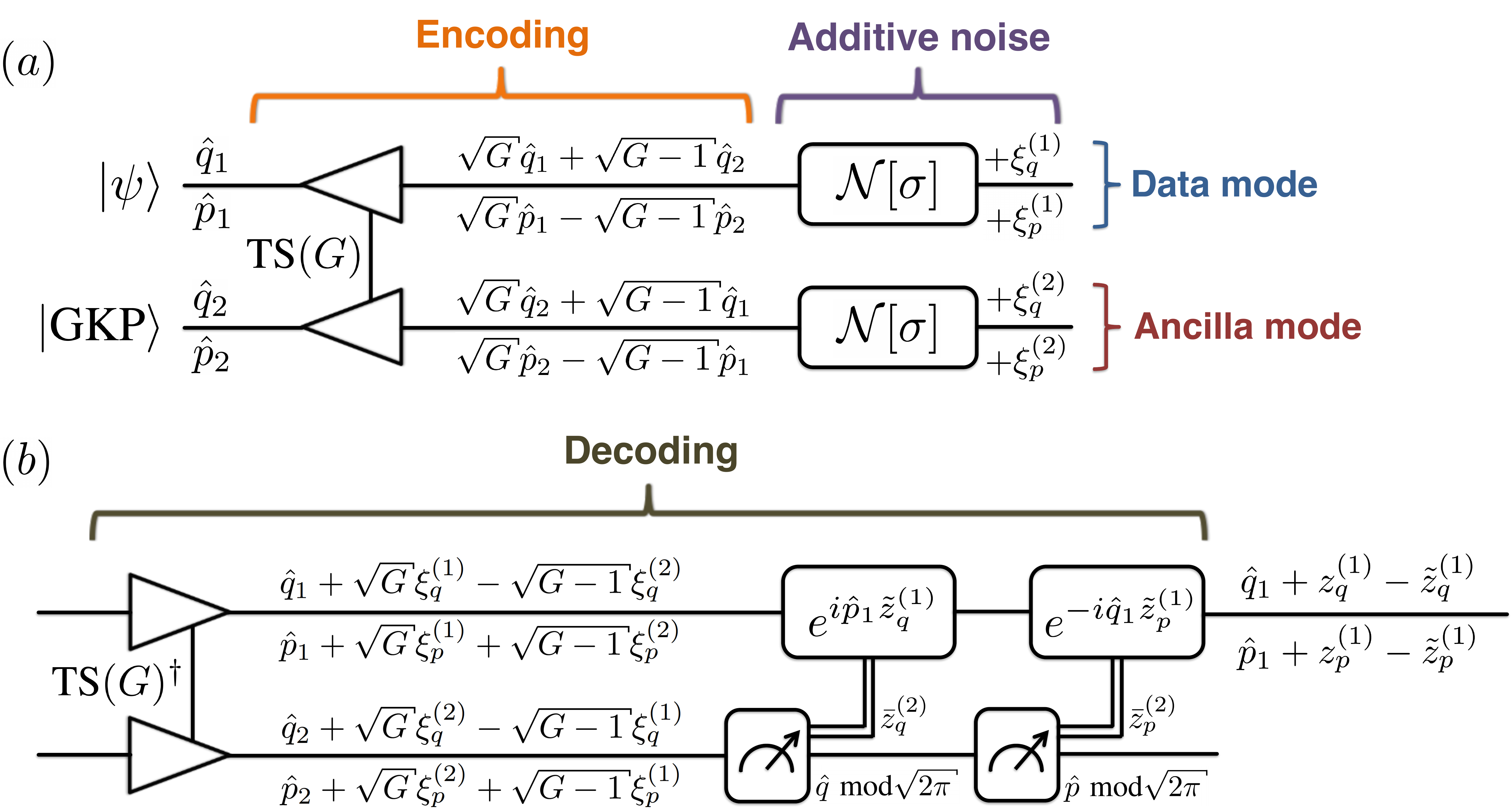}
\caption{(a) Encoding circuit of the GKP-two-mode-squeezing code subject to independent and identically distributed additive Gaussian noise errors. (b) Decoding circuit of the GKP-two-mode-squeezing code. Note that the circuits for the measurements of the position and the momentum operators modulo $\sqrt{2\pi}$ (at the end of the decoding) are defined in Fig.\ \ref{fig:GKP state and measurements} (b). }
\label{fig:GKP-two-mode-squeezing code}
\end{figure*}

%Before formally introducing the two-mode GKP-squeezed-repetition code in Subsection \ref{subsection:The two-mode GKP-squeezed-repetition code}, we stress that only two bosonic modes are sufficient for the quadratic suppression of both the position and momentum quadrature noise, while in the DV case at least five qubits are needed to suppress both the bit-flip and phase-flip errors quadratically \cite{Laflamme1996,Bennett1996}.  

%show that by introducing single-mode squeezing operations, it is indeed possible to modify the presented two-mode GKP-repetition code such that the modified code, the two-mode GKP-squeezed-repetition code, can suppress the quadrature noise quadratically. Before formally introducing the two-mode GKP-squeezed-repetition code in Subsection \ref{subsection:The two-mode GKP-squeezed-repetition code}, we stress that only two bosonic modes are sufficient for the quadratic suppression of both the position and momentum quadrature noise, while in the DV case at least five qubits are needed to suppress both the bit-flip and phase-flip errors quadratically \cite{Laflamme1996,Bennett1996}.     

\section{GKP-stabilizer formalism}
\label{section:Generalization of GKP-repetition codes}

In this section, we generalize our GKP-repetition code. Specifically, in Subsection \ref{subsection:The GKP-two-mode-squeezing code}, we introduce the GKP-two-mode-squeezing code and show that it can suppress both the position and momentum quadrature noise quadratically up to a small logarithmic correction. In Subsection \ref{subsection:GKP-stabilizer codes}, we further generalize our non-Gaussian QEC schemes and provide an even broader class of non-Gaussian oscillator(s)-into-oscillators codes, namely, GKP-stabilizer codes. We also show that logical Gaussian operations can be readily implemented by using only physical Gaussian operations for any GKP-stabilizer code.

%\textcolor{red}{In this section, we provide generalizations of our GKP-repetition code design. In Subsection \ref{subsection:GKP-Shor code}, motivated by the design principle of the nine-qubit Shor code \cite{Shor1995}, we concatenate two-mode GKP-repetition codes and construct a four-mode GKP-Shor code. We then show that the four-mode GKP-Shor code can quadratically suppress the variances of both the position and the momentum quadrature noise. Moreover, in Subsection \ref{subsection:GKP-stabilizer codes} we further generalize our coding schemes to an even broader class of non-Gaussian QEC schemes and introduce GKP-stabilizer codes. In Subsection \ref{subsection:Logical operations and applications}, we show that for the class of GKP-stabilizer codes any logical Gaussian operations can be implemented by using only physical Gaussian operations and discuss potential applications of our scheme. }

%\subsection{The two-mode GKP-squeezed-repetition code}
%\label{subsection:The two-mode GKP-squeezed-repetition code}

\subsection{The GKP-two-mode-squeezing code}
\label{subsection:The GKP-two-mode-squeezing code}

%\textit{Generalizations of GKP-repetition codes--}

%We define the two-mode GKP-squeezed-repetition code as follows:
%\begin{align}
%|\psi_{L}\rangle &=  \textrm{Sq}_{1}\Big{(} \frac{1}{\lambda} \Big{)}\textrm{Sq}_{2} ( \lambda ) \textrm{SUM}_{1\rightarrow 2} |\psi\rangle|\textrm{GKP}_{2}\rangle, \label{eq:logical states two-mode GKP-squeezed-repetition code}
%\end{align}
%where $|\psi\rangle =\int dq\psi(q)|\hat{q}_{1} =q \rangle$ is an arbitrary bosonic state and $\textrm{Sq}_{k}(\lambda)$ is the single-mode squeezing operation acting on the $k^{\textrm{th}}$ mode and transforms $\hat{q}_{k}$ and $\hat{p}_{k}$ into $\lambda\hat{q}_{k}$ and $\hat{p}_{k}/\lambda$ (see Fig. \ref{fig:Two-mode GKP-squeezed-repetition code} (a) for the encoding circuit). Note that the squeezing parameter $\lambda$ is a free parameter that can be chosen at will. 

We define the GKP-two-mode-squeezing code as follows:
\begin{align}
|\psi_{L}\rangle &=  \textrm{TS}_{1,2}(G) |\psi\rangle\otimes |\textrm{GKP}\rangle, \label{eq:logical states two-mode GKP-squeezed-repetition code}
\end{align}
where $|\psi\rangle =\int dq\psi(q)|\hat{q}_{1} =q \rangle$ is an arbitrary bosonic state of the first mode, and $|\textrm{GKP}\rangle$ is the canonical GKP state of the second mode. Also, $\textrm{TS}_{1,2}(G)$ is the two-mode squeezing operation acting on the modes $1$ and $2$ with a gain $G\ge 1$ (hence the name of the code; see Fig.\ \ref{fig:GKP-two-mode-squeezing code} (a)). In the Heisenberg picture, the two-mode squeezing operation $\textrm{TS}_{1,2}(G)$ transforms the quadrature operators $\boldsymbol{x}=(\hat{q}_{1},\hat{p}_{1},\hat{q}_{2},\hat{p}_{2})^{T}$ into $\boldsymbol{x'}=(\hat{q}'_{1},\hat{p}'_{1},\hat{q}'_{2},\hat{p}'_{2})^{T} = \boldsymbol{S_{\textrm{TS}}}(G)\boldsymbol{x}$, where the $4\times 4$ symplectic matrix $\boldsymbol{S_{\textrm{TS}}}(G)$ associated with $\textrm{TS}_{1,2}(G)$ is given by   
\begin{align}
\boldsymbol{S_{\textrm{TS}}}(G) = \begin{bmatrix}
\sqrt{G}\boldsymbol{I}&\sqrt{G-1}\boldsymbol{Z}\\
\sqrt{G-1}\boldsymbol{Z}&\sqrt{G}\boldsymbol{I}
\end{bmatrix}. 
\end{align}  
Here, $\boldsymbol{I}=\textrm{diag}(1,1)$ is the $2\times 2$ identity matrix and $\boldsymbol{Z}=\textrm{diag}(1,-1)$ is the Pauli $Z$ matrix.  

The two-mode squeezing operation $\textrm{TS}_{1,2}(G)$ can be decomposed into a sequence of $50:50$ beam splitter operations and single-mode squeezing operations: 
\begin{align}
\textrm{Tr}_{1,2}(G) = \textrm{BS}_{1,2}\Big{(}\frac{1}{2}\Big{)}  \textrm{Sq}_{1}\Big{(}\frac{1}{\lambda}\Big{)}\textrm{Sq}_{2}(\lambda) \Big{[}\textrm{BS}_{1,2}\Big{(}\frac{1}{2}\Big{)}\Big{]}^{\dagger}. \label{eq:decomposition of two mode squeezing}
\end{align}
where $\lambda = \sqrt{G}+\sqrt{G-1}$. Here, $\textrm{Sq}_{k}(\lambda)$ is the single-mode squeezing operation that transforms $\hat{q}_{k}$ and $\hat{p}_{k}$ into $\lambda\hat{q}_{k}$ and $\hat{p}_{k}/\lambda$, respectively. Also, $\textrm{BS}_{1,2}(\eta)$ is the beam splitter interaction between the modes $1$ and $2$ with a transmissivity $\eta\in [0,1]$ and is associated with a $4\times 4$ symplectic matrix
\begin{align}
\boldsymbol{S_{\textrm{BS}}}(\eta) &= \begin{bmatrix}
\sqrt{\eta}\boldsymbol{I}&\sqrt{1-\eta}\boldsymbol{I}\\
-\sqrt{1-\eta}\boldsymbol{I}&\sqrt{\eta}\boldsymbol{I}
\end{bmatrix}.  
\end{align}
Note that the squeezing parameter $\lambda$ (or the gain $G$) can be chosen at will to optimize the performance of the error correction scheme.

%The action of the encoding circuit of the two-mode GKP-squeezed-repetition code can be described by a symplectic transformation $\boldsymbol{\hat{x}'} = \boldsymbol{S}\boldsymbol{\hat{x}}$, where $\boldsymbol{\hat{x}} = (\hat{q}_{1},\hat{p}_{1},\hat{q}_{2},\hat{p}_{2})^{T}$ and $\boldsymbol{\hat{x}'} = (\hat{q}'_{1},\hat{p}'_{1},\hat{q}'_{2},\hat{p}'_{2})^{T}$ are the original and transformed quadrature operators, and the symplectic matrix $\boldsymbol{S}$ is given by  
%\begin{align}
%\boldsymbol{S} = \begin{bmatrix}
%1/\lambda  &0&0&0\\
%0&\lambda&0&-\lambda\\
%\lambda&0&\lambda&0\\
%0&0&0& 1/\lambda
%\end{bmatrix}. 
%\end{align}

Upon the addition of independent and identically distributed Gaussian noise errors, the quadrature operator $\boldsymbol{\hat{x}'}$ is further transformed into $\boldsymbol{\hat{x}''} = \boldsymbol{\hat{x}'} + \boldsymbol{\xi}$, where $\boldsymbol{\xi} = (\xi_{q}^{(1)},\xi_{p}^{(1)},\xi_{q}^{(2)},\xi_{p}^{(2)})^{T}$ is the quadrature noise vector obeying $(\xi_{q}^{(1)},\xi_{p}^{(1)},\xi_{q}^{(2)},\xi_{p}^{(2)})\sim_{\textrm{iid}}\mathcal{N}(0,\sigma^{2})$. Then, the inverse of the encoding circuit $(\textrm{TS}_{1,2}(G))^{\dagger}$ in the decoding procedure (shown in Fig.\ \ref{fig:GKP-two-mode-squeezing code} (b)) transforms the quadrature operator into $
\boldsymbol{\hat{x}'''} = \boldsymbol{(S_{\textrm{TS}}}(G)\boldsymbol{)^{-1}}\boldsymbol{\hat{x}''} = \boldsymbol{\hat{x}} + \boldsymbol{z}$, where $\boldsymbol{z}\equiv (z_{q}^{(1)},z_{p}^{(1)},z_{q}^{(2)},z_{p}^{(2)})^{T}$ is the reshaped quadrature noise vector which is given by
\begin{align}
\boldsymbol{z} &= \boldsymbol{(S_{\textrm{TS}}}(G)\boldsymbol{)^{-1}}\boldsymbol{\xi} 
\nonumber\\
&= \begin{bmatrix}
\sqrt{G}\xi_{q}^{(1)}-\sqrt{G-1}\xi_{q}^{(2)} \\
\sqrt{G}\xi_{p}^{(1)}+\sqrt{G-1}\xi_{p}^{(2)} \\
\sqrt{G}\xi_{q}^{(2)}-\sqrt{G-1}\xi_{q}^{(1)}  \\
\sqrt{G}\xi_{p}^{(2)}+\sqrt{G-1}\xi_{p}^{(1)}
\end{bmatrix} \equiv \begin{bmatrix}
z_{q}^{(1)}\\
z_{p}^{(1)}\\
z_{q}^{(2)}\\
z_{p}^{(2)}
\end{bmatrix}. \label{eq:reshaped quadrature noise two-mode GKP-squeezed-repetition code}
\end{align}
As in the case of GKP-repetition code, information about the reshaped ancilla quadrature noise $z_{q}^{(2)}$ and $z_{p}^{(2)}$ is extracted by simultaneously measuring the position and momentum quadrature operators of the ancilla mode modulo $\sqrt{2\pi}$. These modular measurement outcomes are then used to estimate the reshaped data quadrature noise $z_{q}^{(1)}$ and $z_{p}^{(1)}$.  

Before elaborating the detailed estimation strategy from the obtained syndrome measurement outcomes, let us explain the key idea behind it. Note that the bare quadrature noise $\boldsymbol{\xi} = (\xi_{q}^{(1)},\xi_{p}^{(1)},\xi_{q}^{(2)},\xi_{p}^{(2)})^{T}$ is uncorrelated and thus its covariance matrix is proportional to the $4\times 4$ identity matrix: $V_{\boldsymbol{\xi}} = \sigma^{2}\textrm{diag}(1,1,1,1)$. On the other hand, the covariance matrix of the reshaped quadrature noise $\boldsymbol{z} = \boldsymbol{(S_{\textrm{TS}}}(G)\boldsymbol{)^{-1}}\boldsymbol{\xi}$ is given by
\begin{align}
V_{\boldsymbol{z}} &= \boldsymbol{(S_{\textrm{TS}}}(G)\boldsymbol{)^{-1}} V_{\boldsymbol{\xi}} \boldsymbol{((S_{\textrm{TS}}}(G)\boldsymbol{)^{-1})^{T}}
\nonumber\\
&= \sigma^{2}\begin{bmatrix}
(2G-1)\boldsymbol{I}&-2\sqrt{G(G-1)}\boldsymbol{Z}\\
-2\sqrt{G(G-1)}\boldsymbol{Z}&(2G-1)\boldsymbol{I}
\end{bmatrix}. 
\end{align}
Thus, the reshaped data and ancilla quadrature noise are correlated whenever $G>1$. In other words, the inverse of the encoding circuit $(\textrm{TS}(G))^{\dagger}$ converts the uncorrelated quadrature noise $\boldsymbol{\xi}$ into a correlated noise $\boldsymbol{z}$. In particular, in the $G\gg 1$ limit, the covariance matrix $V_{\boldsymbol{z}}$ is asymptotically given by 
\begin{align}
V_{\boldsymbol{z}} &\xrightarrow{G\gg 1} 2G\sigma^{2} \begin{bmatrix}
\boldsymbol{I}&-\boldsymbol{Z}\\
-\boldsymbol{Z}&\boldsymbol{I}
\end{bmatrix}, 
\end{align}    
and therefore the data and ancilla position quadrature noise are perfectly anti-correlated (i.e., $z_{q}^{(1)} =-z_{q}^{(2)}$) and the data and ancilla momentum quadrature noise are perfectly correlated (i.e., $z_{p}^{(1)} = z_{p}^{(2)}$). These strong correlations in the $G\gg 1$ limit allow us to reliably estimate the data quadrature noise $z_{q}^{(1)}$ and $z_{p}^{(1)}$ based solely on the knowledge of the ancilla quadrature noise $z_{q}^{(2)}$ and $z_{p}^{(2)}$.  

We may be tempted to choose as large gain $G$ as possible since the data and ancilla quadrature noise are perfectly correlated (or anti-correlated) in the $G\gg 1$ limit. However, we cannot increase the gain $G$ indefinitely because then the reshaped ancilla quadrature noise $z_{q}^{(2)}$ and $z_{p}^{(2)}$ (whose variances are given by $(2G-1)\sigma^{2}$) are not contained within the unambiguously distinguishable range $[-\sqrt{\pi/2},\sqrt{\pi/2}]$. Therefore, we have to choose the gain $G$ such that the reshaped ancilla quadrature noise lie mostly in the unambiguously distinguishable range. Below, we describe the decoding strategy and the parameter optimization in more detail.

Based on the outcomes of the measurements of the ancilla quadrature operators modulo $\sqrt{2\pi}$, we estimate that the reshaped ancilla noise $z_{q}^{(2)}$ and $z_{p}^{(2)}$ are the smallest ones that are compatible with the modular measurement outcomes, i.e.,  
\begin{align}
\bar{z}_{q}^{(2)} &= R_{\sqrt{2\pi}}(z_{q}^{(2)}) \,\,\,\textrm{and}\,\,\,\bar{z}_{p}^{(2)} = R_{\sqrt{2\pi}}(z_{p}^{(2)}),  \label{eq:estimates of the ancilla quadrature noise two-mode GKP-squeezed-repetition code}
\end{align}  
similarly as in Eq. \eqref{eq:estimation of the reshaped ancilla noise GKP repetition}. Then, we further estimate that the reshaped data position and momentum quadrature noise $z_{q}^{(1)}$ and $z_{p}^{(1)}$ are 
\begin{align}
\tilde{z}_{q}^{(1)} &=  - \frac{2\sqrt{G(G-1)}}{2G-1} \bar{z}_{q}^{(2)}, 
\nonumber\\
\tilde{z}_{p}^{(1)} &=   \frac{2\sqrt{G(G-1)}}{2G-1} \bar{z}_{p}^{(2)}.  \label{eq:estimates of the data quadrature noise two-mode GKP-squeezed-repetition code}
\end{align}   
The underlying reasons behind our choice of these estimates are explained in detail in Appendix \ref{appendix:Probability density of the logical quadrature noise: The two-mode GKP-squeezed-repetition code}. In the $G\gg 1$ limit, the estimates $\tilde{z}_{q}^{(1)}$ and $\tilde{z}_{p}^{(1)}$ in Eq. \eqref{eq:estimates of the data quadrature noise two-mode GKP-squeezed-repetition code} are respectively reduced to $-\bar{z}_{q}^{(2)}$ and $\bar{z}_{p}^{(2)}$, which are reasonable because then the data and ancilla position quadrature noise are perfectly anti-correlated whereas the data and ancilla momentum quadrature noise are perfectly correlated. 

\begin{figure*}
\centering
\includegraphics[width=6.8in]{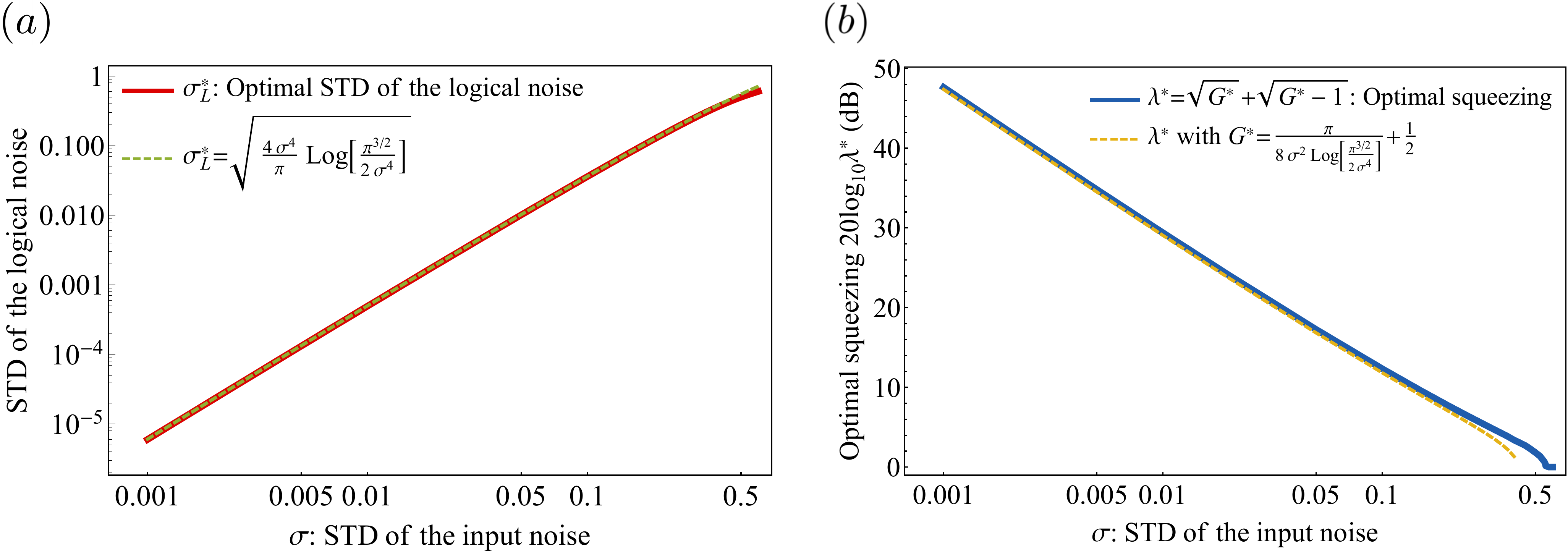}
\caption{(a) The minimum standard deviations of the output logical quadrature noise $\sigma_{q}=\sigma_{p}=\sigma_{L}^{\star}$ as a function of the input standard deviation $\sigma$ for the GKP-two-mode-squeezing code and (b) the optimal two-mode squeezing gain $G^{\star}$ that achieves $\sigma_{L}^{\star}$, translated to the required single-mode squeezing in the unit of decibel $20\log_{10}\lambda^{\star}$ where $\lambda^{\star} \equiv \sqrt{G^{\star}} + \sqrt{G^{\star}-1}$. The green dashed line in (a) represents $\sigma_{L}^{\star} = \frac{2\sigma^{2}}{\sqrt{\pi}}\sqrt{\log_{e}[\frac{\pi^{3/2}}{2\sigma^{4}}]}$ and the yellow dashed line in (b) represents $G^{\star} = \frac{\pi}{8\sigma^{2}}(\log_{e}[\frac{\pi^{3/2}}{2\sigma^{4}}])^{-1} +\frac{1}{2}$.   }
\label{fig:performance of the two-mode GKP-squeezed-repetition code}
\end{figure*}

Finally, applying the counter displacement operations $\exp[i\hat{p}_{1}\tilde{z}_{q}^{(1)}]$ and $\exp[-i\hat{q}_{1}\tilde{z}_{p}^{(1)}]$, we end up with the following output logical quadrature noise
\begin{align}
\xi_{q} &\equiv z_{q}^{(1)}-\tilde{z}_{q}^{(1)} = z_{q}^{(1)} +\frac{2\sqrt{G(G-1)}}{2G-1} R_{\sqrt{2\pi}}(z_{q}^{(2)}), 
\nonumber\\
\xi_{p} &\equiv z_{p}^{(1)}-\tilde{z}_{p}^{(1)} = z_{p}^{(1)} - \frac{2\sqrt{G(G-1)}}{2G-1} R_{\sqrt{2\pi}}(z_{p}^{(2)}). \label{eq:logical quadrature noise two-mode GKP-squeezed-repetition code}
\end{align} 
As shown in detail in Appendix \ref{appendix:Probability density of the logical quadrature noise: The two-mode GKP-squeezed-repetition code}, the probability density functions of the output logical quadrature noise $\xi_{q}$ and $\xi_{p}$ are identical to each other and are given by a mixture of Gaussian distributions:
\begin{align}
Q(\xi) = \sum_{n\in\mathbb{Z}} q_{n} \cdot p\Big{[}\frac{\sigma}{\sqrt{2G-1}}\Big{]}(\xi-\mu_{n}). \label{eq:probability density function of the logical quadrature noise}
\end{align}
Here, $p[\sigma](z) \equiv \frac{1}{\sqrt{2\pi\sigma^{2}}}\exp[-z^{2}/2\sigma^{2}]$ is the probability density function of the Gaussian distribution $\mathcal{N}(0,\sigma^{2})$ and $q_{n}$ and $\mu_{n}$ are given by
\begin{align}
q_{n} &= \int_{(n-\frac{1}{2})\sqrt{2\pi}}^{(n+\frac{1}{2})\sqrt{2\pi}} dz p[\sqrt{2G-1}\sigma](z), 
\nonumber\\
\mu_{n} &= \frac{2\sqrt{G(G-1)}}{2G-1}\sqrt{2\pi}n. \label{eq:qn and mun}
\end{align} 
Thus, the mixture probabilities $q_{n}$ sum up to unity, i.e., $\sum_{n\in\mathbb{Z}}q_{n}=1$. One can understand the probability density function in Eq. \eqref{eq:probability density function of the logical quadrature noise} in the following way: First of all, $q_{n}$ is the probability that the ancilla position quadrature noise $z_{q}^{(2)}$ (whose standard deviation is given by $\sqrt{2G-1}\sigma$) lies in the range $[(n-1/2)\sqrt{2\pi},(n+1/2)\sqrt{2\pi}]$. In this case, we have $R_{\sqrt{2\pi}}(z_{q}^{(2)}) = z_{q}^{(2)} - \sqrt{2\pi}n$ and therefore misidentify $z_{q}^{(2)}$ by a finite shift $\sqrt{2\pi}n$. Secondly, such a misidentification results in an undesired shift $\mu_{n}$ to the data mode through a miscalibrated counter displacement (see Eq. \eqref{eq:estimates of the data quadrature noise two-mode GKP-squeezed-repetition code}).

Now, we consider the variance $(\sigma_{L})^{2}$ of the output logical quadrature noise. By using the probability density function in Eq. \eqref{eq:probability density function of the logical quadrature noise}, we find that the variance $(\sigma_{L})^{2}$ is given by 
\begin{align}
(\sigma_{L})^{2} = \frac{\sigma^{2}}{2G-1} + \sum_{n\in\mathbb{Z}} q_{n}(\mu_{n})^{2}. 
\end{align}   
Recall that we can freely choose the gain $G$ to optimize the performance of the GKP-two-mode-squeezing code. Here, we choose $G$ such that the standard deviation of the output logical quadrature noise $\sigma_{L}$ is minimized. In Fig.\ \ref{fig:performance of the two-mode GKP-squeezed-repetition code}, we plot the minimum standard deviation of the output logical quadrature noise $\sigma_{L}^{\star}$ (see Fig.\ \ref{fig:performance of the two-mode GKP-squeezed-repetition code} (a)) and the optimal gain $G^{\star}$ that achieves the minimum standard deviation (see Fig.\ \ref{fig:performance of the two-mode GKP-squeezed-repetition code} (b)). These optimal values are obtained via a brute-force numerical optimization. Note that in Fig.\ \ref{fig:performance of the two-mode GKP-squeezed-repetition code} (b), we show the strength of the required single-mode squeezing operations to achieve the optimal gain $G^{\star}$ in the unit of decibel (i.e., $20\log_{10}\lambda^{\star}$ where $\lambda^{\star} = \sqrt{G^{\star}}+\sqrt{G^{\star}-1}$; see Eq.\ \eqref{eq:decomposition of two mode squeezing}).

We find that for $\sigma\ge 0.558$, the optimal gain $G^{\star}$ is trivially given by $G^{\star}=1$ and thus the GKP-two-mode-squeezing code cannot reduce the noise: $\sigma_{L}^{\star} = \sigma$. On the other hand, if the standard deviation of the input noise is small enough, i.e., $\sigma < 0.558$, the optimal gain $G^{\star}$ is strictly larger than $1$ and the minimum standard deviation of the output logical quadrature noise $\sigma_{L}$ can be made smaller than the standard deviation of the input quadrature noise $\sigma$: $\sigma_{L}^{\star}< \sigma$. Note that the condition $\sigma < 0.558$ corresponds to $\gamma = \sigma^{2} < 0.311$ for the pure excitation loss error where $\gamma$ is the loss probability (see Lemma 6 and Table 1 in Ref. \cite{Noh2019}). Below, we provide asymptotic expressions for $G^{\star}$ and $\sigma_{L}^{\star}$ in the $\sigma\ll 1$ limit and show that the GKP-two-mode-squeezing code can \textit{quadratically} suppress additive Gaussian noise errors up to a small logarithmic correction.

\begin{figure*}
\centering
\includegraphics[width=6.0in]{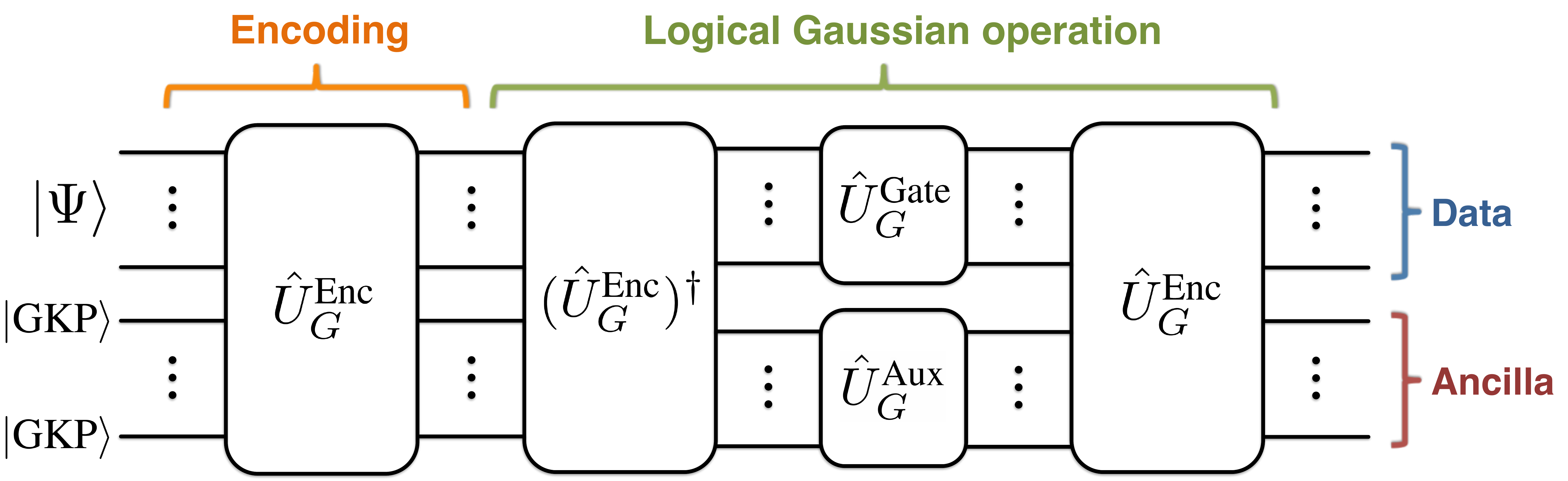}
\caption{Encoding circuit of a general GKP-stabilizer code and implementation of a general logical Gaussian operation $(\hat{U}_{G}^{\textrm{Gate}})_{L}$.  }
\label{fig:GKP-stabilizer codes and logical Gaussian operations}
\end{figure*}

We observe that the optimal solutions are found in the regime where $\sqrt{2G-1}\sigma\ll \sqrt{2\pi}$ holds and thus the ancilla quadrature noise $z_{q}^{(2)}$ and $z_{p}^{(2)}$ are contained within the range $[-\sqrt{\pi/2},\sqrt{\pi/2}]$ with a very high probability. In this regime, the variance $(\sigma_{L})^{2}$ is approximately given by 
\begin{align}
(\sigma_{L})^{2} \simeq \frac{\sigma^{2}}{2G-1} + \frac{8\pi G(G-1)}{(2G-1)^{2}}\textrm{erfc}\Big{(} \frac{\sqrt{\pi}}{2\sqrt{2G-1}\sigma} \Big{)}, \label{eq:variance of the logical quadrature noise asymptotic}
\end{align} 
where $\textrm{erfc}(x) \equiv \frac{2}{\sqrt{\pi}}\int_{x}^{\infty}dz e^{-z^{2}}$ is the complementary error function. Further assuming $\sigma\ll 1$, we find that the variance $(\sigma_{L})^{2}$ in Eq. \eqref{eq:variance of the logical quadrature noise asymptotic} is minimized when the gain $G$ is given by 
\begin{align}
G^{\star} \xrightarrow{\sigma\ll 1} \frac{\pi}{8\sigma^{2}} \Big{(} \log_{e}\Big{[}\frac{\pi^{3/2}}{2\sigma^{4}}\Big{]} \Big{)}^{-1} +\frac{1}{2}. \label{eq:optimal G asymptotic}
\end{align}
(See Appendix \ref{appendix:Probability density of the logical quadrature noise: The two-mode GKP-squeezed-repetition code} for the derivation.) Then, the optimal standard deviation of the output logical quadrature noise $\sigma_{L}^{\star}$ is given by
\begin{align}
\sigma_{L}^{\star} \xrightarrow{\sigma \ll 1} \frac{2\sigma^{2}}{\sqrt{\pi}} \sqrt{ \log_{e}\Big{[}\frac{\pi^{3/2}}{2\sigma^{4}}\Big{]} }.  \label{eq:optimal STD asymptotic}
\end{align}
As can be seen from Fig.\ \ref{fig:performance of the two-mode GKP-squeezed-repetition code}, these asymptotic expressions agree well with the exact numerical results. In particular, the minimum standard deviation of the logical quadrature noise $\sigma_{L}^{\star}$ decreases quadratically as $\sigma$ decreases (i.e., $\sigma_{L}^{\star}\propto \sigma^{2}$) up to a small logarithmic correction. For example, when $\sigma = 0.1$ (which corresponds to $\gamma = \sigma^{2}=0.01$ for the pure excitation loss error), the optimal gain is given by $G^{\star} = 4.806$ which requires $20\log_{10}\lambda^{\star} = 12.35\textrm{dB}$ single-mode squeezing operations. In this case, the resulting standard deviation of the output noise is given by $\sigma_{L}^{\star} = 0.036$ which corresponds to the loss probability $0.13\%$. This corresponds to a QEC ``gain'' for the protocol of $1/0.13\simeq 7.7$ in terms of the loss probability and $0.1/0.036\simeq 2.8$ in terms of displacement errors.

We emphasize again that our scheme is hardware-efficient because only two bosonic modes (one data mode and one ancilla mode) are needed to achieve the quadratic suppression of both quadrature noise up to a small logarithmic correction. In contrast, in the case of multi-qubit QEC, at least five qubits and high-weight multi-qubit operations are needed to suppress both the bit-flip and phase-flip errors quadratically \cite{Laflamme1996,Bennett1996}.   

On the other hand, while the GKP-two-mode-squeezing code is relatively simple and achieves a quadratic noise suppression, one might at some point want to have a more complicated but even more powerful codes that can achieve a higher (than second) order noise suppression. Below, having this in mind, we generalize our non-Gaussian quantum error-correcting codes to an even broader class of non-Gaussian quantum error-correcting codes.

%Similar to the nine-qubit Shor code design \cite{Shor1995}, we can concatenate the two-mode GKP-repetition codes and define a four-mode GKP-Shor code. Such four-mode GKP-Shor codes will be able to quadratically suppress the variances of both the position and the momentum quadrature noise in a symmetric way. 

\subsection{GKP-stabilizer codes}
\label{subsection:GKP-stabilizer codes}

%\begin{figure*}
%\centering
%\includegraphics[width=5.8in]{Fig_6_GKP_stabilizer_codes_and_GKP_squeezed_repetition_codes.pdf}
%\caption{(a) Encoding circuit of a general GKP-stabilizer code and implementation of a general logical Gaussian operation. (b) The encoding Gaussian circuit $\hat{U}_{G}^{\textrm{Enc}} = \hat{U}_{\textrm{Sq-Rep}}^{[2]}(\lambda)$ of the two-mode GKP-squeezed-repetition code. (c) The encoding Gaussian circuit $\hat{U}_{G}^{\textrm{Enc}} = \hat{U}_{\textrm{Sq-Rep}}^{[N]}(\lambda)$ of the $N$-mode GKP-squeezed-repetition code. Note that only the encoding Gaussian circuits are presented in (b) and (c) and the non-Gaussian input ancilla GKP states are omitted.    }
%\label{fig:GKP-stabilizer codes and general GKP-squeezed-repetition codes}
%\end{figure*}

In this subsection, we further generalize the GKP-repetition code (Section \ref{section:GKP-repetition codes}) and the GKP-two-mode-squeezing code (Subsection \ref{subsection:The GKP-two-mode-squeezing code}) to an even broader class of non-Gaussian oscillator codes, namely, GKP-stabilizer codes: We define logical code states of a general GKP-stabilizer code (encoding $M$ oscillator modes into $N$ oscillator modes) as
\begin{align}
|\Psi_{L}\rangle = \hat{U}_{G}^{\textrm{Enc}} |\Psi\rangle \otimes |\textrm{GKP}\rangle^{\otimes N-M}, 
\end{align} 
where $|\Psi\rangle$ is an arbitrary $M$-mode bosonic state in the first $M$ modes (data modes) and $|\textrm{GKP}\rangle^{\otimes N-M}$ is the $N-M$ copies of the canonical GKP states in the last $N-M$ modes (ancilla modes) and $\hat{U}_{G}^{\textrm{Enc}}$ is an encoding Gaussian unitary operation (see Fig.\ \ref{fig:GKP-stabilizer codes and logical Gaussian operations} for the encoding circuit). For example, for the two-mode GKP-repetition code we have $M=1$, $N=2$ and $\hat{U}_{G}^{\textrm{Enc}} = \textrm{SUM}_{1\rightarrow 2}$. For the GKP-two-mode-squeezing code, we have $M=1$, $N=2$ and $\hat{U}_{G}^{\textrm{Enc}} = \textrm{TS}_{1,2}(G)$.

\begin{figure*}
\centering
\includegraphics[width=7.0in]{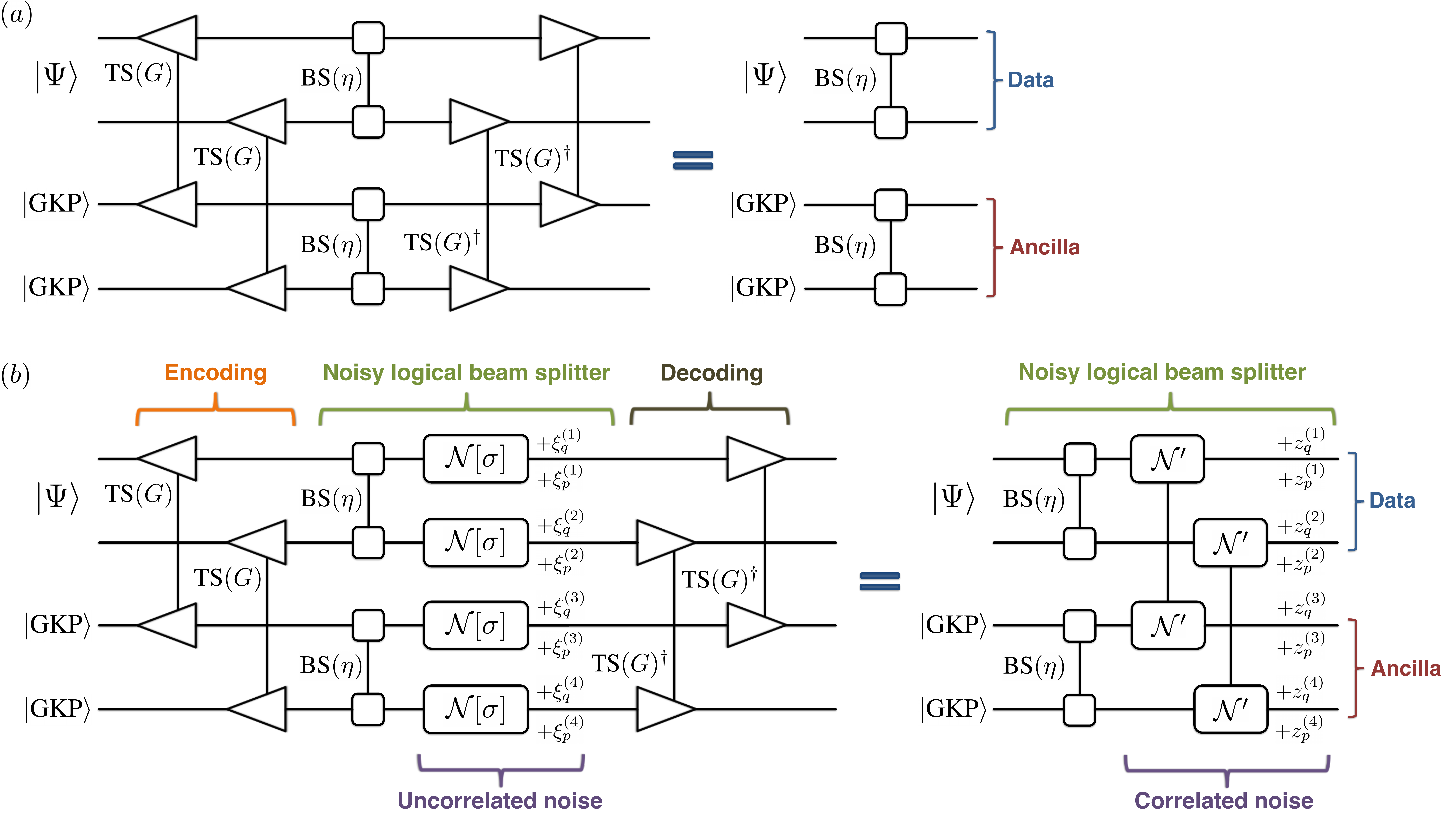}
\caption{Implementation of a logical beam splitter operation for the GKP-two-mode-squeezing code (a) in the absence (b) in the presence of iid additive Gaussian noise error during the beam splitter interaction.  }
\label{fig:GKP-two-mode-squeezing code logical BS transversality}
\end{figure*}

Since one can choose any Gaussian operation as an encoding circuit $\hat{U}_{G}^{\textrm{Enc}}$, our GKP-stabilizer formalism is at least as flexible as the stabilizer formalism for qubit-based QEC \cite{Gottesman1997} which encompasses almost all conventional multi-qubit QEC schemes. Therefore, our GKP-stabilizer formalism has a great deal of potential. As an illustration, in Appendix \ref{appendix:Generalized GKP-squeezed-repetition codes}, we introduce a family of GKP-stabilizer codes, GKP-squeezed-repetition codes, and show that the $N$-mode GKP-squeezed-repetition code can suppress additive Gaussian noise errors to the $N^{\textrm{th}}$ order, as opposed to the second order for the GKP-two-mode-squeezing code. In addition to the GKP-squeezed-repetition codes, there will be many other interesting families of GKP-stabilizer codes. For example, it will be an interesting future research direction to consider an encoding circuit $\hat{U}_{G}^{\textrm{Enc}}$ which is analogous to the encoding circuit of a multi-qubit surface code \cite{Kitaev1997,Fowler2012} and define an oscillator-into-oscillators GKP-surface code which can be implemented locally on a $2$-dimensional plane (see Refs. \cite{Vuillot2019,Noh2019b} for the qubit-into-oscillators toric-GKP code and the surface-GKP code). In the following subsection, we will discuss implementation of logical Gaussian operations for a general GKP-stabilizer code, while leaving the problem of finding useful GKP-stabilizer codes as a future research direction. 
   
\subsection{Logical Gaussian operations}
\label{subsection:Logical Gaussian operations}

Here, we discuss implementation of logical Gaussian operations. In general, for any GKP-stabilizer code with a Gaussian encoding circuit $\hat{U}_{G}^{\textrm{Enc}}$, a logical Gaussian operation $(\hat{U}_{G}^{\textrm{Gate}})_{L}$ can be readily implemented by using only physical Gaussian operations: $\hat{U}_{G}^{\textrm{Enc}}(\hat{U}_{G}^{\textrm{Gate}}\otimes \hat{U}_{G}^{\textrm{Aux}})(\hat{U}_{G}^{\textrm{Enc}})^{\dagger}$, where $\hat{U}_{G}^{\textrm{Gate}}$ is the Gaussian operation that we want to apply to the data modes and $\hat{U}_{G}^{\textrm{Aux}}$ is an auxiliary Gaussian operation to the ancilla modes (see Fig.\ \ref{fig:GKP-stabilizer codes and logical Gaussian operations}). Physical Gaussian operations are sufficient for the implementation of logical Gaussian operations because the input ancilla GKP state $|\textrm{GKP}\rangle^{\otimes N-M}$ is the only non-Gaussian resource in the encoding scheme and the remaining circuit $\hat{U}_{G}^{\textrm{Enc}}$ is Gaussian. We remark that auxiliary Gaussian operations can be used to simplify the form of the required physical Gaussian operation $\hat{U}_{G}^{\textrm{Enc}}(\hat{U}_{G}^{\textrm{Gate}}\otimes \hat{U}_{G}^{\textrm{Aux}})(\hat{U}_{G}^{\textrm{Enc}})^{\dagger}$ (see below for an illustration). 

Many interesting CV quantum information processing tasks such as the KLM protocol \cite{Knill2001}, boson sampling \cite{Aaronson2011} and simulation of vibrational quantum dynamics of molecules \cite{Huh2015,Sparrow2018} require Gaussian operations and, as non-Gaussian resources, input single- or multi-photon Fock states and photon number measurements of the output modes. Since the non-Gaussian resources are consumed locally only in the beginning and the end of the computation, any imperfections in these non-Gaussian resources can be addressed independently to the imperfections in the Gaussian operations. Below, we focus on improving the quality of Gaussian operations, beam splitter interactions specifically, having the KLM protocol and boson sampling in mind.  

Let us recall the GKP-two-mode-squeezing code introduced in Subsection \ref{subsection:The GKP-two-mode-squeezing code}. For simplicity, we consider two logical oscillator modes encoded in four physical oscillator modes via the encoding Gaussian circuit $\hat{U}_{G}^{\textrm{Enc}} = \textrm{TS}_{1,3}(G)\textrm{TS}_{2,4}(G)$. Here, the modes $1,2$ are the data modes and the modes $3,4$ are the ancilla modes which are used for error syndrome detection. Note that the encoding Gaussian circuit of the GKP-two-mode-squeezing code has the following nice property: 
\begin{align}
&\textrm{BS}_{1,2}(\eta)\textrm{BS}_{3,4}(\eta) = \hat{U}_{G}^{\textrm{Enc}}\textrm{BS}_{1,2}(\eta)\textrm{BS}_{3,4}(\eta) \big{(}\hat{U}_{G}^{\textrm{Enc}}\big{)}^{\dagger}, 
\label{eq:transversality GKP-two-mode-squeezing code}
\end{align}
where $\hat{U}_{G}^{\textrm{Enc}} = \textrm{TS}_{1,3}(G)\textrm{TS}_{2,4}(G)$. That is, by choosing $\hat{U}_{G}^{\textrm{Aux}} = \textrm{BS}_{3,4}(\eta)$, we can can implement the logical beam splitter interaction transversally, simply by applying a pair of beam splitter interactions $\textrm{BS}_{1,2}(\eta)$ and $\textrm{BS}_{3,4}(\eta)$ to the data modes and the ancilla modes (see Fig.\ \ref{fig:GKP-two-mode-squeezing code logical BS transversality} (a)).

We now discuss how the transversality in Eq. \eqref{eq:transversality GKP-two-mode-squeezing code} can be used to correct additive Gaussian noise errors that occur during the implementation of a beam splitter interaction. Note that any passive beam splitter interactions commute with iid additive Gaussian noise errors (see Appendix \ref{appendix:commutativity of beam splitter interactions and iid additive Gaussian noise errors} for more details). Therefore, the beam splitter interaction $\textrm{BS}_{1,2}(\eta)\textrm{BS}_{3,4}(\eta)$ continuously corrupted by iid additive Gaussian noise errors can be understood as the ideal beam splitter interaction $\textrm{BS}_{1,2}(\eta)\textrm{BS}_{3,4}(\eta)$ followed by an iid additive Gaussian noise channel $\bigotimes_{k=1}^{4} \mathcal{N}^{(k)}[\sigma]$ where the variance of the additive noise $\sigma^{2}$ is proportional to the time needed to complete the beam splitter interaction.

As shown in Fig.\ \ref{fig:GKP-two-mode-squeezing code logical BS transversality} (b), the uncorrelated additive Gaussian noise error that has been accumulated during the logical beam splitter interaction is converted into a correlated additive Gaussian noise error through the inverse of the encoding circuit $(\hat{U}_{G}^{\textrm{Enc}})^{\dagger} = \textrm{TS}_{1,3}(G)^{\dagger}\textrm{TS}_{2,4}(G)^{\dagger}$ in the decoding procedure: That is, after the noise reshaping, the quadrature noise in the data modes $1$ and $2$ are correlated with the ones in the ancilla modes $3$ and $4$, respectively. Thanks to the correlation between the data modes and the ancilla modes, we can learn about the reshaped data quadrature noise $z_{q/p}^{(1)}$ and $z_{q/p}^{(2)}$ solely by extracting information about the reshaped ancilla quadrature noise $z_{q/p}^{(3)}$ and $z_{q/p}^{(4)}$. This information can then be used to suppress the additive Gaussian noise accumulated in the data mode quadratically up to a small logarithmic correction, similarly as in Subsection \ref{subsection:The GKP-two-mode-squeezing code}.

Note that we have so far assumed ideal GKP states to clearly demonstrate the error-correcting capability of GKP-stabilizer codes. In the following section, we discuss experimental realization of GKP-stabilizer codes. Especially, we address issues related to the use of realistic noisy GKP states and discuss a key factor that should be taken into account in the design of practical GKP-stabilizer codes.

\section{Discussion}
\label{section:Discussion}

\begin{figure*}
\centering
\includegraphics[width=5.8in]{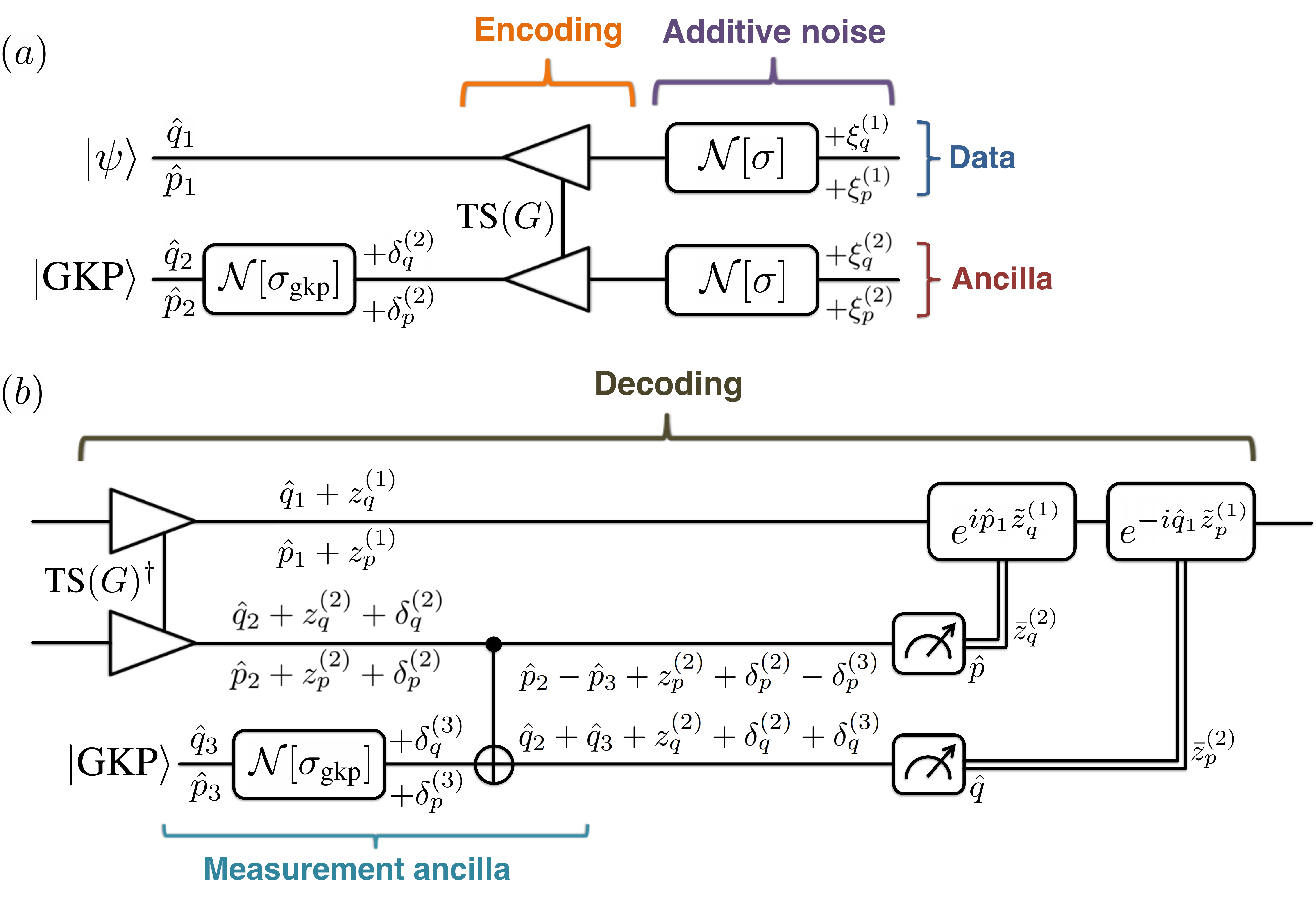}
\caption{(a) Encoding circuit of the GKP-two-mode-squeezing code subject to independent and identically distributed additive Gaussian noise errors. The input GKP state in the ancilla mode is assumed to be a noisy canonical GKP state with a noise standard deviation $\sigma_{\textrm{gkp}}$. (b) Decoding circuit of the GKP-two-mode-squeezing code. Note that the simultaneous position and momentum quadrature measurement modulo $\sqrt{2\pi}$ is implemented by using a third ancilla mode (i.e., the measurement ancilla mode) initialized to a noisy GKP state with a noise standard deviation $\sigma_{\textrm{gkp}}$.} 
\label{fig:GKP-two-mode-squeezing code with noisy GKP states}
\end{figure*}

Here, we discuss experimental realization of our GKP-stabilizer codes and the effects of realistic imperfections. Recall that the only required non-Gaussian resource for implementing GKP-stabilizer codes is the preparation of a canonical GKP state. While Gaussian resources are readily available in many realistic bosonic systems, preparing a canonical GKP state is not strictly possible because it would require infinite squeezing. Recently, however, finitely-squeezed approximate GKP states have been realized in a trapped ion system \cite{Fluhmann2018,Fluhmann2019,Fluhmann2019b} by using a heralded preparation scheme with post-selection \cite{Travaglione2002} and in a circuit QED system by using a deterministic scheme \cite{Terhal2016,Campagne2019}. Thus, our GKP-two-mode-squeezing code can in principle be implemented in the state-of-the-art quantum computing platforms.  

Imperfections in realistic GKP states such as finite squeezing will add additional quadrature noise to the system. Therefore in near-term experiments, the performance of the GKP-two-mode-squeezing code will be mainly limited by the finite squeezing of the approximate GKP states. Indeed, we show below that a non-trivial QEC gain $\sigma^{2} / (\sigma_{L}^{\star})^{2} > 1$ is achievable with the GKP-two-mode-squeezing code only when the supplied GKP states have a squeezing larger than $11.0$dB. On the other hand, the squeezing of the experimentally realized GKP states ranges from $5.5$dB to $9.5$dB \cite{Fluhmann2019,Campagne2019}. In this regard, we emphasize that our oscillator encoding scheme is compatible with \textit{non-deterministic} GKP state preparation schemes. This is because the required GKP states can be prepared offline and then supplied to the error correction circuit in the middle of the decoding procedure (similar to the magic state injection for the qubit-based universal quantum computation \cite{Bravyi2005}). Thus in near-term experiments, it will be more advantageous to sacrifice the success probability of the GKP state preparation schemes and aim to prepare a GKP state of higher quality (with a squeezing larger than the critical value $11.0$dB) by using post-selection.    

%we remark that the critical GKP squeezing value $11.0$dB is close to the fault-tolerant threshold values of the GKP squeezing (e.g., $11.2$dB for the surface-GKP code \cite{Noh2019b}) established in the context of qubit(s)-into-oscillators bosonic QEC.  

In general, the imperfections in GKP states may be especially detrimental to a GKP-stabilizer code involving a large squeezing parameter. This is because such imperfections may be significantly amplified by the large squeezing operations. With this concern in mind, let us revisit the GKP-two-mode-squeezing code in Subsection \ref{subsection:The GKP-two-mode-squeezing code} and recall that the optimal gain $G^{\star}$ is asymptotically given by $G^{\star} \propto 1/\sigma^{2}$ in the $\sigma\ll 1$ limit. Therefore, if the standard deviation of the input noise is very small, we indeed have a huge gain parameter $G^{\star} \gg 1$ (or $\lambda^{\star} = \sqrt{G^{\star}}+\sqrt{G^{\star}-1} \gg 1$). However, we explain in detail below that we have designed the GKP-two-mode-squeezing code very carefully so that any imperfections in GKP states are not amplified by the large squeezing operations. 

With these potential issues in mind, let us now analyze the adverse effects of the finite squeezing in a rigorous and quantitative manner. Recall that an approximate GKP state with a finite squeezing can be modeled by $|\textrm{GKP}_{\Delta}\rangle \propto \exp[-\Delta \hat{n}]|\textrm{GKP}\rangle$. As shown in Ref.\ \cite{Noh2019b}, one can convert the finitely-squeezed GKP state $|\textrm{GKP}_{\Delta}\rangle$ via a noise twirling into 
\begin{align}
\mathcal{N}[\sigma_{\textrm{gkp}}] ( |\textrm{GKP}\rangle\langle \textrm{GKP}| ), \label{eq:noisy GKP state after the twirling} 
\end{align}
i.e., an ideal canonical GKP state corrupted by an incoherent random shift error $\mathcal{N}[\sigma_{\textrm{gkp}}]$. Here, $\sigma_{\textrm{gkp}}^{2} = (1-e^{-\Delta}) / (1+e^{-\Delta})$ is the variance of the additive noise associated with the finite GKP squeezing. The noise standard deviation $\sigma_{\textrm{gkp}}$ characterizes the width of each peak in the Wigner function of an approximate GKP state. The GKP squeezing is then defined as $s_{\textrm{gkp}}\equiv -10\log_{10}(2\sigma_{\textrm{gkp}}^{2})$. The GKP squeezing $s_{\textrm{gkp}}$ quantifies how much an approximate GKP state is squeezed in both the position and the momentum quadrature in comparison to the vacuum noise variance $1/2$.  

In Fig.\ \ref{fig:GKP-two-mode-squeezing code with noisy GKP states}, we present the full circuit for the implementation of the GKP-two-mode-squeezing code. Note that the third mode (or the measurement mode) in the decoding scheme is introduced to simultaneously measure the position and momentum operators of the ancilla mode modulo $\sqrt{2\pi}$. That is, we consume one GKP state to perform the simultaneous and modular position and momentum measurements. 

We remark that we would have consumed two GKP states for the simultaneous and modular quadrature measurements if we were to use the measurement circuits in Fig.\ \ref{fig:GKP state and measurements} (b) (i.e., one for the modular position measurement and the other for the modular momentum measurement). While this scheme certainly works, it is not the most efficient strategy. This is because the measurement circuits in Fig.\ \ref{fig:GKP state and measurements} (b) are for \textit{non-destructive} measurements. While the first measurement (e.g., the modular position measurement) has to be performed in a non-destructive way, the following measurement (e.g., the modular momentum measurement) can be done in a destructive way since we no longer need the quantum state in the ancilla mode and instead only need the classical measurement outcomes $\bar{z}_{q}^{(2)}$ and $\bar{z}_{p}^{(2)}$. This is the reason why we simply measure the momentum quadrature of the second mode (i.e., the ancilla mode) in a destructive way after the modular position measurement (see Fig.\ \ref{fig:GKP-two-mode-squeezing code with noisy GKP states} (b)). Such a non-Gaussian resource saving is especially crucial in the regime where the finite squeezing of an approximate GKP state is the limiting factor.  

Thanks to the resource saving described above, we only need to supply two GKP states to implement the GKP-two-mode-squeezing code (one in the input of the ancilla mode and the other for the simultaneous and modular ancilla quadrature measurements). We assume that these two GKP states are corrupted by an additive Gaussian noise channel $\mathcal{N}[\sigma_{\textrm{gkp}}]$, i.e., $(\delta_{q}^{(2)},\delta_{p}^{(2)},\delta_{q}^{(3)},\delta_{p}^{(3)}) \sim_{\textrm{iid}}\mathcal{N}(0, \sigma_{\textrm{gkp}}^{2})$ (see Eq.\ \eqref{eq:noisy GKP state after the twirling} and Fig.\ \ref{fig:GKP-two-mode-squeezing code with noisy GKP states}). Due to this additional noise associated with the finite squeezing of the GKP states, the estimated reshaped ancilla quadrature noise in Eq. \eqref{eq:estimates of the ancilla quadrature noise two-mode GKP-squeezed-repetition code} is corrupted as follows: 
\begin{align}
\bar{z}_{q}^{(2)} &= R_{\sqrt{2\pi}}( z_{q}^{(2)} + \xi_{q}^{(\textrm{gkp})} ) , 
\nonumber\\
\bar{z}_{p}^{(2)} &= R_{\sqrt{2\pi}}( z_{p}^{(2)} + \xi_{p}^{(\textrm{gkp})} ) , \label{eq:estimated ancilla quadrature noise noisy GKP states}
\end{align}
Here, $\xi_{q}^{(\textrm{gkp})} \equiv\delta_{q}^{(2)} + \delta_{q}^{(3)} $ and $\xi_{p}^{(\textrm{gkp})} \equiv \delta_{p}^{(2)} - \delta_{p}^{(3)}$ are the additional noise due to the finite GKP squeezing and follow $(\xi_{q}^{(\textrm{gkp})}, \xi_{p}^{(\textrm{gkp})} ) \sim_{\textrm{iid}} \mathcal{N}( 0, 2\sigma_{\textrm{gkp}}^{2} )$. Such additional noise will then be propagated to the data mode through the miscalibrated counter displacement operations based on noisy estimates. In the presence of additional GKP noise, the sizes of the optimal counter displacements $\exp[i\hat{p}_{1}\tilde{z}_{q}^{(1)}]$ and $\exp[-i\hat{q}_{1}\tilde{z}_{p}^{(1)}]$ are given by 
\begin{align}
\tilde{z}_{q}^{(1)} &=  - \frac{2\sqrt{G(G-1)} \sigma^{2} }{(2G-1)\sigma^{2} + 2\sigma_{\textrm{gkp}}^{2} } \bar{z}_{q}^{(2)} \xrightarrow{G\gg 1} -\bar{z}_{q}^{(2)}, 
\nonumber\\
\tilde{z}_{p}^{(1)} &=   \frac{2\sqrt{G(G-1)} \sigma^{2} }{(2G-1)\sigma^{2} + 2\sigma_{\textrm{gkp}}^{2} } \bar{z}_{p}^{(2)} \xrightarrow{G\gg 1} \bar{z}_{p}^{(2)} 
\end{align}
and do not explicitly depend on $G$ in the $G \gg 1$ limit (see Eq. \eqref{eq:estimates of the data quadrature noise two-mode GKP-squeezed-repetition code}). Therefore, the additional GKP noise $\xi_{q}^{(\textrm{gkp})}$ and $\xi_{p}^{(\textrm{gkp})}$ will simply be added to the data quadrature operators \textit{without} being amplified by the large gain parameter $G \gg 1$. This absence of the noise amplification is a critically important feature of our scheme and is generally not the case for a generic GKP-stabilizer code involving large squeezing operations (see Appendix \ref{appendix:Generalized GKP-squeezed-repetition codes} for an illustration). 

\begin{figure*}[t!]
\centering
\includegraphics[width=6.8in]{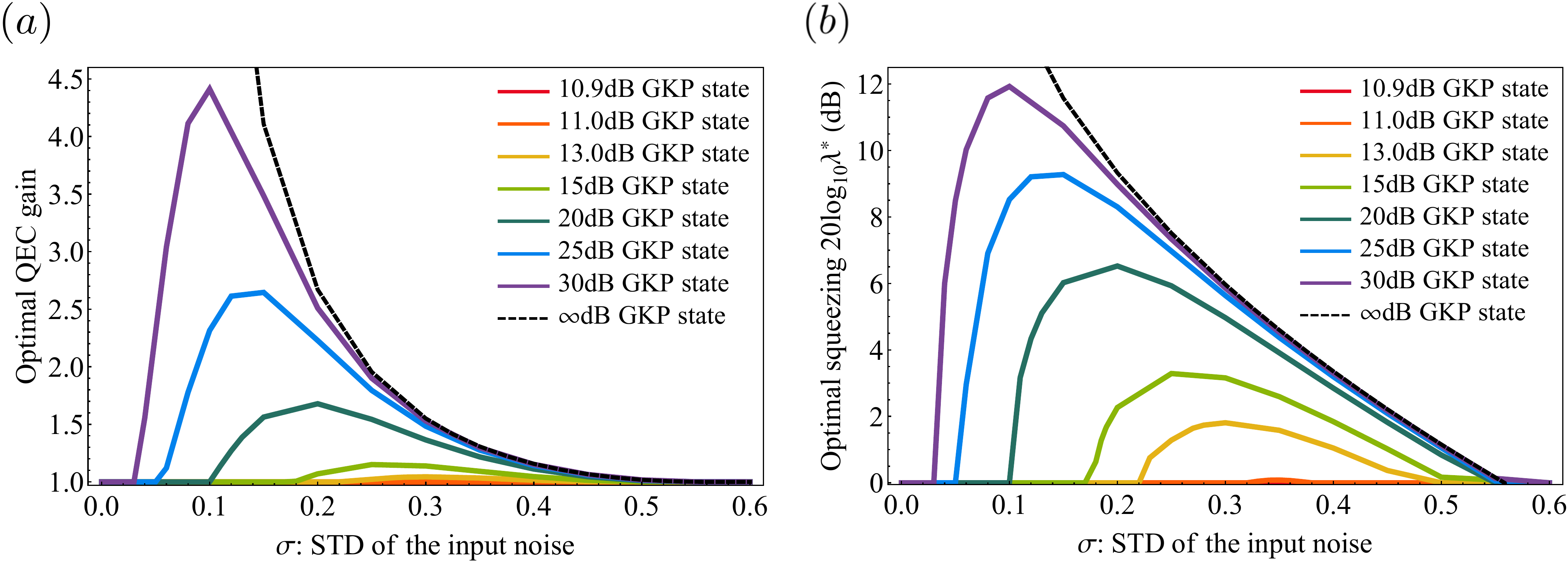}
\caption{(a) The optimal QEC gain $\sigma^{2} / (\sigma_{L}^{\star})^{2}$ as a function of input noise standard deviation $\sigma$ for various values of the GKP squeezing $s_{\textrm{gkp}} = -10\log_{10}(2\sigma_{\textrm{gkp}}^{2})$ ranging from $12.8$dB to $30$dB and (b) the optimal two-mode squeezing gain $G^{\star}$, translated to the required single-mode squeezing in the unit of decibel $20\log_{10}\lambda^{\star}$ where $\lambda^{\star} \equiv \sqrt{G^{\star}} +\sqrt{G^{\star}-1 }$. The non-trivial QEC gain $\sigma^{2} / (\sigma_{L}^{\star})^{2}>1$ is achievable only when the squeezing of the supplied approximate GKP states is larger than the critical squeezing $11.0$dB. The dashed black lines represent the asymptotic results for the infinitely squeezed canonical GKP states which are shown in Fig.\ \ref{fig:performance of the two-mode GKP-squeezed-repetition code}.  }
\label{fig:performance of the two-mode GKP-two-mode-squeezing code noisy GKP states}
\end{figure*}

In Appendix \ref{appendix:Probability density of the logical quadrature noise: The two-mode GKP-squeezed-repetition code}, we provide a through analysis of the adverse effects of the finitely-squeezed GKP states. In particular, we derive the variance of the output logical quadrature noise $(\sigma_{L})^{2}$ as a function of the input noise standard deviation $\sigma$, the GKP noise standard deviation $\sigma_{\textrm{gkp}}$, and the gain of the two-mode squeezing $G$ (see Eq. \eqref{eq:variance of the output logical quadrature noise noisy GKP states better}). Similarly as in Subsection \ref{subsection:The GKP-two-mode-squeezing code}, we optimize the gain $G$ of the two-mode squeezing to minimize the output logical noise standard deviation $\sigma_{L}$ for given $\sigma$ and $\sigma_{\textrm{gkp}}$.   

In Fig.\ \ref{fig:performance of the two-mode GKP-two-mode-squeezing code noisy GKP states} (a), we plot the maximum achievable QEC gain $\sigma^{2} / (\sigma_{L}^{\star})^{2}$ as a function of the input noise standard deviation $\sigma$ for various values of the GKP squeezing ranging from $10.9$dB to $30$dB. In Fig.\ \ref{fig:performance of the two-mode GKP-two-mode-squeezing code noisy GKP states} (b), we plot the optimal gain $G^{\star}$ of the two-mode squeezing, translated to the required single-mode squeezing in the unit of decibel. We observe that the non-trivial QEC gain $\sigma^{2} / (\sigma_{L}^{\star})^{2}>1$ can be achieved only when the supplied GKP states have a squeezing larger than the critical value $11.0$dB. Also, when the supplied GKP states have a squeezing of $30$dB, the maximum QEC gain is given by $\sigma^{2} / (\sigma_{L}^{\star})^{2} = 4.41$, which is achieved when $\sigma = 0.1$. For comparison, the QEC gain at the same input noise standard deviation $\sigma=0.1$ is given by $\sigma^{2} / (\sigma_{L}^{\star})^{2} = 7.7$ when the ideal canonical GKP states are used to implement the GKP-two-mode-squeezing code (see Subsection \ref{subsection:The GKP-two-mode-squeezing code}). The fact that these two values ($4.41$ verses $7.7$) are close to each other is an indicative of the fact that the additional GKP noise is not catastrophically amplified by the large ($12.3$dB) single-mode squeezing operations needed in this regime.    

%The avoidance of the GKP noise amplification is essential in the design of practical GKP-stabilizer codes that can be experimentally realized.  

Lastly, we outline several open questions. First, note that our scheme works with imperfect GKP states but is not truly fault tolerant, in the sense that the added noise due to the finitely-squeezed GKP states is not suppressed to an arbitrarily small value. 
%That is, we have found that the added noise due to GKP imperfections is not amplified.  
An important open question is whether a fault-tolerant scheme can be found in which the effects of added noise due to non-ideal GKP ancilla states can be made arbitrarily small.
%continuing the discussion of imperfect GKP states, it remains to be answered if there exists a family of fault-tolerant GKP-stabilizer codes where the added noise from GKP states mentioned above can be suppressed arbitrarily as the system size increases when the imperfections are below a certain noise threshold. 
Moreover, it will be an interesting research avenue to search for a family of efficient GKP-stabilizer codes by exploring various encoding circuits $\hat{U}_{G}^{\textrm{Enc}}$. For example, one can look for GKP-stabilizer codes that can be implemented locally in a low dimensional space, or for ones with low resource overheads, or ones with high fault-tolerant threshold, if any. 
%a family of GKP-stabilizer codes that can be implemented locally on a $1$-dimensional line or a $2$-dimensional plane. 
In addition, while logical Gaussian operations can be readily implemented by using only Gaussian operations for any GKP-stabilizer codes, implementation of logical non-Gaussian operations will require some non-Gaussian resources. Thus, it will also be interesting to explore whether logical non-Gaussian operations can be implemented efficiently by using, e.g., GKP states or cubic phase states as non-Gaussian resources.

%we remark that the quality (e.g., size and fidelity) of the canonical GKP state will be the dominant imperfection in the experimental implementation of our GKP-repetition or GKP-stabilizer codes. However, such GKP states can be prepared offline and then be supplied to the error correction circuit during the correction procedure (see Fig. \ref{fig:GKP state and measurements} (b)--(d)). Therefore, deterministic preparation of GKP states is not necessary and instead heralded preparation using post-selection is sufficient. Thus, our general GKP-stabilizer QEC schemes are compatible with the heralded GKP state generation scheme recently realized in a trapped ion system \cite{Travaglione2002,Fluhmann2018,Fluhmann2019}, as well as the deterministic GKP state generation scheme realized in circuit QED systems \cite{Terhal2016,Touzard2019}.    

\section{Conclusion}
\label{section:Conclusion}

We have worked around the previous no-go theorems on Gaussian QEC schemes \cite{Eisert2002,Niset2009,Vuillot2019} and proposed several non-Gaussian oscillator-into-oscillators codes, i.e., the two-mode GKP-repetition code and the GKP-two-mode-squeezing code, that can correct additive Gaussian noise errors. We generalized them to an even broader class of non-Gaussian oscillator codes, namely, GKP-stabilizer codes. In particular, we explicitly constructed a family of $N$-mode GKP-stabilizer codes that can efficiently suppress additive Gaussian noise errors to the $N^{\textrm{th}}$ order. We also showed that our proposed QEC schemes can also correct excitation loss and thermal noise errors as well as additive Gaussian noise errors by a suitable noise conversion through a quantum-limited amplification channel. The only required non-Gaussian resource for our GKP-stabilizer QEC schemes is the preparation of the canonical GKP state. We showed that, for any GKP-stabilizer codes, logical Gaussian operations can be readily implemented by using only physical Gaussian operations. Therefore, our GKP-stabilizer QEC schemes will be useful for realizing error-corrected boson sampling and simulation of bosonic systems. In addition, our GKP-stabilizer QEC schemes may also be able to suppress errors for quantum metrology with bosonic sensors. Indeed a recent follow-up work \cite{Zhuang2019} has theoretically demonstrated that the GKP-two-mode-squeezing code can be used to enhance the robustness of CV distributed sensing protocols. Moreover, our oscillator codes may remove the need for CV-DV-CV conversion for quantum communication over bosonic channels. Lastly, our oscillator encoding schemes can be useful for overcoming loss errors in transduction protocols \cite{Zhang2018,Lau2019} where a microwave-frequency bosonic mode should be transferred to an optical-frequency bosonic mode.

\section*{Acknowledgments}
\label{section:Acknowledgments}

%\textit{Acknowledgment--}
We would like to thank Rob Schoelkopf for helpful discussions. We acknowledge support from the ARL-CDQI (W911NF-15-2-0067, W911NF-18-2-0237), ARO (W911NF-18-1-0020, W911NF-18-1-0212), ARO MURI (W911NF-16-1-0349 ), AFOSR MURI (FA9550-14-1-0052, FA9550-15-1-0015), DOE (DE-SC0019406), NSF (EFMA-1640959, DMR-1609326), and the Packard Foundation (2013-39273). K.N. acknowledges support through the Korea Foundation for Advanced
Studies. 

\appendix

\section{Gaussian-repetition codes}
\label{appendix:Gaussian-repetition codes}

Here, we introduce and analyze the $N$-mode Gaussian-repetition code, which is a straightforward generalization of the three-mode Gaussian-repetition code introduced in Refs. \cite{Lloyd1998,Braunstein1998}. In particular, we introduce the maximum likelihood estimation of the data position quadrature noise $\xi_{q}^{(1)}$ for the $N$-mode Gaussian-repetition code (Eq. \eqref{supp_eq:estimation of position noise Gaussian repetition codes}), which is a key motivation behind our choice of the estimate $\tilde{\xi}_{q}^{(1)}$ in Eq. \eqref{eq:GKP-rep estimates for quadrature noise}.

\begin{figure*}[t!]
\centering
\includegraphics[width=6.8in]{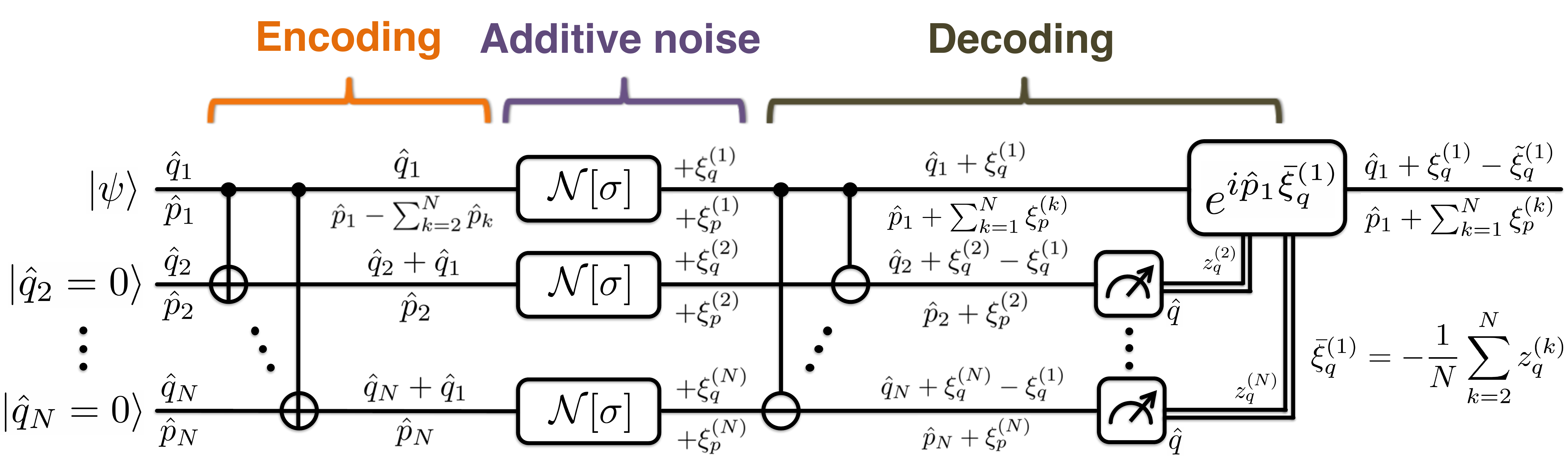}
\caption{Encoding and decoding circuits of the $N$-mode Gaussian-repetition code, subject to independent and identically distributed additive Gaussian noise errors.}
\label{supp_fig:Gaussian repetition codes}
\end{figure*} 

Consider an arbitrary oscillator state $|\psi\rangle = \int dq \psi(q)|\hat{q}_{1}=q\rangle$. The state $|\psi\rangle$ can be embedded in $N$ oscillator modes via the $N$-mode Gaussian-repetition code as follows: 
\begin{equation}
|\psi_{L}\rangle = \int dq \psi(q) \bigotimes_{k=1}^{N} |\hat{q}_{k}= q\rangle, 
\end{equation}
where the first mode is the data mode and the rest are the ancilla modes. Note that the data position eigenstate $|\hat{q}_{1}=q\rangle$ is mapped into $\bigotimes_{k=1}^{N} |\hat{q}_{k}= q\rangle$ through the encoding procedure. As shown in Fig.\ \ref{supp_fig:Gaussian repetition codes}, this encoding can be realized by applying a sequence of Gaussian $\textrm{SUM}$ gates $\textrm{SUM}_{1\rightarrow k}\equiv \exp[-i\hat{q}_{1}\hat{p}_{k}]$ where $k\in \lbrace 2,\cdots, N\rbrace$. Upon the encoding circuit, the quadrature operators are transformed into 
\begin{alignat}{2}
&\hat{q}_{1}  \rightarrow \hat{q}'_{1}\equiv \hat{q}_{1},\quad && \hat{p}_{1}\rightarrow \hat{p}'_{1}\equiv  \hat{p}_{1} - \sum_{k=2}^{N}\hat{p}_{k} , 
\nonumber\\
&\hat{q}_{k} \rightarrow \hat{q}'_{k} \equiv \hat{q}_{k}+\hat{q}_{1}, \quad && \hat{p}_{k} \rightarrow \hat{p}'_{k}\equiv \hat{p}_{k}, \label{eq:quadrature transformation due to encoding circuit Gaussian repetition}
\end{alignat}
as in Eq. \eqref{eq:quadrature transformation due to encoding circuit} where $k\in\lbrace 2,\cdots, N \rbrace$. We then assume that the oscillator modes undergo independent and identically distributed additive Gaussian noise errors $\mathcal{N} = \bigotimes_{k=1}^{N} \mathcal{N}^{(k)}[\sigma]$, i.e., 
\begin{align}
\hat{q}'_{k}\rightarrow \hat{q}''_{k} \equiv \hat{q}'_{k} + \xi_{q}^{(k)} \,\,\,\textrm{and}\,\,\, \hat{p}'_{k}\rightarrow \hat{p}''_{k} \equiv \hat{p}'_{k} + \xi_{p}^{(k)}, 
\end{align}
as in Eq. \eqref{eq:quadrature transformation due to added noise}. The added noise $\xi_{q/p}^{(1)},\cdots,\xi_{q/p}^{(N)}$ follow an independent and identically distributed Gaussian random distribution $(\xi_{q}^{(1)},\xi_{p}^{(1)},\cdots,\xi_{q}^{(N)},\xi_{p}^{(N)})\sim_{\textrm{iid}}\mathcal{N}(0,\sigma^{2})$.

The goal of the decoding procedure (shown in Fig.\ \ref{supp_fig:Gaussian repetition codes}) is to extract some information about the added noise $\xi_{q}^{(k)}$ and $\xi_{p}^{(k)}$ ($k\in\lbrace 1,\cdots,N \rbrace$) through a set of syndrome measurements. The decoding procedure begins with the inverse of the encoding circuit, i.e., by a sequence of inverse-$\textrm{SUM}$ gates $\textrm{SUM}^{\dagger}_{1\rightarrow k} \equiv \exp[i\hat{q}_{1}\hat{p}_{k}]$ for $k\in\lbrace 2,\cdots, N\rbrace$. Upon the inverse of the encoding circuit, the quadrature operators are transformed into $\hat{q}''_{k} \rightarrow \hat{q}_{k} + z_{q}^{(k)}$ and $\hat{p}''_{k} \rightarrow \hat{p}_{k} + z_{p}^{(k)}$, where the reshaped quadrature noise is given by
\begin{alignat}{2}
z_{q}^{(1)}&\equiv  \xi_{q}^{(1)},\qquad  && z_{p}^{(1)}\equiv \sum_{k=1}^{N}\xi_{p}^{(k)},
\nonumber\\
z_{q}^{(k)}&\equiv \xi_{q}^{(k)}-\xi_{q}^{(1)},\qquad  &&  z_{p}^{(k)}\equiv \xi_{p}^{(k)} ,  \label{eq:noise quadrature transformation Gaussian repetition}
\end{alignat}
as in Eq. \eqref{eq:noise quadrature transformation} where $k\in\lbrace 2,\cdots, N \rbrace$. Then, by performing homodyne measurements of the ancilla position quadrature operators, we can exactly extract the values of 
\begin{align}
z_{q}^{(k)} = \xi_{q}^{(k)}-\xi_{q}^{(1)} 
\end{align}
for all $k\in\lbrace 2,\cdots,N \rbrace$. This is because the ancilla modes are initially in the position eigenstates $|\hat{q}_{k}=0\rangle$ and thus measuring $\hat{q}'''_{k}  =\hat{q}_{k}+z_{q}^{(k)}$ is equivalent to measuring $z_{q}^{(k)}$ for all $k\in\lbrace 2,\cdots, N\rbrace$. Note, however, that we cannot extract any information about the reshaped momentum quadrature noise $z_{p}^{(1)},\cdots,z_{p}^{(N)}$. This will later turn out to be the key limitation of Gaussian-repetition codes.

From the extracted values of $z_{q}^{(k)} = \xi_{q}^{(k)}-\xi_{q}^{(1)}$, we can infer that position quadrature noise is given by $\vec{\xi}_{q}\equiv (\xi_{q}^{(1)},\xi_{q}^{(2)},\cdots,\xi_{q}^{(N)}) =  (\xi_{q}^{(1)},\xi_{q}^{(1)}+z_{q}^{(2)},\cdots,\xi_{q}^{(1)}+z_{q}^{(N)})$. Then, the undetermined data position quadrature noise $\xi_{q}^{(1)}$ can be estimated by a maximum likelihood estimation: Since noise with smaller $|\vec{\xi}_{q}|^{2} \equiv \sum_{k=1}^{N}(\xi_{q}^{(k)})^{2}$ are more likely, we estimate that $\xi_{q}^{(1)}$ is 
\begin{align}
\bar{\xi}_{q}^{(1)}
&= \textrm{argmin}_{\xi_{q}^{(1)}}\Big{[} (\xi_{q}^{(1)})^{2} +\sum_{k=2}^{N} (\xi_{q}^{(1)}+z_{q}^{(k)})^{2} \Big{]} 
\nonumber\\
&=  -\frac{1}{N}\sum_{k=2}^{N}z_{q}^{(k)} \label{supp_eq:estimation of position noise Gaussian repetition codes}
\end{align}
from the syndrome measurement outcomes $z_{q}^{(2)},\cdots,z_{q}^{(N)}$. Note that for $N=2$, Eq. \eqref{supp_eq:estimation of position noise Gaussian repetition codes} reduces to $\bar{\xi}_{q}^{(1)} = -(z_{q}^{(1)}+z_{q}^{(2)})/2$: This is the main reason why we chose $\tilde{\xi}_{q}^{(1)}$ in Eq. \eqref{eq:GKP-rep estimates for quadrature noise} as the estimate of $\xi_{q}^{(1)}$. Finally, the decoding procedure ends with an application of the counter displacement operation $\exp[i\hat{p}_{1}\bar{\xi}_{q}^{(1)}]$ to the data oscillator mode (see Fig.\ \ref{supp_fig:Gaussian repetition codes}).

As a result of the encoding and decoding procedures, we end up with a logical additive noise error $\hat{q}_{1}\rightarrow \hat{q}_{1} + \xi_{q}$ and $\hat{p}_{1}\rightarrow \hat{p}_{1} + \xi_{p}$ of the data oscillator mode, where the output logical position and momentum quadrature noise are given by 
\begin{align}
\xi_{q} &\equiv z_{q}^{(1)} - \bar{\xi}_{q}^{(1)} = \frac{1}{N}\sum_{k=1}^{N}\xi_{q}^{(k)}, 
\nonumber\\
\xi_{p}&\equiv z_{p}^{(1)} = \sum_{k=1}^{N}\xi_{p}^{(k)}. 
\end{align}
Since $(\xi_{q}^{(1)},\xi_{p}^{(1)},\cdots,\xi_{q}^{(N)},\xi_{p}^{(N)})\sim_{\textrm{iid}}\mathcal{N}(0,\sigma^{2})$, we have
\begin{align}
\xi_{q} &= \frac{1}{N}\sum_{k=1}^{N}\xi_{q}^{(k)} \sim \mathcal{N}\Big{(}0,\sigma_{q}^{2}\equiv\frac{1}{N}\sigma^{2}\Big{)}, 
\nonumber\\
\xi_{p}&= \sum_{k=1}^{N}\xi_{p}^{(k)} \sim \mathcal{N} \Big{(} 0, \sigma_{p}^{2}\equiv N \sigma^{2} \Big{)} . \label{supp_eq:Gaussian repetition code doesnt help}
\end{align}
Thus, the variance of the position quadrature noise is reduced by a factor of $N$, but the variance of the momentum quadrature noise is increased by the same factor. The latter increase is due to the fact that the ancilla momentum quadrature noise which are transferred to the data momentum quadrature (see $+\sum_{k=2}^{N}\xi_{p}^{(k)}$ in $z_{p}^{(1)}$) are left completely undetected by the position homodyne measurements during the syndrome extraction stage. As a result, the product of the noise standard deviations remains unchanged at the end of the error correction procedure (i.e., $\sigma_{q}\sigma_{p} = \sigma^{2}$). This implies that Gaussian-repetition codes can only squeeze the Gaussian quadrature noise but cannot actually correct them. This reaffirms the previous no-go results \cite{Eisert2002,Niset2009,Vuillot2019}.

In Section \ref{section:GKP-repetition codes} in the main text, we modify the two-mode Gaussian-repetition code (i.e., $N=2$) and introduce the two-mode GKP-repetition code that can indeed correct additive Gaussian noise errors. Specifically, we replace several Gaussian elements in the Gaussian-repetition code by non-Gaussian ones involving the canonical GKP state, such that we can only benefit from the decreased position noise variance by a factor of $N$, while preventing the momentum noise variance from increasing by the same factor.

\section{Supplementary material for the two-mode GKP-repetition code}
\label{appendix:Probability density of the logical quadrature noise: GKP-repetition codes}

Here, we provide explicit expressions for the probability density functions of the logical quadrature noise $\xi_{q}$ and $\xi_{p}$ for the two-mode GKP-repetition code. Recall Eq. \eqref{eq:logical quadrature noise} and note that the logical quadrature noise $\xi_{q}$ and $\xi_{p}$ for the two-mode GKP-repetition are given by 
\begin{align}
\xi_{q} & = \xi_{q}^{(1)} + \frac{1}{2}R_{\sqrt{2\pi}}(\xi_{q}^{(2)}-\xi_{q}^{(1)}) ,
\nonumber\\
\xi_{p} & =  \xi_{p}^{(1)} +\xi_{p}^{(2)}-R_{\sqrt{2\pi}}(\xi_{p}^{(2)}), 
\end{align}
where $(\xi_{q}^{(1)},\xi_{p}^{(1)},\xi_{q}^{(2)},\xi_{p}^{(2)})\sim_{\textrm{iid}} \mathcal{N}(0,\sigma^{2})$ and $R_{s}(z)\equiv z-n^{\star}(z)s$ and $n^{\star}(z)\equiv \textrm{argmin}_{n\in\mathbb{Z}}|z-ns|$. Let $p_{\sigma}(z)$ denote the probability density function of a Gaussian distribution with zero mean and variance $\sigma^{2}$, i.e., $p_{\sigma}(z)\equiv (1/\sqrt{2\pi\sigma^{2}})\exp[-z^{2}/(2\sigma^{2})]$. Then, the probability density functions $Q(\xi_{q})$ and $P(\xi_{p})$ of the logical quadrature noise $\xi_{q}$ and $\xi_{p}$ are given by   
\begin{align}
Q(\xi_{q})&\equiv \int_{-\infty}^{\infty}d\xi_{q}^{(1)}\int_{-\infty}^{\infty}d\xi_{q}^{(2)} p_{\sigma}(\xi_{q}^{(1)})p_{\sigma}(\xi_{q}^{(2)}) 
\nonumber\\
&\times \delta\Big{(} \xi_{q}-\xi_{q}^{(1)} - \frac{1}{2}R_{\sqrt{2\pi}}(\xi_{q}^{(2)}-\xi_{q}^{(1)}) \Big{)} ,  \label{supp_eq:probability density functions of logical position noise}
\end{align}
and
\begin{align}
P(\xi_{p})&\equiv \int_{-\infty}^{\infty}d\xi_{p}^{(1)}\int_{-\infty}^{\infty}d\xi_{p}^{(2)} p_{\sigma}(\xi_{p}^{(1)})p_{\sigma}(\xi_{p}^{(2)})
\nonumber\\
&\quad\times  \delta\Big{(} \xi_{p} -\xi_{p}^{(1)} - \xi_{p}^{(2)}+R_{\sqrt{2\pi}}(\xi_{p}^{(2)}) \Big{)}, \label{supp_eq:probability density functions of logical momentum noise}
\end{align}
where $\delta(x)$ is the Dirac delta function. Note that $R_{s}(z)$ can be expressed as
\begin{align}
R_{s}(z) \equiv \sum_{n\in\mathbb{Z}} (z-ns)\cdot  I\Big{\lbrace} z\in\Big{[}\Big{(}n-\frac{1}{2}\Big{)}s,\Big{(}n+\frac{1}{2}\Big{)}s\Big{]} \Big{\rbrace},  \label{supp_eq:identity on Rs function}
\end{align}
where $I\lbrace C \rbrace$ is an indicator function, i.e., $I\lbrace C \rbrace=1$ if $C$ is true $I\lbrace C \rbrace=0$ if $C$ is false. Then, using Eq. \eqref{supp_eq:identity on Rs function}, we can make $Q(\xi_{q})$ and $P(\xi_{p})$ in Eqs. \eqref{supp_eq:probability density functions of logical position noise},\eqref{supp_eq:probability density functions of logical momentum noise} more explicit as follows: 
\begin{widetext}
\begin{align}
Q(\xi_{q})&= \sum_{n_{2}\in\mathbb{Z}} \int_{-\infty}^{\infty}d\xi_{q}^{(1)}\int_{-\infty}^{\infty}d\xi_{q}^{(2)} p_{\sigma}(\xi_{q}^{(1)})p_{\sigma}(\xi_{q}^{(2)})  \delta\Big{(} \xi_{q}-\xi_{q}^{(1)} -\frac{1}{2} \big{(} \xi_{q}^{(2)}-\xi_{q}^{(1)} -\sqrt{2\pi}n_{2} \big{)} \Big{)}
\nonumber\\
&\qquad\qquad\qquad\qquad\qquad\qquad\qquad\quad\times  I\Big{\lbrace} \xi_{q}^{(2)}-\xi_{q}^{(1)}\in\Big{[}\Big{(}n_{2}-\frac{1}{2}\Big{)}\sqrt{2\pi},\Big{(}n_{2}+\frac{1}{2}\Big{)}\sqrt{2\pi}\Big{]} \Big{\rbrace} 
\nonumber\\
&= \sum_{n_{2}\in\mathbb{Z}} \int_{-\sqrt{\pi/2}}^{\sqrt{\pi/2}}d\xi_{q}^{(2)} p_{\sigma}\Big{(} \xi_{q} -\frac{1}{2} \xi_{q}^{(2)} \Big{)}  p_{\sigma}\Big{(} \xi_{q} +\frac{1}{2} \xi_{q}^{(2)} +\sqrt{2\pi}n_{2} \Big{)} ,  \label{supp_eq:Prob density logical position}
\end{align}
and 
\begin{align}
P(\xi_{p})&= \sum_{n_{2}\in\mathbb{Z}} \int_{-\infty}^{\infty}d\xi_{p}^{(1)}\int_{-\infty}^{\infty}d\xi_{p}^{(2)} p_{\sigma}(\xi_{p}^{(1)})p_{\sigma}(\xi_{p}^{(2)})   \delta\Big{(} \xi_{p} - \xi_{p}^{(1)} - \sqrt{2\pi}  n_{2} \Big{)}
\nonumber\\
&\qquad\qquad\qquad\qquad\qquad\qquad\times  I\Big{\lbrace} \xi_{p}^{(2)}\in\Big{[}\Big{(}n_{2}-\frac{1}{2}\Big{)}\sqrt{2\pi},\Big{(}n_{2}+\frac{1}{2}\Big{)}\sqrt{2\pi}\Big{]} \Big{\rbrace}  
\nonumber\\
&= \sum_{n_{2}\in\mathbb{Z}} \Big{[} \int_{\sqrt{2\pi}(n_{2}-\frac{1}{2})}^{\sqrt{2\pi}(n_{2}+\frac{1}{2})} d\xi_{p}^{(2)} \cdot  p_{\sigma}(\xi_{p}^{(2)})  \Big{]} p_{\sigma}\Big{(} \xi_{p}-\sqrt{2\pi}n_{2}\Big{)} .\label{supp_eq:Prob density logical momentum} 
\end{align}
\end{widetext}

Fig.\ \ref{fig:performance of GKP repetition codes} in the main text is obtained by numerically computing these probability density functions and then evaluating the standard deviations of the obtained probability density functions.

\section{Supplementary material for the GKP-two-mode-squeezing code}
\label{appendix:Probability density of the logical quadrature noise: The two-mode GKP-squeezed-repetition code}

Here, we first explain the underlying reasons behind our choice of the estimates $\tilde{z}_{q}^{(1)}$ and $\tilde{z}_{p}^{(1)}$ in Eq. \eqref{eq:estimates of the data quadrature noise two-mode GKP-squeezed-repetition code} for the GKP-two-mode-squeezing code. Recall that the covariance matrix of the reshaped noise $\boldsymbol{z} = (z_{q}^{(1)},z_{p}^{(1)},z_{q}^{(2)},z_{p}^{(2)})^{T}$ is given by 
\begin{align}
V_{\boldsymbol{z}} = \sigma^{2}\begin{bmatrix}
(2G-1)\boldsymbol{I}&-2\sqrt{G(G-1)}\boldsymbol{Z}\\
-2\sqrt{G(G-1)}\boldsymbol{Z}&(2G-1)\boldsymbol{I}
\end{bmatrix}. 
\end{align}
For now, let us ignore the fact that we can measure $z_{q}^{(2)}$ and $z_{p}^{(2)}$ only modulo $\sqrt{2\pi}$ and instead assume that we know their exact values. Note that $z_{q}^{(1)}$ is only correlated with $z_{q}^{(2)}$, whereas $z_{p}^{(1)}$ is only correlated with $z_{p}^{(2)}$. Consider the estimates of the form $\bar{z}_{q}^{(1)} = c_{q}z_{q}^{(2)}$ and $\bar{z}_{p}^{(1)} = c_{p}z_{p}^{(2)}$, where $c_{q}$ and $c_{p}$ are constants. We choose $c_{q}$ and $c_{p}$ such that the variances of $z_{q}^{(1)}-\bar{z}_{q}^{(1)}$ and $z_{p}^{(1)}-\bar{z}_{p}^{(1)}$ are minimized: Since $\textrm{Var}(z_{q}^{(1)}-\bar{z}_{q}^{(1)})$ and $\textrm{Var}(z_{p}^{(1)}-\bar{z}_{p}^{(1)})$ are given by
\begin{align}
\textrm{Var}(z_{q}^{(1)}-\bar{z}_{q}^{(1)}) &= \textrm{Var}(z_{q}^{(1)}) - 2c_{q} \cdot \textrm{Cov}(z_{q}^{(1)},z_{q}^{(2)}) 
\nonumber\\
&\quad  + c_{q}^{2} \textrm{Var}(z_{q}^{(2)}) , 
\nonumber\\
\textrm{Var}(z_{p}^{(1)}-\bar{z}_{p}^{(1)}) &= \textrm{Var}(z_{p}^{(1)}) - 2c_{p} \cdot \textrm{Cov}(z_{p}^{(1)},z_{p}^{(2)}) 
\nonumber\\
&\quad  + c_{p}^{2} \textrm{Var}(z_{p}^{(2)}),  
\end{align} 
they are minimized when 
\begin{align}
c_{q} &= \frac{\textrm{Cov}(z_{q}^{(1)},z_{q}^{(2)})}{\textrm{Var}(z_{q}^{(2)})} = -\frac{2\sqrt{G(G-1)}}{2G-1} ,
\nonumber\\
c_{q} &= \frac{\textrm{Cov}(z_{p}^{(1)},z_{p}^{(2)})}{\textrm{Var}(z_{p}^{(2)})} = \frac{2\sqrt{G(G-1)}}{2G-1}. 
\end{align} 
Therefore, if both $z_{q}^{(2)}$ and $z_{p}^{(2)}$ are precisely known, the optimal estimates of $z_{q}^{(1)}$ and $z_{p}^{(1)}$ are given by 
\begin{align}
\bar{z}_{q}^{(1)} &=  -\frac{2\sqrt{G(G-1)}}{2G-1} z_{q}^{(2)}, 
\nonumber\\
\bar{z}_{p}^{(1)} &=  \frac{2\sqrt{G(G-1)}}{2G-1} z_{p}^{(2)}. 
\end{align}  
Since, however, we can only measure $z_{q}^{(2)}$ and $z_{p}^{(2)}$ modulo $\sqrt{2\pi}$, we replace $z_{q}^{(2)}$ and $z_{p}^{(2)}$ by $\bar{z}_{q}^{(2)} = R_{\sqrt{2\pi}}(z_{q}^{(2)})$ and $\bar{z}_{p}^{(2)} = R_{\sqrt{2\pi}}(z_{p}^{(2)})$ and get the estimates $\tilde{z}_{q}^{(1)}$ and $\tilde{z}_{p}^{(1)}$ in Eq. \eqref{eq:estimates of the data quadrature noise two-mode GKP-squeezed-repetition code}.

Now we provide explicit expression for the probability density functions of the logical quadrature noise $\xi_{q}$ and $\xi_{p}$ for the GKP-two-mode-squeezing code. Recall Eq. \eqref{eq:logical quadrature noise two-mode GKP-squeezed-repetition code}: 
\begin{align}
\xi_{q} & = z_{q}^{(1)} - \tilde{z}_{q}^{(1)} = z_{q}^{(1)} + \frac{2\sqrt{G(G-1)}}{2G-1}R_{\sqrt{2\pi}}(z_{q}^{(2)}),
\nonumber\\
\xi_{p} & = z_{p}^{(1)} - \tilde{z}_{p}^{(1)} = z_{p}^{(1)} - \frac{2\sqrt{G(G-1)}}{2G-1}R_{\sqrt{2\pi}}(z_{p}^{(2)}), 
\end{align}
where $\boldsymbol{z} = (z_{q}^{(1)},z_{p}^{(1)},z_{q}^{(2)},z_{p}^{(2)})^{T}$ follows a Gaussian distribution with zero means and the covariance matrix $V_{\boldsymbol{z}}$. By using Eq. \eqref{supp_eq:identity on Rs function}, we find that the probability density functions of the quadrature noise are given by  
\begin{widetext}
\begin{align}
Q(\xi_{q})&\equiv \int_{-\infty}^{\infty}dz_{q}^{(1)}\int_{-\infty}^{\infty}dz_{q}^{(2)} \frac{1}{2\pi\sigma^{2}}\exp\Big{[} -\frac{(2G-1)}{2\sigma^{2}}\big{(}(z_{q}^{(1)})^{2}+(z_{q}^{(2)})^{2}\big{)} - \frac{2\sqrt{G(G-1)}}{\sigma^{2}}z_{q}^{(1)}z_{q}^{(2)} \Big{]}
\nonumber\\
&\qquad\qquad\qquad\qquad\qquad\qquad\qquad\times  \delta\Big{(} \xi_{q}-z_{q}^{(1)} - \frac{2\sqrt{G(G-1)}}{2G-1}R_{\sqrt{2\pi}}(z_{q}^{(2)}) \Big{)} 
\nonumber\\
&= \sum_{n\in\mathbb{Z}}\int_{-\infty}^{\infty}dz_{q}^{(1)}\int_{-\infty}^{\infty}dz_{q}^{(2)} \frac{1}{2\pi\sigma^{2}}\exp\Big{[} -\frac{(2G-1)}{2\sigma^{2}}\Big{(} z_{q}^{(1)} + \frac{2\sqrt{G(G-1)}}{2G-1}z_{q}^{(2)} \Big{)}^{2} - \frac{1}{2(2G-1)\sigma^{2}}(z_{q}^{(2)})^{2}\Big{]}
\nonumber\\
&\qquad\qquad\qquad\times  \delta\Big{(} \xi_{q}-z_{q}^{(1)} - \frac{2\sqrt{G(G-1)}}{2G-1}(z_{q}^{(2)}-\sqrt{2\pi}n) \Big{)}I\Big{\lbrace} z_{q}^{(2)}\in \Big{[} \Big{(} n-\frac{1}{2} \Big{)}\sqrt{2\pi},\Big{(} n+\frac{1}{2} \Big{)}\sqrt{2\pi} \Big{]} \Big{\rbrace} 
\nonumber\\
&= \sum_{n\in\mathbb{Z}} \int_{(n-\frac{1}{2})\sqrt{2\pi}}^{(n+\frac{1}{2})\sqrt{2\pi}}dz_{q}^{(2)} \frac{1}{\sqrt{2\pi(2G-1)\sigma^{2}}}\exp\Big{[} - \frac{1}{2(2G-1)\sigma^{2}}(z_{q}^{(2)})^{2}\Big{]}
\nonumber\\
&\qquad\qquad\qquad\qquad\qquad\times \sqrt{\frac{2G-1}{2\pi\sigma^{2}}}\exp\Big{[} -\frac{(2G-1)}{2\sigma^{2}}\Big{(} \xi_{q} + \frac{2\sqrt{G(G-1)}}{2G-1}\sqrt{2\pi}n \Big{)}^{2} \Big{]}
\nonumber\\
&= \sum_{n\in\mathbb{Z}}q_{n}\cdot p\Big{[}\frac{\sigma}{\sqrt{2G-1}}\Big{]}(\xi_{q}-\mu_{n}) , \label{appendix_eq:probability density functions of the position quadrature noise two-mode GKP-squeezed-repetition code without GKP noise}
\end{align}
and similarly
\begin{align}
P(\xi_{p})&\equiv \int_{-\infty}^{\infty}dz_{p}^{(1)}\int_{-\infty}^{\infty}dz_{p}^{(2)} \frac{1}{2\pi\sigma^{2}}\exp\Big{[} -\frac{(2G-1)}{2\sigma^{2}}\big{(}(z_{p}^{(1)})^{2}+(z_{p}^{(2)})^{2}\big{)} + \frac{2\sqrt{G(G-1)}}{\sigma^{2}}z_{p}^{(1)}z_{p}^{(2)} \Big{]}
\nonumber\\
&\qquad\qquad\qquad\qquad\qquad\qquad\qquad\times  \delta\Big{(} \xi_{p}-z_{p}^{(1)} + \frac{2\sqrt{G(G-1)}}{2G-1}R_{\sqrt{2\pi}}(z_{p}^{(2)}) \Big{)} 
\nonumber\\
&= \sum_{n\in\mathbb{Z}}q_{n}\cdot p\Big{[}\frac{\sigma}{\sqrt{2G-1}}\Big{]}(\xi_{q}-\mu_{n}),    \label{appendix_eq:probability density functions of the momentum quadrature noise two-mode GKP-squeezed-repetition code without GKP noise}
\end{align}
where $q_{n}(=q_{-n})$ and $\mu_{n}$ are as defined in Eq. \eqref{eq:qn and mun}. 
\end{widetext} 

Finally, we derive the asymptotic expressions for the optimal gain $G^{\star}$ and the minimum standard deviation $\sigma_{L}^{\star}$ given in Eqs. \eqref{eq:optimal G asymptotic}, \eqref{eq:optimal STD asymptotic}. Recall that assuming $\sqrt{2G-1}\sigma \ll 1$, we have Eq. \eqref{eq:variance of the logical quadrature noise asymptotic}: 
\begin{align}
(\sigma_{L})^{2} \simeq \frac{\sigma^{2}}{2G-1} + \frac{8\pi\sqrt{G(G-1)}}{(2G-1)^{2}}\textrm{erfc}\Big{(} \frac{\sqrt{\pi}}{2\sqrt{2G-1}\sigma} \Big{)}. \label{appendix_eq:variance of the logical quadrature noise asymptotic} 
\end{align}
Assuming further that $G\gg 1$ (which is relevant when $\sigma\ll 1$) and using the asymptotic formula for the complementary error function, i.e.,  
\begin{align}
\textrm{erfc}(x) \xrightarrow{x\rightarrow \infty} \frac{1}{x\sqrt{\pi}}\exp[-x^{2}], 
\end{align}
we can simplify Eq. \eqref{appendix_eq:variance of the logical quadrature noise asymptotic} as 
\begin{align}
(\sigma_{L})^{2} \simeq \sigma^{2}x + \frac{4\sigma}{\sqrt{x} }\exp\Big{[} -\frac{\pi}{4\sigma^{2}}x \Big{]} \equiv f(x), 
\end{align}
where $x\equiv 1/(2G-1)$. The optimum $x^{\star}$ can be found by solving $f'(x^{\star})=0$. Note that $f'(x)$ is given by
\begin{align}
f'(x) = \sigma^{2} - \Big{(} \frac{\pi}{\sigma \sqrt{x}} + \frac{2\sigma}{\sqrt{x^{3}}} \Big{)}\exp\Big{[} -\frac{\pi}{4\sigma^{2}}x \Big{]} . 
\end{align} 
Thus, $x^{\star}$ should satisfy 
\begin{align}
x^{\star} &= \frac{4\sigma^{2}}{\pi} \log_{e}\Big{(} \frac{\pi}{\sigma^{3}\sqrt{x^{\star}}} + \frac{2}{\sigma \sqrt{(x^{\star})^{3}}} \Big{)}
\nonumber\\
&= \frac{4\sigma^{2}}{\pi} \log_{e}\Big{(} \frac{\pi^{3/2}}{2\sigma^{4}\sqrt{\log_{e}(\cdots)}} + \frac{\pi^{3/2}}{4\sigma^{4} \sqrt{(\log_{e}(\cdots))^{3}}} \Big{)} . 
\end{align} 
where we iteratively plugged in the first equation into itself to get the second equation. Since $\log_{e}(\cdots) \gg 1$, we can disregard the second term in the second line. By further neglecting the logarithmic factor $\sqrt{\log_{e}(\cdots)}$, we get 
\begin{align}
x^{\star} \simeq \frac{4\sigma^{2}}{\pi} \log_{e}\Big{(} \frac{\pi^{3/2}}{2\sigma^{4}} \Big{)}. 
\end{align}
Since $G^{\star} = \frac{1}{2x^{\star}} + \frac{1}{2}$, Eq. \eqref{eq:optimal G asymptotic} follows: 
\begin{align}
G^{\star}  \xrightarrow{\sigma \ll 1} \frac{\pi}{8\sigma^{2}} \Big{(} \log_{e}\Big{[} \frac{\pi^{3/2}}{2\sigma^{4}} \Big{]} \Big{)}^{-1} + \frac{1}{2}. 
\end{align}
The optimal value $(\sigma^{\star}_{L})^{2} = f(x^{\star})$ is then approximately given by 
\begin{align}
(\sigma^{\star}_{L})^{2} \simeq \frac{4\sigma^{4}}{\pi}\log_{e}\Big{[} \frac{\pi^{3/2}}{2\sigma^{4}} \Big{]}  + \frac{4\sigma^{4}}{\pi}  \Big{(}\log_{e}\Big{[} \frac{\pi^{3/2}}{2\sigma^{4}} \Big{]}\Big{)}^{-\frac{1}{2}}
\end{align}
Since $\log_{e}(\pi^{3/2}/(2\sigma^{4}))\gg 1$, we can disregard the second term and obtain Eq. \eqref{eq:optimal STD asymptotic}: 
\begin{align}
\sigma^{\star}_{L} \xrightarrow{\sigma \ll 1} \frac{2\sigma^{2}}{\sqrt{\pi}}\Big{(}\log_{e}\Big{[} \frac{\pi^{3/2}}{2\sigma^{4}} \Big{]} \Big{)}^{\frac{1}{2}}. 
\end{align}

\begin{widetext}
Let us now consider the case with noisy GKP states (see Fig.\ \ref{fig:GKP-two-mode-squeezing code with noisy GKP states} and Eq.\ \eqref{eq:estimated ancilla quadrature noise noisy GKP states}). Then, we have 
\begin{align}
\xi_{q} &\equiv z_{q}^{(1)} +\frac{2\sqrt{G(G-1)} \sigma^{2} }{ (2G-1)\sigma^{2} + 2\sigma_{\textrm{gkp}}^{2} } R_{\sqrt{2\pi}}(z_{q}^{(2)} +\xi_{q}^{(\textrm{gkp})} ) , 
\end{align}
where the GKP noise $\xi_{q}^{(\textrm{gkp})}$ is independent of $z_{q}^{(1)}$ and $z_{q}^{(2)}$ and follows $\xi_{q}^{(\textrm{gkp})}\sim \mathcal{N}( 0, 2\sigma_{\textrm{gkp}}^{2} )$. Then, the probability density function $Q(\xi_{q})$ is given by
\begin{align}
Q(\xi_{q})&\equiv \int_{ \mathbb{R}^{3} }dz_{q}^{(1)}dz_{q}^{(2)}  d\xi_{q}^{(\textrm{gkp})} \frac{1}{\sqrt{ 16\pi^{3}\sigma^{4}\sigma_{\textrm{gkp}}^{2} }} \exp\Big{[} -\frac{(2G-1)}{2\sigma^{2}}\big{(}(z_{q}^{(1)})^{2}+(z_{q}^{(2)})^{2}\big{)} - \frac{2\sqrt{G(G-1)}  }{\sigma^{2}}z_{q}^{(1)}z_{q}^{(2)}   \Big{]}
\nonumber\\
&\qquad\qquad\qquad\qquad\qquad \times  \exp\Big{[} - \frac{1}{4\sigma_{\textrm{gkp}}^{2}}  ( \xi_{q}^{(\textrm{gkp})} )^{2}  \Big{]} \delta\Big{(} \xi_{q}-z_{q}^{(1)} - \frac{2\sqrt{G(G-1)} \sigma^{2} }{ (2G-1)\sigma^{2} +2\sigma_{\textrm{gkp}}^{2} }R_{\sqrt{2\pi}}(z_{q}^{(2)} + \xi_{q}^{(\textrm{gkp})} ) \Big{)} 
\nonumber\\
&=\sum_{n\in\mathbb{Z}} \int_{ \mathbb{R}^{3} }dz_{q}^{(1)}dz_{q}^{(2)}  d\xi_{q}^{(\textrm{gkp})} \frac{1}{\sqrt{ 16\pi^{3}\sigma^{4}\sigma_{\textrm{gkp}}^{2} }} \exp\Big{[} -\frac{(2G-1)}{2\sigma^{2}}\Big{(} z_{q}^{(1)} + \frac{2\sqrt{G(G-1)}}{2G-1}z_{q}^{(2)} \Big{)}^{2} - \frac{1}{2(2G-1)\sigma^{2}}(z_{q}^{(2)})^{2}\Big{]}
\nonumber\\
&\qquad\qquad\qquad\qquad\qquad \times \exp\Big{[} - \frac{1}{4\sigma_{\textrm{gkp}}^{2}}  ( \xi_{q}^{(\textrm{gkp})} )^{2}  \Big{]}  \delta\Big{(} \xi_{q}-z_{q}^{(1)} - \frac{2\sqrt{G(G-1)} \sigma^{2} }{ (2G-1)\sigma^{2} + 2\sigma_{\textrm{gkp}}^{2} }(z_{q}^{(2)} + \xi_{q}^{(\textrm{gkp})} -\sqrt{2\pi}n) \Big{)} 
\nonumber\\
&\qquad\qquad\qquad\qquad\qquad \times I\Big{\lbrace} z_{q}^{(2)} + \xi_{q}^{(\textrm{gkp})} \in \Big{[} \Big{(} n-\frac{1}{2} \Big{)}\sqrt{2\pi},\Big{(} n+\frac{1}{2} \Big{)}\sqrt{2\pi} \Big{]} \Big{\rbrace} 
\nonumber\\
&= \sum_{n\in\mathbb{Z}}\int_{\mathbb{R}^{2}} dz_{q}^{(2)}  d\xi_{q}^{(\textrm{gkp})}  \frac{1}{\sqrt{ 16 \pi^{3}\sigma^{4}\sigma_{\textrm{gkp}}^{2} }} \exp\Big{[} - \frac{1}{2(2G-1)\sigma^{2}}(z_{q}^{(2)})^{2}\Big{]} \exp\Big{[} - \frac{1}{4\sigma_{\textrm{gkp}}^{2}}  ( \xi_{q}^{(\textrm{gkp})} )^{2}  \Big{]}  
\nonumber\\
&\qquad\qquad \times \exp\Big{[} -\frac{(2G-1)}{2\sigma^{2}}\Big{(}  \xi_{q} - \frac{2\sqrt{G(G-1)}\sigma^{2}}{ (2G-1)\sigma^{2} +2\sigma_{\textrm{gkp}}^{2} }( \xi_{q}^{(\textrm{gkp})} -\sqrt{2\pi}n) + \frac{2\sqrt{G(G-1)} 2\sigma_{\textrm{gkp}}^{2} }{ (2G-1)( (2G-1)\sigma^{2} + 2\sigma_{\textrm{gkp}}^{2} ) } z_{q}^{(2)}  \Big{)}^{2} \Big{]}
\nonumber\\
&\qquad\qquad\qquad\qquad\qquad \times I\Big{\lbrace} z_{q}^{(2)} +\xi_{q}^{(\textrm{gkp})} \in \Big{[} \Big{(} n-\frac{1}{2} \Big{)}\sqrt{2\pi},\Big{(} n+\frac{1}{2} \Big{)}\sqrt{2\pi} \Big{]} \Big{\rbrace} . 
\end{align}
Thus, the variance of the output logical quadrature noise $(\sigma_{L})^{2} = \textrm{Var}[\xi_{q}] = \mathbb{E}[ (\xi_{q})^{2} ]$ is given by 
\begin{align}
(\sigma_{L})^{2} &= \sum_{n\in\mathbb{Z}}\int_{\mathbb{R}^{3}} dz_{q}^{(2)}  d\xi_{q}^{(\textrm{gkp})}  d\xi_{q}  \frac{1}{\sqrt{ 16\pi^{3}\sigma^{4}\sigma_{\textrm{gkp}}^{2} }}  \exp\Big{[} - \frac{1}{2(2G-1)\sigma^{2}}(z_{q}^{(2)})^{2}\Big{]} \exp\Big{[} - \frac{1}{4\sigma_{\textrm{gkp}}^{2}}  ( \xi_{q}^{(\textrm{gkp})} )^{2}  \Big{]}  
\nonumber\\
&\qquad \times (\xi_{q})^{2} \exp\Big{[} -\frac{(2G-1)}{2\sigma^{2}}\Big{(}  \xi_{q} - \frac{2\sqrt{G(G-1)}\sigma^{2}}{ (2G-1)\sigma^{2} +2\sigma_{\textrm{gkp}}^{2} }( \xi_{q}^{(\textrm{gkp})} -\sqrt{2\pi}n) + \frac{2\sqrt{G(G-1)} 2\sigma_{\textrm{gkp}}^{2} }{ (2G-1)( (2G-1)\sigma^{2} + 2\sigma_{\textrm{gkp}}^{2} ) } z_{q}^{(2)}  \Big{)}^{2} \Big{]}
\nonumber\\
&\qquad\qquad\qquad\qquad\qquad \times I\Big{\lbrace} z_{q}^{(2)} +\xi_{q}^{(\textrm{gkp})} \in \Big{[} \Big{(} n-\frac{1}{2} \Big{)}\sqrt{2\pi},\Big{(} n+\frac{1}{2} \Big{)}\sqrt{2\pi} \Big{]} \Big{\rbrace} 
\nonumber\\
&= \sum_{n\in\mathbb{Z}}\int_{\mathbb{R}^{2}} dz_{q}^{(2)}  d\xi_{q}^{(\textrm{gkp})}    p [  \sqrt{2G-1}\sigma ] ( z_{q}^{(2)} ) \cdot p[\sqrt{2}\sigma_{\textrm{gkp}}]( \xi_{q}^{(\textrm{gkp})} ) 
\nonumber\\
&\qquad\qquad\times \Big{[} \frac{\sigma^{2}}{2G-1} +  \Big{(}\frac{2\sqrt{G(G-1)}\sigma^{2}}{ (2G-1)\sigma^{2} +2\sigma_{\textrm{gkp}}^{2} }( \xi_{q}^{(\textrm{gkp})} -\sqrt{2\pi}n) - \frac{2\sqrt{G(G-1)} 2\sigma_{\textrm{gkp}}^{2} }{ (2G-1)( (2G-1)\sigma^{2} + 2\sigma_{\textrm{gkp}}^{2} ) } z_{q}^{(2)} \Big{)}^{2}  \Big{]}
\nonumber\\
&\qquad\qquad\qquad\qquad\qquad \times I\Big{\lbrace} z_{q}^{(2)} +\xi_{q}^{(\textrm{gkp})} \in \Big{[} \Big{(} n-\frac{1}{2} \Big{)}\sqrt{2\pi},\Big{(} n+\frac{1}{2} \Big{)}\sqrt{2\pi} \Big{]} \Big{\rbrace}.  \label{eq:variance of the output logical quadrature noise noisy GKP states better}
\end{align}
\end{widetext}

\section{GKP-squeezed-repetition codes}
\label{appendix:Generalized GKP-squeezed-repetition codes}

\begin{figure*}[t!]
\centering
\includegraphics[width=5.8in]{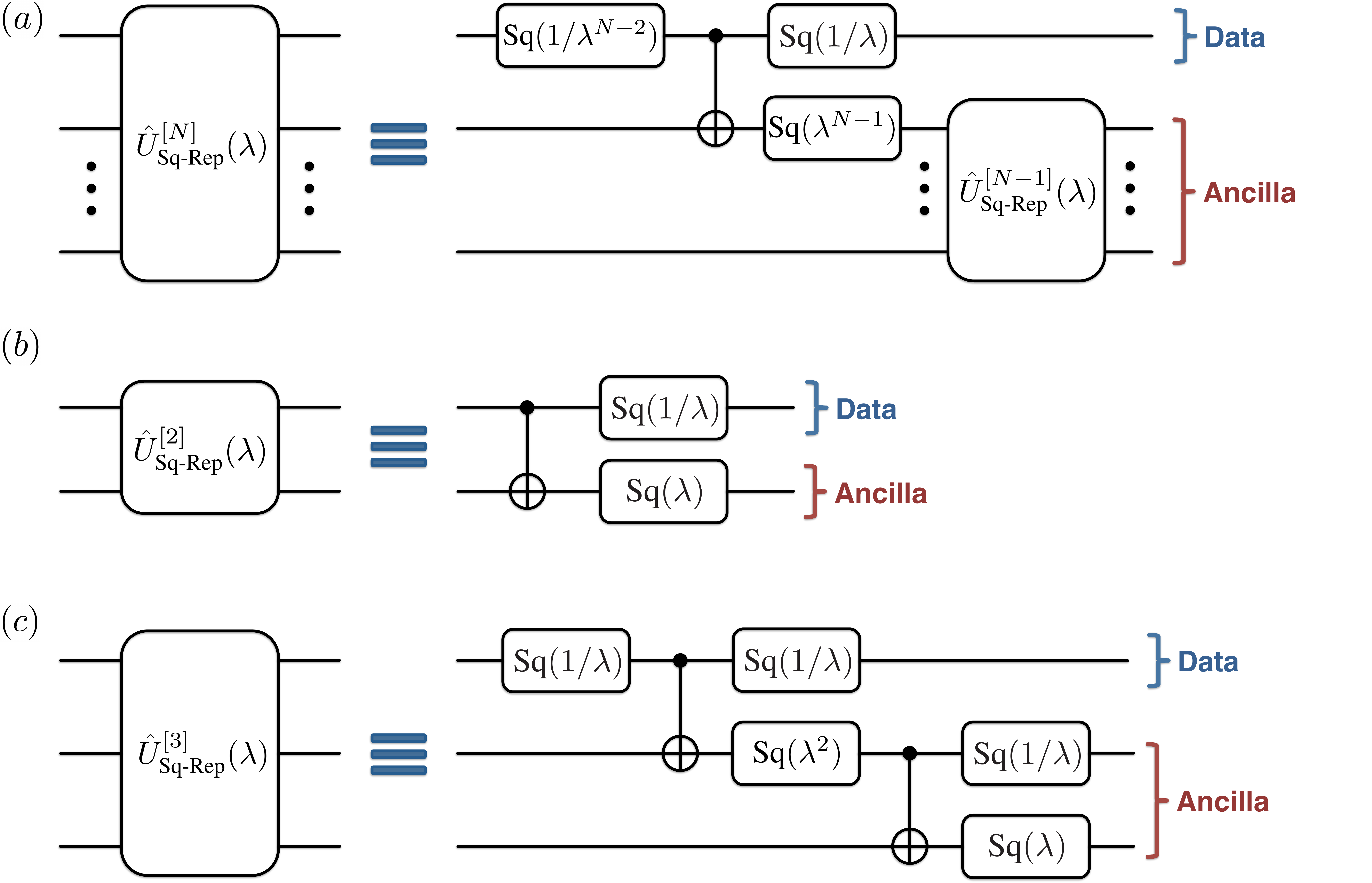}
\caption{Encoding Gaussian circuits $\hat{U}_{\textrm{Sq-Rep}}^{[N]}$ of the $N$-mode GKP-squeezed repetition code for (a) general $N\ge 3$ (b) $N=2$ and (c) $N=3$.}
\label{supp_fig:Generalized GKP-squeezed-repetition codes}
\end{figure*} 

Here, we introduce a family of GKP-stabilizer codes, GKP-squeezed-repetition codes, and briefly explain that the $N$-mode GKP-squeezed-repetition code can suppress additive Gaussian noise errors to the $N^{\textrm{th}}$ order, i.e., $\sigma_{q},\sigma_{p} \propto \sigma^{N}$, where $\sigma$ is the standard deviation of the input additive Gaussian noise and $\sigma_{q}$ and $\sigma_{p}$ are the standard deviations of the output logical position and momentum quadrature noise, respectively.    

Let $|\psi\rangle = \int dq \psi(q)|\hat{q}_{1}=q\rangle$ be an arbitrary one-mode bosonic state. We define the encoded logical state $|\psi_{L}\rangle$ of the $N$-mode GKP-squeezed-repetition code as 
\begin{align}
|\psi_{L}\rangle \equiv \hat{U}_{\textrm{Sq-Rep}}^{[N]}|\psi\rangle \otimes |\textrm{GKP}\rangle^{\otimes N-1}
\end{align}   
where the encoding Gaussian circuit $\hat{U}_{\textrm{Sq-Rep}}^{[N]}$ is recursively defined as 
\begin{align}
\hat{U}_{\textrm{Sq-Rep}}^{[N]} &\equiv \textrm{Sq}_{1}\Big{(}\frac{1}{\lambda^{N-2}}\Big{)} \textrm{SUM}_{1\rightarrow 2} \textrm{Sq}_{1}\Big{(}\frac{1}{\lambda}\Big{)}\textrm{Sq}_{2}(\lambda^{N-1}) 
\nonumber\\
&\quad \times \Big{(} \hat{I}_{1}\otimes \hat{U}_{\textrm{Sq-Rep}}^{[N-1]} \Big{)}
\end{align}
for $N\ge 3$ (see Fig.\ \ref{supp_fig:Generalized GKP-squeezed-repetition codes} (a)). In the base case ($N=2$), $\hat{U}_{\textrm{Sq-Rep}}^{[2]}$ is given by $\hat{U}_{\textrm{Sq-Rep}}^{[2]} \equiv\textrm{Sq}_{1}(1/\lambda)\textrm{Sq}_{2}(\lambda)\textrm{SUM}_{1\rightarrow 2}$ (see Fig.\ \ref{supp_fig:Generalized GKP-squeezed-repetition codes} (b)) and the symplectic matrix $\boldsymbol{S}_{\textrm{Sq-Rep}}^{[2]}$ associated with the encoding Gaussian circuit $\hat{U}_{\textrm{Sq-Rep}}^{[2]}$ is given by  
\begin{align}
\boldsymbol{S}_{\textrm{Sq-Rep}}^{[2]} =  \begin{bmatrix}
1/\lambda& 0 &0 &0 \\
0 & \lambda & 0 & -\lambda \\
\lambda& 0 &\lambda &0\\
0&0&0&1/\lambda
\end{bmatrix} . 
\end{align}  
 
To illustrate how this family of codes works, let us focus on the $N=3$ case: The symplectic matrix $\boldsymbol{S}_{\textrm{Sq-Rep}}^{[3]}$ associated with the encoding Gaussian circuit $\hat{U}_{\textrm{Sq-Rep}}^{[3]}$ (see Fig.\ \ref{supp_fig:Generalized GKP-squeezed-repetition codes} (c)) is explicitly given by  
\begin{align}
\boldsymbol{S}_{\textrm{Sq-Rep}}^{[3]} = \begin{bmatrix}
1/\lambda^{2}&0&0&0&0&0\\
0&\lambda^{2}&0&-\lambda&0&0\\
1&0&\lambda&0&0&0\\
0&0&0&1/\lambda&0&-\lambda\\
\lambda^{2}&0&\lambda^{3}&0&\lambda&0\\
0&0&0&0&0&1/\lambda
\end{bmatrix}. 
\end{align}
Thus, the reshaped quadrature noise vector $\boldsymbol{z}=\boldsymbol{(S}_{\textrm{Sq-Rep}}^{[3]}\boldsymbol{)^{-1}}\boldsymbol{\xi}$ is given by 
\begin{align}
\boldsymbol{z} = \begin{bmatrix}
\lambda^{2}\xi_{q}^{(1)}\\
\xi_{p}^{(1)}/\lambda^{2} + \xi_{p}^{(2)}+\lambda^{2}\xi_{p}^{(3)} \\
-\lambda\xi_{q}^{(1)} + \xi_{q}^{(2)}/\lambda \\
\lambda\xi_{p}^{(2)} + \lambda^{3}\xi_{p}^{(3)}\\
-\lambda\xi_{q}^{(2)}+\xi_{q}^{(3)}/\lambda \\
\lambda\xi_{p}^{(3)}
\end{bmatrix} \equiv \begin{bmatrix}
z_{q}^{(1)} \\
z_{p}^{(1)} \\
z_{q}^{(2)} \\
z_{p}^{(2)} \\
z_{q}^{(3)} \\
z_{p}^{(3)}
\end{bmatrix},
\end{align}
where $\boldsymbol{\xi}\equiv (\xi_{q}^{(1)},\xi_{p}^{(1)},\xi_{q}^{(2)},\xi_{p}^{(2)},\xi_{q}^{(3)},\xi_{p}^{(3)})^{T}$ is the original quadrature noise vector whose elements follow independent and identically distributed Gaussian random distributions with variance $\sigma^{2}$.

Through the noise reshaping, the data position quadrature noise is amplified by a factor of $\lambda^{2}$ (i.e., $z_{q}^{(1)} = \lambda^{2} \xi_{q}^{(1)}$). Note that, by choosing $\lambda = \sqrt{2\pi}c/\sigma$ with a small constant $c\ll 1$, we can make sure that the ancilla position quadrature noise $z_{q}^{(2)}$ and $z_{q}^{(3)}$ are contained within the unambiguously distinguishable range $[-\sqrt{\pi/2},\sqrt{\pi/2}]$ with a very high probability. Then, by measuring the reshaped ancilla position quadrature noise of the second mode $z_{q}^{(2)} = -\lambda \xi_{q}^{(1)} + \xi_{q}^{(2)}/\lambda$, (modulo $\sqrt{2\pi}$) we can learn about the amplified position quadrature noise $z_{q}^{(1)} = \lambda^{2}\xi_{q}^{(1)}$ up to an error $\xi_{q}^{(2)}$. Then, by further measuring the ancilla position quadrature noise of the third mode, $z_{q}^{(3)} = -\lambda \xi_{q}^{(2)} + \xi_{q}^{(3)}/\lambda$, (modulo $\sqrt{2\pi}$) we can learn about the residual error $\xi_{q}^{(2)}$ up to an even smaller error $\xi_{q}^{(3)}/\lambda^{2}$. As a result, we can reduce the position quadrature noise of the data mode by a factor of $\lambda^{2}$ despite the temporary increase by the same factor before the correction. Since $\lambda \propto 1/\sigma$, this implies that the standard deviation of the position quadrature noise $\sigma_{q}$ is suppressed \textit{cubically}, i.e., $\sigma_{q} = \sigma / \lambda^{2} \propto \sigma^{3}$.

Through the noise reshaping, the data momentum quadrature noise is immediately reduced by a factor of $\lambda^{2}$ (see $\xi_{p}^{(1)}/\lambda^{2}$ in $z_{p}^{(1)}$) but there are transferred ancilla momentum quadrature noise (see $+\xi_{p}^{(2)}$ and $+\lambda^{2}\xi_{p}^{(3)}$ in $z_{p}^{(1)}$). Note that by measuring the ancilla momentum quadrature noise of the third mode, $z_{p}^{(3)} = \lambda\xi_{p}^{(3)}$ (which is contained within the range $[-\sqrt{\pi/2},\sqrt{\pi/2}]$ with a very high probability), we can precisely extract the value of $\xi_{p}^{(3)}$ and then eliminate the transferred ancilla momentum quadrature noise $+\lambda^{2}\xi_{p}^{(3)}$ in $z_{p}^{(1)}$ and also $+\lambda^{3}\xi_{p}^{(3)}$ in $z_{p}^{(2)}$. Then, by further measuring the reshaped ancilla momentum quadrature noise of the second mode (after eliminating $\lambda^{3}\xi_{p}^{(2)}$), we can precisely extract the value of $z_{p}^{(2)}-\lambda^{3}\xi_{p}^{(3)} = \lambda \xi_{p}^{(2)}$ and then eliminate the transferred ancilla momentum quadrature noise $+\xi_{p}^{(2)}$ in $z_{p}^{(1)}$. As a result, we end up with the reduced date momentum quadrature noise $\xi_{p}^{(1)}/\lambda^{2}$ without any transferred ancilla momentum quadrature noise. Similarly as above, since $\lambda \propto 1/\sigma$, we can suppress the momentum quadrature noise cubically, i.e., $\sigma_{p} = \sigma / \lambda^{2} \propto \sigma^{3}$.   

Generalizing these arguments, one can inductively show that the $N$-mode GKP-squeezed-repetition code can suppress additive quadrature noise errors to the $N^{\textrm{th}}$ order, i.e., $\sigma_{q},\sigma_{p} = \sigma / \lambda^{N-1} \propto \sigma^{N}$ for any $N\ge 2$. We emphasize, however, these codes are susceptible to the realistic noise in GKP states (except for the $N=2$ case), because the GKP noise can be amplified by large squeezing operations. For example, in the case of the three-mode GKP-squeezed-repetition code discussed above, one has to multiply a factor of $\lambda$ into the obtained measurement outcome of $z_{p}^{(3)} = \lambda \xi_{p}^{(3)}$ to eliminate the transferred noise $+\lambda^{2}\xi_{p}^{(3)}$ in the reshaped data momentum quadrature noise $z_{p}^{(1)}$. Therefore, the GKP noise that corrupt the measurement outcome of $z_{p}^{(3)}$ are amplified by a large squeezing parameter $\lambda = \sqrt{2\pi}c/\sigma \xrightarrow{\sigma\rightarrow 0} \infty $ when they are propagated to the data mode via miscalibrated counter displacement operations.  

Thus, although the $N$-mode GKP-squeezed-repetition code can correct additive Gaussian noise errors to the $N^{\textrm{th}}$ order, it is sensitive to the realistic noise in GKP states. It will therefore be interesting to look for a family of GKP-stabilizer codes that can achieve higher order error suppression ($N \ge 3$) while not amplifying the GKP noise.   

\section{Commutativity of beam splitter interactions and iid additive Gaussian noise errors}
\label{appendix:commutativity of beam splitter interactions and iid additive Gaussian noise errors}

Here, we show that any passive beam splitter interactions commute with iid additive Gaussian noise errors. Consider $n$ bosonic modes described by bosonic annihilation operators $\hat{a}_{1},\cdots,\hat{a}_{n}$. A general passive beam splitter interaction among these $n$ modes is generated by a Hamiltonian of the following form:
\begin{align}
\hat{H}_{\textrm{BS}} = \sum_{k,l=1}^{n}g_{kl}\hat{a}_{k}^{\dagger}\hat{a}_{l}, 
\end{align}
where $g_{kl}=g_{lk}^{*}$ for all $k,l\in \lbrace 1,\cdots,n \rbrace$. Note that the total excitation number is conserved under a general beam splitter interaction: $[\hat{H}_{\textrm{BS}},\sum_{j=1}^{n}\hat{a}_{j}^{\dagger}\hat{a}_{j}]=0$. Also, iid additive Gaussian noise errors are generated by the Lindbladian 
\begin{align}
\mathcal{L} \equiv \sum_{j=1}^{n}\big{(}\mathcal{D}[\hat{a}_{j}] + \mathcal{D}[\hat{a}^{\dagger}_{j}] \big{)}, 
\end{align} 
where $\mathcal{D}[\hat{A}](\hat{\rho})\equiv \hat{A}\hat{\rho}\hat{A}^{\dagger}-\frac{1}{2}\lbrace \hat{A}^{\dagger}\hat{A},\hat{\rho} \rbrace$. Specifically, the Lindbladian generator $\mathcal{L}$ is related to the additive Gaussian noise error $\mathcal{N}[\sigma]$ by the relation 
\begin{align}
\exp\big{[} D t \cdot \mathcal{L}  \big{]} = \bigotimes_{k=1}^{n} \mathcal{N}^{(k)}[\sigma= \sqrt{Dt} ] ,  
\end{align}  
where $D$ is the diffusion rate and $t$ is the time elapsed.

Having introduced the generators of the beam splitter interactions and iid additive Gaussian noise errors, we now prove the following commutation relation:  
\begin{align}
[\mathcal{V}_{\textrm{BS}},\mathcal{L}]=0, \label{appendix_eq:commutation relation between bs and iid additive noise}
\end{align} 
where $\mathcal{V}_{\textrm{BS}}(\hat{\rho}) \equiv -i[\hat{H}_{\textrm{BS}},\hat{\rho}]$ is the Lindbladian superoperator associated with a beam splitter Hamiltonian $\hat{H}_{\textrm{BS}} = \sum_{k,l=1}^{n}g_{kl}\hat{a}_{k}^{\dagger}\hat{a}_{l}$. Note that for a general $\mathcal{V} = -i[\hat{H},\bullet]$ and $\mathcal{D}[\hat{A}]$ we have 
\begin{align}
\mathcal{V} \cdot \mathcal{D}[\hat{A}](\hat{\rho}) &=  -i\Big{[}\hat{H},\hat{A}\hat{\rho}\hat{A}^{\dagger} - \frac{1}{2}\hat{A}^{\dagger}\hat{A}\hat{\rho} -\frac{1}{2}\hat{\rho}\hat{A}^{\dagger}\hat{A} \Big{]} 
\nonumber\\
&= -i\hat{H}\hat{A}\hat{\rho}\hat{A}^{\dagger} +\frac{i}{2}\hat{H} \hat{A}^{\dagger}\hat{A}\hat{\rho} +\frac{i}{2}\hat{H}\hat{\rho}\hat{A}^{\dagger}\hat{A}
\nonumber\\
&\quad +i\hat{A}\hat{\rho}\hat{A}^{\dagger}\hat{H} -\frac{i}{2} \hat{A}^{\dagger}\hat{A}\hat{\rho} \hat{H} -\frac{i}{2}\hat{\rho}\hat{A}^{\dagger}\hat{A}\hat{H}, 
\end{align}
and similarly 
\begin{align}
\mathcal{D}[\hat{A}] \cdot \mathcal{V} (\hat{\rho}) &=\mathcal{D}[\hat{A}](-i[\hat{H},\hat{\rho}]) 
\nonumber\\
&= -i\hat{A}\hat{H}\hat{\rho}\hat{A}^{\dagger}+\frac{i}{2}\hat{A}^{\dagger}\hat{A}\hat{H}\hat{\rho} +\frac{i}{2}\hat{H}\hat{\rho} \hat{A}^{\dagger}\hat{A}
\nonumber\\
&\quad+i\hat{A}\hat{\rho}\hat{H}\hat{A}^{\dagger}-\frac{i}{2}\hat{A}^{\dagger}\hat{A}\hat{\rho}\hat{H} -\frac{i}{2}\hat{\rho}\hat{H} \hat{A}^{\dagger}\hat{A}. 
\end{align}
Therefore, we have 
\begin{align}
\big{[}\mathcal{V},\mathcal{D}[\hat{A}]\big{]}(\hat{\rho})&= -i[\hat{H},\hat{A}]\hat{\rho}\hat{A}^{\dagger} +i\hat{A}\hat{\rho}[\hat{A}^{\dagger},\hat{H}] 
\nonumber\\
&\quad + \frac{i}{2}[\hat{H},\hat{A}^{\dagger}\hat{A}]\hat{\rho}   - \frac{i}{2}\hat{\rho}[\hat{A}^{\dagger}\hat{A},\hat{H}] . 
\end{align}
Using this general relation, we find 
\begin{align}
&\Big{[} \mathcal{V}_{\textrm{BS}},\sum_{j=1}^{n}\mathcal{D}[\hat{a}_{j}] \Big{]}(\hat{\rho}) 
\nonumber\\
&=\sum_{j=1}^{n}\Big{[}-i[\hat{H}_{\textrm{BS}},\hat{a}_{j}]\hat{\rho}\hat{a}_{j}^{\dagger} +i\hat{a}_{j}\hat{\rho}[\hat{a}_{j}^{\dagger},\hat{H}_{\textrm{BS}}] 
\nonumber\\
&\qquad\quad + \frac{i}{2}[\hat{H}_{\textrm{BS}},\hat{a}_{j}^{\dagger}\hat{a}_{j}]\hat{\rho}   - \frac{i}{2}\hat{\rho}[\hat{a}_{j}^{\dagger}\hat{a}_{j},\hat{H}_{\textrm{BS}}] \Big{]}
\nonumber\\
&= \sum_{j,k,l=1}^{n}g_{kl}\Big{[} i\delta_{jk}\hat{a}_{l}\hat{\rho}\hat{a}_{j}^{\dagger} -i\delta_{jl}\hat{a}_{j}\hat{\rho}\hat{a}_{k}^{\dagger} \Big{]}
\nonumber\\
&= i\sum_{k,l=1}^{n}g_{kl}\hat{a}_{l}\hat{\rho}\hat{a}_{k}^{\dagger} - i\sum_{k,l=1}^{n}g_{kl} \hat{a}_{l}\hat{\rho}\hat{a}_{k}^{\dagger} =0. 
\end{align}
Similarly, we also have
\begin{align}
&\Big{[} \mathcal{V}_{\textrm{BS}},\sum_{j=1}^{n}\mathcal{D}[\hat{a}_{j}^{\dagger}] \Big{]}(\hat{\rho}) 
\nonumber\\
&=\sum_{j=1}^{n}\Big{[}-i[\hat{H}_{\textrm{BS}},\hat{a}_{j}^{\dagger}]\hat{\rho}\hat{a}_{j} +i\hat{a}_{j}^{\dagger}\hat{\rho}[\hat{a}_{j},\hat{H}_{\textrm{BS}}] 
\nonumber\\
&\qquad\quad + \frac{i}{2}[\hat{H}_{\textrm{BS}},\hat{a}_{j}\hat{a}_{j}^{\dagger}]\hat{\rho}   - \frac{i}{2}\hat{\rho}[\hat{a}_{j}\hat{a}_{j}^{\dagger},\hat{H}_{\textrm{BS}}] \Big{]}
\nonumber\\
&= \sum_{j,k,l=1}^{n}g_{kl}\Big{[} -i\delta_{jl}\hat{a}_{k}^{\dagger}\hat{\rho}\hat{a}_{j}+ i\delta_{jk}\hat{a}_{j}^{\dagger}\hat{\rho}\hat{a}_{l} \Big{]}
\nonumber\\
&= -i\sum_{k,l=1}^{n}g_{kl}\hat{a}_{k}^{\dagger}\hat{\rho}\hat{a}_{l} + i\sum_{k,l=1}^{n}g_{kl} \hat{a}_{k}^{\dagger}\hat{\rho}\hat{a}_{l} =0. 
\end{align} 
Then, since $\mathcal{L} = \sum_{j=1}^{n}(\mathcal{D}[\hat{a}_{j}]+\mathcal{D}[\hat{a}_{j}^{\dagger}])$, Eq. \eqref{appendix_eq:commutation relation between bs and iid additive noise} follows: 
\begin{align}
[\mathcal{V}_{\textrm{BS}},\mathcal{L}] = \sum_{j=1}^{n}\Big{[}\mathcal{V}_{\textrm{BS}},\mathcal{D}[\hat{a}_{j}]+\mathcal{D}[\hat{a}_{j}^{\dagger}]\Big{]}=0. 
\end{align}   

Since the generators of a beam splitter interaction and an iid additive Gaussian noise error commute with each other at the superoperator level, we have 
\begin{align}
\exp\big{[} t(\mathcal{V}_{\textrm{BS}} + D\cdot \mathcal{L}) \big{]} &= \exp\big{[} Dt \cdot \mathcal{L} \big{]} \cdot \exp\big{[} t\cdot  \mathcal{V}_{\textrm{BS}}  \big{]} 
\nonumber\\  
&=\bigotimes_{k=1}^{n}\mathcal{N}^{(k)}[\sigma]\cdot \mathcal{U}_{\textrm{BS}}, \label{eq:loss+diffusion}
\end{align}
where $\sigma^{2} =Dt$ and $\mathcal{U}_{\textrm{BS}} \equiv \hat{U}_{\textrm{BS}}\bullet \hat{U}_{\textrm{BS}}^{\dagger}$ and $\hat{U}_{\textrm{BS}} = \exp[-i\hat{H}_{\textrm{BS}}t]$ is the desired $n$-mode beam splitter unitary operation. This implies that the noisy beam splitter interaction continuously corrupted by iid additive noise errors can be understood as the desired noiseless beam splitter interaction followed by an iid additive Gaussian noise channel with variance $\sigma^{2} = Dt$, where $D$ is the diffusion rate and $t$ is the time needed to complete the beam splitter interaction.    

%%%%%%%%%%%%%%%%%%%%%%%%%%%%%%%%%%%%
%&=  \sum_{n_{2}\in\mathbb{Z}}\int_{-\infty}^{\infty}d\xi_{q}^{1}\int_{-\infty}^{\infty}d\xi_{q}^{2} p_{\sigma}(\xi_{q}^{1})p_{\sigma}(\xi_{q}^{2}) \delta\Big{(} \xi_{q}-\lambda\xi_{q}^{1} - \frac{\lambda^{4}}{1+\lambda^{4}} \Big{(}\frac{\xi_{q}^{2}}{\lambda}-\lambda\xi_{q}^{1} -\sqrt{2\pi}n_{2} \Big{)} \Big{)}
%\nonumber\\
%& \qquad\qquad\qquad\qquad\qquad\qquad\qquad\qquad \times I\Big{\lbrace} \frac{\xi_{q}^{2}}{\lambda}-\lambda\xi_{q}^{1}\in \Big{[} \Big{(} n_{2}-\frac{1}{2} \Big{)}\sqrt{2\pi} , \Big{(} n_{2}+\frac{1}{2} \Big{)}\sqrt{2\pi} \Big{]} \Big{\rbrace}
%\nonumber\\
%%%%%%%%%%%%%%%%%%%%%%%%%%%%%%%%%%%%

%%%%%%%%%%%%%%%%%%%%%%%%%%%%%%%%%%%%
%&= \sum_{n_{2}\in\mathbb{Z}} \int_{-\infty}^{\infty}d\xi_{p}^{1}\int_{-\infty}^{\infty}d\xi_{p}^{2} p_{\sigma}(\xi_{p}^{1})p_{\sigma}(\xi_{p}^{2})  \delta\Big{(} \xi_{p} -\frac{\xi_{p}^{1}}{\lambda} - \sqrt{2\pi}n_{2}  \Big{)} 
%\nonumber\\
%&\qquad\qquad\qquad\qquad\qquad\qquad\qquad  \times I\Big{\lbrace} \lambda\xi_{p}^{2} \in \Big{[} \Big{(} n_{2}-\frac{1}{2} \Big{)}\sqrt{2\pi} , \Big{(} n_{2}+\frac{1}{2} \Big{)}\sqrt{2\pi} \Big{]} \Big{\rbrace}
%\nonumber\\
%%%%%%%%%%%%%%%%%%%%%%%%%%%%%%%%%%%%

\bibliography{Oscillator_encoding_arxiv_v29}

\begin{thebibliography}{79}%
\makeatletter
\providecommand \@ifxundefined [1]{%
 \@ifx{#1\undefined}
}%
\providecommand \@ifnum [1]{%
 \ifnum #1\expandafter \@firstoftwo
 \else \expandafter \@secondoftwo
 \fi
}%
\providecommand \@ifx [1]{%
 \ifx #1\expandafter \@firstoftwo
 \else \expandafter \@secondoftwo
 \fi
}%
\providecommand \natexlab [1]{#1}%
\providecommand \enquote  [1]{``#1''}%
\providecommand \bibnamefont  [1]{#1}%
\providecommand \bibfnamefont [1]{#1}%
\providecommand \citenamefont [1]{#1}%
\providecommand \href@noop [0]{\@secondoftwo}%
\providecommand \href [0]{\begingroup \@sanitize@url \@href}%
\providecommand \@href[1]{\@@startlink{#1}\@@href}%
\providecommand \@@href[1]{\endgroup#1\@@endlink}%
\providecommand \@sanitize@url [0]{\catcode `\\12\catcode `\$12\catcode
  `\&12\catcode `\#12\catcode `\^12\catcode `\_12\catcode `\%12\relax}%
\providecommand \@@startlink[1]{}%
\providecommand \@@endlink[0]{}%
\providecommand \url  [0]{\begingroup\@sanitize@url \@url }%
\providecommand \@url [1]{\endgroup\@href {#1}{\urlprefix }}%
\providecommand \urlprefix  [0]{URL }%
\providecommand \Eprint [0]{\href }%
\providecommand \doibase [0]{http://dx.doi.org/}%
\providecommand \selectlanguage [0]{\@gobble}%
\providecommand \bibinfo  [0]{\@secondoftwo}%
\providecommand \bibfield  [0]{\@secondoftwo}%
\providecommand \translation [1]{[#1]}%
\providecommand \BibitemOpen [0]{}%
\providecommand \bibitemStop [0]{}%
\providecommand \bibitemNoStop [0]{.\EOS\space}%
\providecommand \EOS [0]{\spacefactor3000\relax}%
\providecommand \BibitemShut  [1]{\csname bibitem#1\endcsname}%
\let\auto@bib@innerbib\@empty
%</preamble>
\bibitem [{\citenamefont {Braunstein}\ and\ \citenamefont {van
  Loock}(2005)}]{Braunstein2005}%
  \BibitemOpen
  \bibfield  {author} {\bibinfo {author} {\bibfnamefont {S.~L.}\ \bibnamefont
  {Braunstein}}\ and\ \bibinfo {author} {\bibfnamefont {P.}~\bibnamefont {van
  Loock}},\ }\bibfield  {title} {\enquote {\bibinfo {title} {Quantum
  information with continuous variables},}\ }\href {\doibase
  10.1103/RevModPhys.77.513} {\bibfield  {journal} {\bibinfo  {journal} {Rev.
  Mod. Phys.}\ }\textbf {\bibinfo {volume} {77}},\ \bibinfo {pages} {513--577}
  (\bibinfo {year} {2005})}\BibitemShut {NoStop}%
\bibitem [{\citenamefont {Weedbrook}\ \emph {et~al.}(2012)\citenamefont
  {Weedbrook}, \citenamefont {Pirandola}, \citenamefont {Garc\'{\i}a-Patr\'on},
  \citenamefont {Cerf}, \citenamefont {Ralph}, \citenamefont {Shapiro},\ and\
  \citenamefont {Lloyd}}]{Weedbrook2012}%
  \BibitemOpen
  \bibfield  {author} {\bibinfo {author} {\bibfnamefont {C.}~\bibnamefont
  {Weedbrook}}, \bibinfo {author} {\bibfnamefont {S.}~\bibnamefont
  {Pirandola}}, \bibinfo {author} {\bibfnamefont {R.}~\bibnamefont
  {Garc\'{\i}a-Patr\'on}}, \bibinfo {author} {\bibfnamefont {N.~J.}\
  \bibnamefont {Cerf}}, \bibinfo {author} {\bibfnamefont {T.~C.}\ \bibnamefont
  {Ralph}}, \bibinfo {author} {\bibfnamefont {J.~H.}\ \bibnamefont {Shapiro}},
  \ and\ \bibinfo {author} {\bibfnamefont {S.}~\bibnamefont {Lloyd}},\
  }\bibfield  {title} {\enquote {\bibinfo {title} {Gaussian quantum
  information},}\ }\href {\doibase 10.1103/RevModPhys.84.621} {\bibfield
  {journal} {\bibinfo  {journal} {Rev. Mod. Phys.}\ }\textbf {\bibinfo {volume}
  {84}},\ \bibinfo {pages} {621--669} (\bibinfo {year} {2012})}\BibitemShut
  {NoStop}%
\bibitem [{\citenamefont {Aaronson}\ and\ \citenamefont
  {Arkhipov}(2011)}]{Aaronson2011}%
  \BibitemOpen
  \bibfield  {author} {\bibinfo {author} {\bibfnamefont {S.}~\bibnamefont
  {Aaronson}}\ and\ \bibinfo {author} {\bibfnamefont {A.}~\bibnamefont
  {Arkhipov}},\ }\bibfield  {title} {\enquote {\bibinfo {title} {The
  computational complexity of linear optics},}\ }in\ \href {\doibase
  10.1145/1993636.1993682} {\emph {\bibinfo {booktitle} {Proceedings of the
  Forty-third Annual ACM Symposium on Theory of Computing}}},\ \bibinfo {series
  and number} {STOC '11}\ (\bibinfo  {publisher} {ACM},\ \bibinfo {address}
  {New York, NY, USA},\ \bibinfo {year} {2011})\ pp.\ \bibinfo {pages}
  {333--342}\BibitemShut {NoStop}%
\bibitem [{\citenamefont {Hamilton}\ \emph {et~al.}(2017)\citenamefont
  {Hamilton}, \citenamefont {Kruse}, \citenamefont {Sansoni}, \citenamefont
  {Barkhofen}, \citenamefont {Silberhorn},\ and\ \citenamefont
  {Jex}}]{Hamilton2017}%
  \BibitemOpen
  \bibfield  {author} {\bibinfo {author} {\bibfnamefont {C.~S.}\ \bibnamefont
  {Hamilton}}, \bibinfo {author} {\bibfnamefont {R.}~\bibnamefont {Kruse}},
  \bibinfo {author} {\bibfnamefont {L.}~\bibnamefont {Sansoni}}, \bibinfo
  {author} {\bibfnamefont {S.}~\bibnamefont {Barkhofen}}, \bibinfo {author}
  {\bibfnamefont {C.}~\bibnamefont {Silberhorn}}, \ and\ \bibinfo {author}
  {\bibfnamefont {I.}~\bibnamefont {Jex}},\ }\bibfield  {title} {\enquote
  {\bibinfo {title} {Gaussian boson sampling},}\ }\href {\doibase
  10.1103/PhysRevLett.119.170501} {\bibfield  {journal} {\bibinfo  {journal}
  {Phys. Rev. Lett.}\ }\textbf {\bibinfo {volume} {119}},\ \bibinfo {pages}
  {170501} (\bibinfo {year} {2017})}\BibitemShut {NoStop}%
\bibitem [{\citenamefont {Garc{\'{i}}a-Patr{\'{o}}n}\ \emph
  {et~al.}(2019)\citenamefont {Garc{\'{i}}a-Patr{\'{o}}n}, \citenamefont
  {Renema},\ and\ \citenamefont {Shchesnovich}}]{GarciaPatron2019}%
  \BibitemOpen
  \bibfield  {author} {\bibinfo {author} {\bibfnamefont {R.}~\bibnamefont
  {Garc{\'{i}}a-Patr{\'{o}}n}}, \bibinfo {author} {\bibfnamefont {J.~J.}\
  \bibnamefont {Renema}}, \ and\ \bibinfo {author} {\bibfnamefont
  {V.}~\bibnamefont {Shchesnovich}},\ }\bibfield  {title} {\enquote {\bibinfo
  {title} {Simulating boson sampling in lossy architectures},}\ }\href
  {\doibase 10.22331/q-2019-08-05-169} {\bibfield  {journal} {\bibinfo
  {journal} {{Quantum}}\ }\textbf {\bibinfo {volume} {3}},\ \bibinfo {pages}
  {169} (\bibinfo {year} {2019})}\BibitemShut {NoStop}%
\bibitem [{\citenamefont {Wang}\ \emph {et~al.}(2019)\citenamefont {Wang},
  \citenamefont {Qin}, \citenamefont {Ding}, \citenamefont {Chen},
  \citenamefont {Chen}, \citenamefont {You}, \citenamefont {He}, \citenamefont
  {Jiang}, \citenamefont {You}, \citenamefont {Wang}, \citenamefont
  {Schneider}, \citenamefont {Renema}, \citenamefont {H\"ofling}, \citenamefont
  {Lu},\ and\ \citenamefont {Pan}}]{Wang2019b}%
  \BibitemOpen
  \bibfield  {author} {\bibinfo {author} {\bibfnamefont {H.}~\bibnamefont
  {Wang}}, \bibinfo {author} {\bibfnamefont {J.}~\bibnamefont {Qin}}, \bibinfo
  {author} {\bibfnamefont {X.}~\bibnamefont {Ding}}, \bibinfo {author}
  {\bibfnamefont {M.-C.}\ \bibnamefont {Chen}}, \bibinfo {author}
  {\bibfnamefont {S.}~\bibnamefont {Chen}}, \bibinfo {author} {\bibfnamefont
  {X.}~\bibnamefont {You}}, \bibinfo {author} {\bibfnamefont {Y.-M.}\
  \bibnamefont {He}}, \bibinfo {author} {\bibfnamefont {X.}~\bibnamefont
  {Jiang}}, \bibinfo {author} {\bibfnamefont {L.}~\bibnamefont {You}}, \bibinfo
  {author} {\bibfnamefont {Z.}~\bibnamefont {Wang}}, \bibinfo {author}
  {\bibfnamefont {C.}~\bibnamefont {Schneider}}, \bibinfo {author}
  {\bibfnamefont {J.~J.}\ \bibnamefont {Renema}}, \bibinfo {author}
  {\bibfnamefont {S.}~\bibnamefont {H\"ofling}}, \bibinfo {author}
  {\bibfnamefont {C.-Y.}\ \bibnamefont {Lu}}, \ and\ \bibinfo {author}
  {\bibfnamefont {J.-W.}\ \bibnamefont {Pan}},\ }\bibfield  {title} {\enquote
  {\bibinfo {title} {Boson sampling with 20 input photons and a 60-mode
  interferometer in a $1{0}^{14}$-dimensional hilbert space},}\ }\href
  {\doibase 10.1103/PhysRevLett.123.250503} {\bibfield  {journal} {\bibinfo
  {journal} {Phys. Rev. Lett.}\ }\textbf {\bibinfo {volume} {123}},\ \bibinfo
  {pages} {250503} (\bibinfo {year} {2019})}\BibitemShut {NoStop}%
\bibitem [{\citenamefont {Huh}\ \emph {et~al.}(2015)\citenamefont {Huh},
  \citenamefont {Guerreschi}, \citenamefont {Peropadre}, \citenamefont
  {McClean},\ and\ \citenamefont {Aspuru-Guzik}}]{Huh2015}%
  \BibitemOpen
  \bibfield  {author} {\bibinfo {author} {\bibfnamefont {J.}~\bibnamefont
  {Huh}}, \bibinfo {author} {\bibfnamefont {G.~G.}\ \bibnamefont {Guerreschi}},
  \bibinfo {author} {\bibfnamefont {B.}~\bibnamefont {Peropadre}}, \bibinfo
  {author} {\bibfnamefont {J.~R.}\ \bibnamefont {McClean}}, \ and\ \bibinfo
  {author} {\bibfnamefont {A.}~\bibnamefont {Aspuru-Guzik}},\ }\bibfield
  {title} {\enquote {\bibinfo {title} {Boson sampling for molecular vibronic
  spectra},}\ }\href {https://doi.org/10.1038/nphoton.2015.153} {\bibfield
  {journal} {\bibinfo  {journal} {Nature Photonics}\ }\textbf {\bibinfo
  {volume} {9}},\ \bibinfo {pages} {615--620} (\bibinfo {year}
  {2015})}\BibitemShut {NoStop}%
\bibitem [{\citenamefont {Sparrow}\ \emph {et~al.}(2018)\citenamefont
  {Sparrow}, \citenamefont {Mart{\'\i}n-L{\'o}pez}, \citenamefont {Maraviglia},
  \citenamefont {Neville}, \citenamefont {Harrold}, \citenamefont {Carolan},
  \citenamefont {Joglekar}, \citenamefont {Hashimoto}, \citenamefont {Matsuda},
  \citenamefont {O'Brien}, \citenamefont {Tew},\ and\ \citenamefont
  {Laing}}]{Sparrow2018}%
  \BibitemOpen
  \bibfield  {author} {\bibinfo {author} {\bibfnamefont {C.}~\bibnamefont
  {Sparrow}}, \bibinfo {author} {\bibfnamefont {E.}~\bibnamefont
  {Mart{\'\i}n-L{\'o}pez}}, \bibinfo {author} {\bibfnamefont {N.}~\bibnamefont
  {Maraviglia}}, \bibinfo {author} {\bibfnamefont {A.}~\bibnamefont {Neville}},
  \bibinfo {author} {\bibfnamefont {C.}~\bibnamefont {Harrold}}, \bibinfo
  {author} {\bibfnamefont {J.}~\bibnamefont {Carolan}}, \bibinfo {author}
  {\bibfnamefont {Y.~N.}\ \bibnamefont {Joglekar}}, \bibinfo {author}
  {\bibfnamefont {T.}~\bibnamefont {Hashimoto}}, \bibinfo {author}
  {\bibfnamefont {N.}~\bibnamefont {Matsuda}}, \bibinfo {author} {\bibfnamefont
  {J.~L.}\ \bibnamefont {O'Brien}}, \bibinfo {author} {\bibfnamefont {D.~P.}\
  \bibnamefont {Tew}}, \ and\ \bibinfo {author} {\bibfnamefont
  {A.}~\bibnamefont {Laing}},\ }\bibfield  {title} {\enquote {\bibinfo {title}
  {Simulating the vibrational quantum dynamics of molecules using photonics},}\
  }\href {\doibase 10.1038/s41586-018-0152-9} {\bibfield  {journal} {\bibinfo
  {journal} {Nature}\ }\textbf {\bibinfo {volume} {557}},\ \bibinfo {pages}
  {660--667} (\bibinfo {year} {2018})}\BibitemShut {NoStop}%
\bibitem [{\citenamefont {Clements}\ \emph {et~al.}(2018)\citenamefont
  {Clements}, \citenamefont {Renema}, \citenamefont {Eckstein}, \citenamefont
  {Valido}, \citenamefont {Lita}, \citenamefont {Gerrits}, \citenamefont {Nam},
  \citenamefont {Kolthammer}, \citenamefont {Huh},\ and\ \citenamefont
  {Walmsley}}]{Clements2018}%
  \BibitemOpen
  \bibfield  {author} {\bibinfo {author} {\bibfnamefont {W.~R.}\ \bibnamefont
  {Clements}}, \bibinfo {author} {\bibfnamefont {J.~J.}\ \bibnamefont
  {Renema}}, \bibinfo {author} {\bibfnamefont {A.}~\bibnamefont {Eckstein}},
  \bibinfo {author} {\bibfnamefont {A.~A.}\ \bibnamefont {Valido}}, \bibinfo
  {author} {\bibfnamefont {A.}~\bibnamefont {Lita}}, \bibinfo {author}
  {\bibfnamefont {T.}~\bibnamefont {Gerrits}}, \bibinfo {author} {\bibfnamefont
  {S.~W.}\ \bibnamefont {Nam}}, \bibinfo {author} {\bibfnamefont {W.~S.}\
  \bibnamefont {Kolthammer}}, \bibinfo {author} {\bibfnamefont
  {J.}~\bibnamefont {Huh}}, \ and\ \bibinfo {author} {\bibfnamefont {I.~A.}\
  \bibnamefont {Walmsley}},\ }\bibfield  {title} {\enquote {\bibinfo {title}
  {Approximating vibronic spectroscopy with imperfect quantum optics},}\ }\href
  {\doibase 10.1088/1361-6455/aaf031} {\bibfield  {journal} {\bibinfo
  {journal} {Journal of Physics B: Atomic, Molecular and Optical Physics}\
  }\textbf {\bibinfo {volume} {51}},\ \bibinfo {pages} {245503} (\bibinfo
  {year} {2018})}\BibitemShut {NoStop}%
\bibitem [{\citenamefont {{Wang}}\ \emph {et~al.}(2019)\citenamefont {{Wang}},
  \citenamefont {{Curtis}}, \citenamefont {{Lester}}, \citenamefont {{Zhang}},
  \citenamefont {{Gao}}, \citenamefont {{Freeze}}, \citenamefont {{Batista}},
  \citenamefont {{Vaccaro}}, \citenamefont {{Chuang}}, \citenamefont
  {{Frunzio}}, \citenamefont {{Jiang}}, \citenamefont {{Girvin}},\ and\
  \citenamefont {{Schoelkopf}}}]{Wang2019}%
  \BibitemOpen
  \bibfield  {author} {\bibinfo {author} {\bibfnamefont {C.~S.}\ \bibnamefont
  {{Wang}}}, \bibinfo {author} {\bibfnamefont {J.~C.}\ \bibnamefont
  {{Curtis}}}, \bibinfo {author} {\bibfnamefont {B.~J.}\ \bibnamefont
  {{Lester}}}, \bibinfo {author} {\bibfnamefont {Y.}~\bibnamefont {{Zhang}}},
  \bibinfo {author} {\bibfnamefont {Y.~Y.}\ \bibnamefont {{Gao}}}, \bibinfo
  {author} {\bibfnamefont {J.}~\bibnamefont {{Freeze}}}, \bibinfo {author}
  {\bibfnamefont {V.~S.}\ \bibnamefont {{Batista}}}, \bibinfo {author}
  {\bibfnamefont {P.~H.}\ \bibnamefont {{Vaccaro}}}, \bibinfo {author}
  {\bibfnamefont {I.~L.}\ \bibnamefont {{Chuang}}}, \bibinfo {author}
  {\bibfnamefont {L.}~\bibnamefont {{Frunzio}}}, \bibinfo {author}
  {\bibfnamefont {L.}~\bibnamefont {{Jiang}}}, \bibinfo {author} {\bibfnamefont
  {S.~M.}\ \bibnamefont {{Girvin}}}, \ and\ \bibinfo {author} {\bibfnamefont
  {R.~J.}\ \bibnamefont {{Schoelkopf}}},\ }\bibfield  {title} {\enquote
  {\bibinfo {title} {{Efficient multiphoton sampling of molecular vibronic
  spectra on a superconducting bosonic processor}},}\ }\href@noop {} {\bibfield
   {journal} {\bibinfo  {journal} {arXiv e-prints}\ ,\ \bibinfo {eid}
  {arXiv:1908.03598}} (\bibinfo {year} {2019})},\ \Eprint
  {http://arxiv.org/abs/1908.03598} {arXiv:1908.03598 [quant-ph]} \BibitemShut
  {NoStop}%
\bibitem [{\citenamefont {{Gottesman}}(2009)}]{Gottesman2009}%
  \BibitemOpen
  \bibfield  {author} {\bibinfo {author} {\bibfnamefont {D.}~\bibnamefont
  {{Gottesman}}},\ }\bibfield  {title} {\enquote {\bibinfo {title} {{An
  Introduction to Quantum Error Correction and Fault-Tolerant Quantum
  Computation}},}\ }\href@noop {} {\bibfield  {journal} {\bibinfo  {journal}
  {arXiv e-prints}\ ,\ \bibinfo {eid} {arXiv:0904.2557}} (\bibinfo {year}
  {2009})},\ \Eprint {http://arxiv.org/abs/0904.2557} {arXiv:0904.2557
  [quant-ph]} \BibitemShut {NoStop}%
\bibitem [{\citenamefont {Albert}\ \emph {et~al.}(2018)\citenamefont {Albert},
  \citenamefont {Noh}, \citenamefont {Duivenvoorden}, \citenamefont {Young},
  \citenamefont {Brierley}, \citenamefont {Reinhold}, \citenamefont {Vuillot},
  \citenamefont {Li}, \citenamefont {Shen}, \citenamefont {Girvin},
  \citenamefont {Terhal},\ and\ \citenamefont {Jiang}}]{Albert2018}%
  \BibitemOpen
  \bibfield  {author} {\bibinfo {author} {\bibfnamefont {V.~V.}\ \bibnamefont
  {Albert}}, \bibinfo {author} {\bibfnamefont {K.}~\bibnamefont {Noh}},
  \bibinfo {author} {\bibfnamefont {K.}~\bibnamefont {Duivenvoorden}}, \bibinfo
  {author} {\bibfnamefont {D.~J.}\ \bibnamefont {Young}}, \bibinfo {author}
  {\bibfnamefont {R.~T.}\ \bibnamefont {Brierley}}, \bibinfo {author}
  {\bibfnamefont {P.}~\bibnamefont {Reinhold}}, \bibinfo {author}
  {\bibfnamefont {C.}~\bibnamefont {Vuillot}}, \bibinfo {author} {\bibfnamefont
  {L.}~\bibnamefont {Li}}, \bibinfo {author} {\bibfnamefont {C.}~\bibnamefont
  {Shen}}, \bibinfo {author} {\bibfnamefont {S.~M.}\ \bibnamefont {Girvin}},
  \bibinfo {author} {\bibfnamefont {B.~M.}\ \bibnamefont {Terhal}}, \ and\
  \bibinfo {author} {\bibfnamefont {L.}~\bibnamefont {Jiang}},\ }\bibfield
  {title} {\enquote {\bibinfo {title} {Performance and structure of single-mode
  bosonic codes},}\ }\href {\doibase 10.1103/PhysRevA.97.032346} {\bibfield
  {journal} {\bibinfo  {journal} {Phys. Rev. A}\ }\textbf {\bibinfo {volume}
  {97}},\ \bibinfo {pages} {032346} (\bibinfo {year} {2018})}\BibitemShut
  {NoStop}%
\bibitem [{\citenamefont {Cochrane}\ \emph {et~al.}(1999)\citenamefont
  {Cochrane}, \citenamefont {Milburn},\ and\ \citenamefont
  {Munro}}]{Cochrane1999}%
  \BibitemOpen
  \bibfield  {author} {\bibinfo {author} {\bibfnamefont {P.~T.}\ \bibnamefont
  {Cochrane}}, \bibinfo {author} {\bibfnamefont {G.~J.}\ \bibnamefont
  {Milburn}}, \ and\ \bibinfo {author} {\bibfnamefont {W.~J.}\ \bibnamefont
  {Munro}},\ }\bibfield  {title} {\enquote {\bibinfo {title} {Macroscopically
  distinct quantum-superposition states as a bosonic code for amplitude
  damping},}\ }\href {\doibase 10.1103/PhysRevA.59.2631} {\bibfield  {journal}
  {\bibinfo  {journal} {Phys. Rev. A}\ }\textbf {\bibinfo {volume} {59}},\
  \bibinfo {pages} {2631--2634} (\bibinfo {year} {1999})}\BibitemShut {NoStop}%
\bibitem [{\citenamefont {Gottesman}\ \emph {et~al.}(2001)\citenamefont
  {Gottesman}, \citenamefont {Kitaev},\ and\ \citenamefont
  {Preskill}}]{Gottesman2001}%
  \BibitemOpen
  \bibfield  {author} {\bibinfo {author} {\bibfnamefont {D.}~\bibnamefont
  {Gottesman}}, \bibinfo {author} {\bibfnamefont {A.}~\bibnamefont {Kitaev}}, \
  and\ \bibinfo {author} {\bibfnamefont {J.}~\bibnamefont {Preskill}},\
  }\bibfield  {title} {\enquote {\bibinfo {title} {Encoding a qubit in an
  oscillator},}\ }\href {\doibase 10.1103/PhysRevA.64.012310} {\bibfield
  {journal} {\bibinfo  {journal} {Phys. Rev. A}\ }\textbf {\bibinfo {volume}
  {64}},\ \bibinfo {pages} {012310} (\bibinfo {year} {2001})}\BibitemShut
  {NoStop}%
\bibitem [{\citenamefont {Leghtas}\ \emph {et~al.}(2013)\citenamefont
  {Leghtas}, \citenamefont {Kirchmair}, \citenamefont {Vlastakis},
  \citenamefont {Schoelkopf}, \citenamefont {Devoret},\ and\ \citenamefont
  {Mirrahimi}}]{Leghtas2013}%
  \BibitemOpen
  \bibfield  {author} {\bibinfo {author} {\bibfnamefont {Z.}~\bibnamefont
  {Leghtas}}, \bibinfo {author} {\bibfnamefont {G.}~\bibnamefont {Kirchmair}},
  \bibinfo {author} {\bibfnamefont {B.}~\bibnamefont {Vlastakis}}, \bibinfo
  {author} {\bibfnamefont {R.~J.}\ \bibnamefont {Schoelkopf}}, \bibinfo
  {author} {\bibfnamefont {M.~H.}\ \bibnamefont {Devoret}}, \ and\ \bibinfo
  {author} {\bibfnamefont {M.}~\bibnamefont {Mirrahimi}},\ }\bibfield  {title}
  {\enquote {\bibinfo {title} {Hardware-efficient autonomous quantum memory
  protection},}\ }\href {\doibase 10.1103/PhysRevLett.111.120501} {\bibfield
  {journal} {\bibinfo  {journal} {Phys. Rev. Lett.}\ }\textbf {\bibinfo
  {volume} {111}},\ \bibinfo {pages} {120501} (\bibinfo {year}
  {2013})}\BibitemShut {NoStop}%
\bibitem [{\citenamefont {Mirrahimi}\ \emph {et~al.}(2014)\citenamefont
  {Mirrahimi}, \citenamefont {Leghtas}, \citenamefont {Albert}, \citenamefont
  {Touzard}, \citenamefont {Schoelkopf}, \citenamefont {Jiang},\ and\
  \citenamefont {Devoret}}]{Mirrahimi2014}%
  \BibitemOpen
  \bibfield  {author} {\bibinfo {author} {\bibfnamefont {M.}~\bibnamefont
  {Mirrahimi}}, \bibinfo {author} {\bibfnamefont {Z.}~\bibnamefont {Leghtas}},
  \bibinfo {author} {\bibfnamefont {V.~V.}\ \bibnamefont {Albert}}, \bibinfo
  {author} {\bibfnamefont {S.}~\bibnamefont {Touzard}}, \bibinfo {author}
  {\bibfnamefont {R.~J.}\ \bibnamefont {Schoelkopf}}, \bibinfo {author}
  {\bibfnamefont {L.}~\bibnamefont {Jiang}}, \ and\ \bibinfo {author}
  {\bibfnamefont {M.~H.}\ \bibnamefont {Devoret}},\ }\bibfield  {title}
  {\enquote {\bibinfo {title} {Dynamically protected cat-qubits: a new paradigm
  for universal quantum computation},}\ }\href {\doibase
  10.1088/1367-2630/16/4/045014} {\bibfield  {journal} {\bibinfo  {journal}
  {New Journal of Physics}\ }\textbf {\bibinfo {volume} {16}},\ \bibinfo
  {pages} {045014} (\bibinfo {year} {2014})}\BibitemShut {NoStop}%
\bibitem [{\citenamefont {Michael}\ \emph {et~al.}(2016)\citenamefont
  {Michael}, \citenamefont {Silveri}, \citenamefont {Brierley}, \citenamefont
  {Albert}, \citenamefont {Salmilehto}, \citenamefont {Jiang},\ and\
  \citenamefont {Girvin}}]{Michael2016}%
  \BibitemOpen
  \bibfield  {author} {\bibinfo {author} {\bibfnamefont {M.~H.}\ \bibnamefont
  {Michael}}, \bibinfo {author} {\bibfnamefont {M.}~\bibnamefont {Silveri}},
  \bibinfo {author} {\bibfnamefont {R.~T.}\ \bibnamefont {Brierley}}, \bibinfo
  {author} {\bibfnamefont {V.~V.}\ \bibnamefont {Albert}}, \bibinfo {author}
  {\bibfnamefont {J.}~\bibnamefont {Salmilehto}}, \bibinfo {author}
  {\bibfnamefont {L.}~\bibnamefont {Jiang}}, \ and\ \bibinfo {author}
  {\bibfnamefont {S.~M.}\ \bibnamefont {Girvin}},\ }\bibfield  {title}
  {\enquote {\bibinfo {title} {New class of quantum error-correcting codes for
  a bosonic mode},}\ }\href {\doibase 10.1103/PhysRevX.6.031006} {\bibfield
  {journal} {\bibinfo  {journal} {Phys. Rev. X}\ }\textbf {\bibinfo {volume}
  {6}},\ \bibinfo {pages} {031006} (\bibinfo {year} {2016})}\BibitemShut
  {NoStop}%
\bibitem [{\citenamefont {Li}\ \emph {et~al.}(2017)\citenamefont {Li},
  \citenamefont {Zou}, \citenamefont {Albert}, \citenamefont {Muralidharan},
  \citenamefont {Girvin},\ and\ \citenamefont {Jiang}}]{Li2017}%
  \BibitemOpen
  \bibfield  {author} {\bibinfo {author} {\bibfnamefont {L.}~\bibnamefont
  {Li}}, \bibinfo {author} {\bibfnamefont {C.-L.}\ \bibnamefont {Zou}},
  \bibinfo {author} {\bibfnamefont {V.~V.}\ \bibnamefont {Albert}}, \bibinfo
  {author} {\bibfnamefont {S.}~\bibnamefont {Muralidharan}}, \bibinfo {author}
  {\bibfnamefont {S.~M.}\ \bibnamefont {Girvin}}, \ and\ \bibinfo {author}
  {\bibfnamefont {L.}~\bibnamefont {Jiang}},\ }\bibfield  {title} {\enquote
  {\bibinfo {title} {Cat codes with optimal decoherence suppression for a lossy
  bosonic channel},}\ }\href {\doibase 10.1103/PhysRevLett.119.030502}
  {\bibfield  {journal} {\bibinfo  {journal} {Phys. Rev. Lett.}\ }\textbf
  {\bibinfo {volume} {119}},\ \bibinfo {pages} {030502} (\bibinfo {year}
  {2017})}\BibitemShut {NoStop}%
\bibitem [{\citenamefont {Chuang}\ \emph {et~al.}(1997)\citenamefont {Chuang},
  \citenamefont {Leung},\ and\ \citenamefont {Yamamoto}}]{Chuang1997}%
  \BibitemOpen
  \bibfield  {author} {\bibinfo {author} {\bibfnamefont {I.~L.}\ \bibnamefont
  {Chuang}}, \bibinfo {author} {\bibfnamefont {D.~W.}\ \bibnamefont {Leung}}, \
  and\ \bibinfo {author} {\bibfnamefont {Y.}~\bibnamefont {Yamamoto}},\
  }\bibfield  {title} {\enquote {\bibinfo {title} {Bosonic quantum codes for
  amplitude damping},}\ }\href {\doibase 10.1103/PhysRevA.56.1114} {\bibfield
  {journal} {\bibinfo  {journal} {Phys. Rev. A}\ }\textbf {\bibinfo {volume}
  {56}},\ \bibinfo {pages} {1114--1125} (\bibinfo {year} {1997})}\BibitemShut
  {NoStop}%
\bibitem [{\citenamefont {Harrington}\ and\ \citenamefont
  {Preskill}(2001)}]{Harrington2001}%
  \BibitemOpen
  \bibfield  {author} {\bibinfo {author} {\bibfnamefont {J.}~\bibnamefont
  {Harrington}}\ and\ \bibinfo {author} {\bibfnamefont {J.}~\bibnamefont
  {Preskill}},\ }\bibfield  {title} {\enquote {\bibinfo {title} {Achievable
  rates for the {G}aussian quantum channel},}\ }\href {\doibase
  10.1103/PhysRevA.64.062301} {\bibfield  {journal} {\bibinfo  {journal} {Phys.
  Rev. A}\ }\textbf {\bibinfo {volume} {64}},\ \bibinfo {pages} {062301}
  (\bibinfo {year} {2001})}\BibitemShut {NoStop}%
\bibitem [{\citenamefont {Bergmann}\ and\ \citenamefont {van
  Loock}(2016)}]{Bergmann2016a}%
  \BibitemOpen
  \bibfield  {author} {\bibinfo {author} {\bibfnamefont {M.}~\bibnamefont
  {Bergmann}}\ and\ \bibinfo {author} {\bibfnamefont {P.}~\bibnamefont {van
  Loock}},\ }\bibfield  {title} {\enquote {\bibinfo {title} {{Quantum error
  correction against photon loss using NOON states}},}\ }\href {\doibase
  10.1103/PhysRevA.94.012311} {\bibfield  {journal} {\bibinfo  {journal} {Phys.
  Rev. A}\ }\textbf {\bibinfo {volume} {94}},\ \bibinfo {pages} {012311}
  (\bibinfo {year} {2016})}\BibitemShut {NoStop}%
\bibitem [{\citenamefont {Fukui}\ \emph {et~al.}(2017)\citenamefont {Fukui},
  \citenamefont {Tomita},\ and\ \citenamefont {Okamoto}}]{Fukui2017}%
  \BibitemOpen
  \bibfield  {author} {\bibinfo {author} {\bibfnamefont {K.}~\bibnamefont
  {Fukui}}, \bibinfo {author} {\bibfnamefont {A.}~\bibnamefont {Tomita}}, \
  and\ \bibinfo {author} {\bibfnamefont {A.}~\bibnamefont {Okamoto}},\
  }\bibfield  {title} {\enquote {\bibinfo {title} {Analog quantum error
  correction with encoding a qubit into an oscillator},}\ }\href {\doibase
  10.1103/PhysRevLett.119.180507} {\bibfield  {journal} {\bibinfo  {journal}
  {Phys. Rev. Lett.}\ }\textbf {\bibinfo {volume} {119}},\ \bibinfo {pages}
  {180507} (\bibinfo {year} {2017})}\BibitemShut {NoStop}%
\bibitem [{\citenamefont {Niu}\ \emph {et~al.}(2018)\citenamefont {Niu},
  \citenamefont {Chuang},\ and\ \citenamefont {Shapiro}}]{Niu2018}%
  \BibitemOpen
  \bibfield  {author} {\bibinfo {author} {\bibfnamefont {M.~Y.}\ \bibnamefont
  {Niu}}, \bibinfo {author} {\bibfnamefont {I.~L.}\ \bibnamefont {Chuang}}, \
  and\ \bibinfo {author} {\bibfnamefont {J.~H.}\ \bibnamefont {Shapiro}},\
  }\bibfield  {title} {\enquote {\bibinfo {title} {Hardware-efficient bosonic
  quantum error-correcting codes based on symmetry operators},}\ }\href
  {\doibase 10.1103/PhysRevA.97.032323} {\bibfield  {journal} {\bibinfo
  {journal} {Phys. Rev. A}\ }\textbf {\bibinfo {volume} {97}},\ \bibinfo
  {pages} {032323} (\bibinfo {year} {2018})}\BibitemShut {NoStop}%
\bibitem [{\citenamefont {Fukui}\ \emph
  {et~al.}(2018{\natexlab{a}})\citenamefont {Fukui}, \citenamefont {Tomita},
  \citenamefont {Okamoto},\ and\ \citenamefont {Fujii}}]{Fukui2018a}%
  \BibitemOpen
  \bibfield  {author} {\bibinfo {author} {\bibfnamefont {K.}~\bibnamefont
  {Fukui}}, \bibinfo {author} {\bibfnamefont {A.}~\bibnamefont {Tomita}},
  \bibinfo {author} {\bibfnamefont {A.}~\bibnamefont {Okamoto}}, \ and\
  \bibinfo {author} {\bibfnamefont {K.}~\bibnamefont {Fujii}},\ }\bibfield
  {title} {\enquote {\bibinfo {title} {High-threshold fault-tolerant quantum
  computation with analog quantum error correction},}\ }\href {\doibase
  10.1103/PhysRevX.8.021054} {\bibfield  {journal} {\bibinfo  {journal} {Phys.
  Rev. X}\ }\textbf {\bibinfo {volume} {8}},\ \bibinfo {pages} {021054}
  (\bibinfo {year} {2018}{\natexlab{a}})}\BibitemShut {NoStop}%
\bibitem [{\citenamefont {Fukui}\ \emph
  {et~al.}(2018{\natexlab{b}})\citenamefont {Fukui}, \citenamefont {Tomita},\
  and\ \citenamefont {Okamoto}}]{Fukui2018b}%
  \BibitemOpen
  \bibfield  {author} {\bibinfo {author} {\bibfnamefont {K.}~\bibnamefont
  {Fukui}}, \bibinfo {author} {\bibfnamefont {A.}~\bibnamefont {Tomita}}, \
  and\ \bibinfo {author} {\bibfnamefont {A.}~\bibnamefont {Okamoto}},\
  }\bibfield  {title} {\enquote {\bibinfo {title} {Tracking quantum error
  correction},}\ }\href {\doibase 10.1103/PhysRevA.98.022326} {\bibfield
  {journal} {\bibinfo  {journal} {Phys. Rev. A}\ }\textbf {\bibinfo {volume}
  {98}},\ \bibinfo {pages} {022326} (\bibinfo {year}
  {2018}{\natexlab{b}})}\BibitemShut {NoStop}%
\bibitem [{\citenamefont {Vuillot}\ \emph {et~al.}(2019)\citenamefont
  {Vuillot}, \citenamefont {Asasi}, \citenamefont {Wang}, \citenamefont
  {Pryadko},\ and\ \citenamefont {Terhal}}]{Vuillot2019}%
  \BibitemOpen
  \bibfield  {author} {\bibinfo {author} {\bibfnamefont {C.}~\bibnamefont
  {Vuillot}}, \bibinfo {author} {\bibfnamefont {H.}~\bibnamefont {Asasi}},
  \bibinfo {author} {\bibfnamefont {Y.}~\bibnamefont {Wang}}, \bibinfo {author}
  {\bibfnamefont {L.~P.}\ \bibnamefont {Pryadko}}, \ and\ \bibinfo {author}
  {\bibfnamefont {B.~M.}\ \bibnamefont {Terhal}},\ }\bibfield  {title}
  {\enquote {\bibinfo {title} {Quantum error correction with the toric
  {G}ottesman-{K}itaev-{P}reskill code},}\ }\href {\doibase
  10.1103/PhysRevA.99.032344} {\bibfield  {journal} {\bibinfo  {journal} {Phys.
  Rev. A}\ }\textbf {\bibinfo {volume} {99}},\ \bibinfo {pages} {032344}
  (\bibinfo {year} {2019})}\BibitemShut {NoStop}%
\bibitem [{\citenamefont {{Fukui}}(2019)}]{Fukui2019}%
  \BibitemOpen
  \bibfield  {author} {\bibinfo {author} {\bibfnamefont {K.}~\bibnamefont
  {{Fukui}}},\ }\bibfield  {title} {\enquote {\bibinfo {title} {{High-threshold
  fault-tolerant quantum computation with the GKP qubit and realistically noisy
  devices}},}\ }\href@noop {} {\bibfield  {journal} {\bibinfo  {journal} {arXiv
  e-prints}\ ,\ \bibinfo {eid} {arXiv:1906.09767}} (\bibinfo {year} {2019})},\
  \Eprint {http://arxiv.org/abs/1906.09767} {arXiv:1906.09767 [quant-ph]}
  \BibitemShut {NoStop}%
\bibitem [{\citenamefont {{Noh}}\ and\ \citenamefont
  {{Chamberland}}(2019)}]{Noh2019b}%
  \BibitemOpen
  \bibfield  {author} {\bibinfo {author} {\bibfnamefont {K.}~\bibnamefont
  {{Noh}}}\ and\ \bibinfo {author} {\bibfnamefont {C.}~\bibnamefont
  {{Chamberland}}},\ }\bibfield  {title} {\enquote {\bibinfo {title}
  {{Fault-tolerant bosonic quantum error correction with the surface-GKP
  code}},}\ }\href@noop {} {\bibfield  {journal} {\bibinfo  {journal} {arXiv
  e-prints}\ ,\ \bibinfo {eid} {arXiv:1908.03579}} (\bibinfo {year} {2019})},\
  \Eprint {http://arxiv.org/abs/1908.03579} {arXiv:1908.03579 [quant-ph]}
  \BibitemShut {NoStop}%
\bibitem [{\citenamefont {Leghtas}\ \emph {et~al.}(2015)\citenamefont
  {Leghtas}, \citenamefont {Touzard}, \citenamefont {Pop}, \citenamefont {Kou},
  \citenamefont {Vlastakis}, \citenamefont {Petrenko}, \citenamefont {Sliwa},
  \citenamefont {Narla}, \citenamefont {Shankar}, \citenamefont {Hatridge},
  \citenamefont {Reagor}, \citenamefont {Frunzio}, \citenamefont {Schoelkopf},
  \citenamefont {Mirrahimi},\ and\ \citenamefont {Devoret}}]{Leghtas2015}%
  \BibitemOpen
  \bibfield  {author} {\bibinfo {author} {\bibfnamefont {Z.}~\bibnamefont
  {Leghtas}}, \bibinfo {author} {\bibfnamefont {S.}~\bibnamefont {Touzard}},
  \bibinfo {author} {\bibfnamefont {I.~M.}\ \bibnamefont {Pop}}, \bibinfo
  {author} {\bibfnamefont {A.}~\bibnamefont {Kou}}, \bibinfo {author}
  {\bibfnamefont {B.}~\bibnamefont {Vlastakis}}, \bibinfo {author}
  {\bibfnamefont {A.}~\bibnamefont {Petrenko}}, \bibinfo {author}
  {\bibfnamefont {K.~M.}\ \bibnamefont {Sliwa}}, \bibinfo {author}
  {\bibfnamefont {A.}~\bibnamefont {Narla}}, \bibinfo {author} {\bibfnamefont
  {S.}~\bibnamefont {Shankar}}, \bibinfo {author} {\bibfnamefont {M.~J.}\
  \bibnamefont {Hatridge}}, \bibinfo {author} {\bibfnamefont {M.}~\bibnamefont
  {Reagor}}, \bibinfo {author} {\bibfnamefont {L.}~\bibnamefont {Frunzio}},
  \bibinfo {author} {\bibfnamefont {R.~J.}\ \bibnamefont {Schoelkopf}},
  \bibinfo {author} {\bibfnamefont {M.}~\bibnamefont {Mirrahimi}}, \ and\
  \bibinfo {author} {\bibfnamefont {M.~H.}\ \bibnamefont {Devoret}},\
  }\bibfield  {title} {\enquote {\bibinfo {title} {Confining the state of light
  to a quantum manifold by engineered two-photon loss},}\ }\href {\doibase
  10.1126/science.aaa2085} {\bibfield  {journal} {\bibinfo  {journal}
  {Science}\ }\textbf {\bibinfo {volume} {347}},\ \bibinfo {pages} {853--857}
  (\bibinfo {year} {2015})}\BibitemShut {NoStop}%
\bibitem [{\citenamefont {Ofek}\ \emph {et~al.}(2016)\citenamefont {Ofek},
  \citenamefont {Petrenko}, \citenamefont {Heeres}, \citenamefont {Reinhold},
  \citenamefont {Leghtas}, \citenamefont {Vlastakis}, \citenamefont {Liu},
  \citenamefont {Frunzio}, \citenamefont {Girvin}, \citenamefont {Jiang},
  \citenamefont {Mirrahimi}, \citenamefont {Devoret},\ and\ \citenamefont
  {Schoelkopf}}]{Ofek2016}%
  \BibitemOpen
  \bibfield  {author} {\bibinfo {author} {\bibfnamefont {N.}~\bibnamefont
  {Ofek}}, \bibinfo {author} {\bibfnamefont {A.}~\bibnamefont {Petrenko}},
  \bibinfo {author} {\bibfnamefont {R.}~\bibnamefont {Heeres}}, \bibinfo
  {author} {\bibfnamefont {P.}~\bibnamefont {Reinhold}}, \bibinfo {author}
  {\bibfnamefont {Z.}~\bibnamefont {Leghtas}}, \bibinfo {author} {\bibfnamefont
  {B.}~\bibnamefont {Vlastakis}}, \bibinfo {author} {\bibfnamefont
  {Y.}~\bibnamefont {Liu}}, \bibinfo {author} {\bibfnamefont {L.}~\bibnamefont
  {Frunzio}}, \bibinfo {author} {\bibfnamefont {S.~M.}\ \bibnamefont {Girvin}},
  \bibinfo {author} {\bibfnamefont {L.}~\bibnamefont {Jiang}}, \bibinfo
  {author} {\bibfnamefont {M.}~\bibnamefont {Mirrahimi}}, \bibinfo {author}
  {\bibfnamefont {M.~H.}\ \bibnamefont {Devoret}}, \ and\ \bibinfo {author}
  {\bibfnamefont {R.~J.}\ \bibnamefont {Schoelkopf}},\ }\bibfield  {title}
  {\enquote {\bibinfo {title} {Extending the lifetime of a quantum bit with
  error correction in superconducting circuits},}\ }\href
  {http://dx.doi.org/10.1038/nature18949} {\bibfield  {journal} {\bibinfo
  {journal} {Nature}\ }\textbf {\bibinfo {volume} {536}},\ \bibinfo {pages}
  {441--445} (\bibinfo {year} {2016})}\BibitemShut {NoStop}%
\bibitem [{\citenamefont {Touzard}\ \emph {et~al.}(2018)\citenamefont
  {Touzard}, \citenamefont {Grimm}, \citenamefont {Leghtas}, \citenamefont
  {Mundhada}, \citenamefont {Reinhold}, \citenamefont {Axline}, \citenamefont
  {Reagor}, \citenamefont {Chou}, \citenamefont {Blumoff}, \citenamefont
  {Sliwa}, \citenamefont {Shankar}, \citenamefont {Frunzio}, \citenamefont
  {Schoelkopf}, \citenamefont {Mirrahimi},\ and\ \citenamefont
  {Devoret}}]{Touzard2017}%
  \BibitemOpen
  \bibfield  {author} {\bibinfo {author} {\bibfnamefont {S.}~\bibnamefont
  {Touzard}}, \bibinfo {author} {\bibfnamefont {A.}~\bibnamefont {Grimm}},
  \bibinfo {author} {\bibfnamefont {Z.}~\bibnamefont {Leghtas}}, \bibinfo
  {author} {\bibfnamefont {S.~O.}\ \bibnamefont {Mundhada}}, \bibinfo {author}
  {\bibfnamefont {P.}~\bibnamefont {Reinhold}}, \bibinfo {author}
  {\bibfnamefont {C.}~\bibnamefont {Axline}}, \bibinfo {author} {\bibfnamefont
  {M.}~\bibnamefont {Reagor}}, \bibinfo {author} {\bibfnamefont
  {K.}~\bibnamefont {Chou}}, \bibinfo {author} {\bibfnamefont {J.}~\bibnamefont
  {Blumoff}}, \bibinfo {author} {\bibfnamefont {K.~M.}\ \bibnamefont {Sliwa}},
  \bibinfo {author} {\bibfnamefont {S.}~\bibnamefont {Shankar}}, \bibinfo
  {author} {\bibfnamefont {L.}~\bibnamefont {Frunzio}}, \bibinfo {author}
  {\bibfnamefont {R.~J.}\ \bibnamefont {Schoelkopf}}, \bibinfo {author}
  {\bibfnamefont {M.}~\bibnamefont {Mirrahimi}}, \ and\ \bibinfo {author}
  {\bibfnamefont {M.~H.}\ \bibnamefont {Devoret}},\ }\bibfield  {title}
  {\enquote {\bibinfo {title} {Coherent oscillations inside a quantum manifold
  stabilized by dissipation},}\ }\href {\doibase 10.1103/PhysRevX.8.021005}
  {\bibfield  {journal} {\bibinfo  {journal} {Phys. Rev. X}\ }\textbf {\bibinfo
  {volume} {8}},\ \bibinfo {pages} {021005} (\bibinfo {year}
  {2018})}\BibitemShut {NoStop}%
\bibitem [{\citenamefont {Hu}\ \emph {et~al.}(2019)\citenamefont {Hu},
  \citenamefont {Ma}, \citenamefont {Cai}, \citenamefont {Mu}, \citenamefont
  {Xu}, \citenamefont {Wang}, \citenamefont {Wu}, \citenamefont {Wang},
  \citenamefont {Song}, \citenamefont {Zou}, \citenamefont {Girvin},
  \citenamefont {Duan},\ and\ \citenamefont {Sun}}]{Hu2019}%
  \BibitemOpen
  \bibfield  {author} {\bibinfo {author} {\bibfnamefont {L.}~\bibnamefont
  {Hu}}, \bibinfo {author} {\bibfnamefont {Y.}~\bibnamefont {Ma}}, \bibinfo
  {author} {\bibfnamefont {W.}~\bibnamefont {Cai}}, \bibinfo {author}
  {\bibfnamefont {X.}~\bibnamefont {Mu}}, \bibinfo {author} {\bibfnamefont
  {Y.}~\bibnamefont {Xu}}, \bibinfo {author} {\bibfnamefont {W.}~\bibnamefont
  {Wang}}, \bibinfo {author} {\bibfnamefont {Y.}~\bibnamefont {Wu}}, \bibinfo
  {author} {\bibfnamefont {H.}~\bibnamefont {Wang}}, \bibinfo {author}
  {\bibfnamefont {Y.~P.}\ \bibnamefont {Song}}, \bibinfo {author}
  {\bibfnamefont {C.-L.}\ \bibnamefont {Zou}}, \bibinfo {author} {\bibfnamefont
  {S.~M.}\ \bibnamefont {Girvin}}, \bibinfo {author} {\bibfnamefont {L.-M.}\
  \bibnamefont {Duan}}, \ and\ \bibinfo {author} {\bibfnamefont
  {L.}~\bibnamefont {Sun}},\ }\bibfield  {title} {\enquote {\bibinfo {title}
  {Quantum error correction and universal gate set operation on a binomial
  bosonic logical qubit},}\ }\href {\doibase 10.1038/s41567-018-0414-3}
  {\bibfield  {journal} {\bibinfo  {journal} {Nature Physics}\ } (\bibinfo
  {year} {2019}),\ 10.1038/s41567-018-0414-3}\BibitemShut {NoStop}%
\bibitem [{\citenamefont {Fl\"uhmann}\ \emph {et~al.}(2018)\citenamefont
  {Fl\"uhmann}, \citenamefont {Negnevitsky}, \citenamefont {Marinelli},\ and\
  \citenamefont {Home}}]{Fluhmann2018}%
  \BibitemOpen
  \bibfield  {author} {\bibinfo {author} {\bibfnamefont {C.}~\bibnamefont
  {Fl\"uhmann}}, \bibinfo {author} {\bibfnamefont {V.}~\bibnamefont
  {Negnevitsky}}, \bibinfo {author} {\bibfnamefont {M.}~\bibnamefont
  {Marinelli}}, \ and\ \bibinfo {author} {\bibfnamefont {J.~P.}\ \bibnamefont
  {Home}},\ }\bibfield  {title} {\enquote {\bibinfo {title} {Sequential modular
  position and momentum measurements of a trapped ion mechanical oscillator},}\
  }\href {\doibase 10.1103/PhysRevX.8.021001} {\bibfield  {journal} {\bibinfo
  {journal} {Phys. Rev. X}\ }\textbf {\bibinfo {volume} {8}},\ \bibinfo {pages}
  {021001} (\bibinfo {year} {2018})}\BibitemShut {NoStop}%
\bibitem [{\citenamefont {Fl{\"u}hmann}\ \emph {et~al.}(2019)\citenamefont
  {Fl{\"u}hmann}, \citenamefont {Nguyen}, \citenamefont {Marinelli},
  \citenamefont {Negnevitsky}, \citenamefont {Mehta},\ and\ \citenamefont
  {Home}}]{Fluhmann2019}%
  \BibitemOpen
  \bibfield  {author} {\bibinfo {author} {\bibfnamefont {C.}~\bibnamefont
  {Fl{\"u}hmann}}, \bibinfo {author} {\bibfnamefont {T.~L.}\ \bibnamefont
  {Nguyen}}, \bibinfo {author} {\bibfnamefont {M.}~\bibnamefont {Marinelli}},
  \bibinfo {author} {\bibfnamefont {V.}~\bibnamefont {Negnevitsky}}, \bibinfo
  {author} {\bibfnamefont {K.}~\bibnamefont {Mehta}}, \ and\ \bibinfo {author}
  {\bibfnamefont {J.~P.}\ \bibnamefont {Home}},\ }\bibfield  {title} {\enquote
  {\bibinfo {title} {Encoding a qubit in a trapped-ion mechanical
  oscillator},}\ }\href {\doibase 10.1038/s41586-019-0960-6} {\bibfield
  {journal} {\bibinfo  {journal} {Nature}\ }\textbf {\bibinfo {volume} {566}},\
  \bibinfo {pages} {513--517} (\bibinfo {year} {2019})}\BibitemShut {NoStop}%
\bibitem [{\citenamefont {{Fl{\"u}hmann}}\ and\ \citenamefont
  {{Home}}(2019)}]{Fluhmann2019b}%
  \BibitemOpen
  \bibfield  {author} {\bibinfo {author} {\bibfnamefont {C.}~\bibnamefont
  {{Fl{\"u}hmann}}}\ and\ \bibinfo {author} {\bibfnamefont {J.~P.}\
  \bibnamefont {{Home}}},\ }\bibfield  {title} {\enquote {\bibinfo {title}
  {{Direct characteristic-function tomography of quantum states of the
  trapped-ion motional oscillator}},}\ }\href@noop {} {\bibfield  {journal}
  {\bibinfo  {journal} {arXiv e-prints}\ ,\ \bibinfo {eid} {arXiv:1907.06478}}
  (\bibinfo {year} {2019})},\ \Eprint {http://arxiv.org/abs/1907.06478}
  {arXiv:1907.06478 [quant-ph]} \BibitemShut {NoStop}%
\bibitem [{\citenamefont {{Grimm}}\ \emph {et~al.}(2019)\citenamefont
  {{Grimm}}, \citenamefont {{Frattini}}, \citenamefont {{Puri}}, \citenamefont
  {{Mundhada}}, \citenamefont {{Touzard}}, \citenamefont {{Mirrahimi}},
  \citenamefont {{Girvin}}, \citenamefont {{Shankar}},\ and\ \citenamefont
  {{Devoret}}}]{Grimm2019}%
  \BibitemOpen
  \bibfield  {author} {\bibinfo {author} {\bibfnamefont {A.}~\bibnamefont
  {{Grimm}}}, \bibinfo {author} {\bibfnamefont {N.~E.}\ \bibnamefont
  {{Frattini}}}, \bibinfo {author} {\bibfnamefont {S.}~\bibnamefont {{Puri}}},
  \bibinfo {author} {\bibfnamefont {S.~O.}\ \bibnamefont {{Mundhada}}},
  \bibinfo {author} {\bibfnamefont {S.}~\bibnamefont {{Touzard}}}, \bibinfo
  {author} {\bibfnamefont {M.}~\bibnamefont {{Mirrahimi}}}, \bibinfo {author}
  {\bibfnamefont {S.~M.}\ \bibnamefont {{Girvin}}}, \bibinfo {author}
  {\bibfnamefont {S.}~\bibnamefont {{Shankar}}}, \ and\ \bibinfo {author}
  {\bibfnamefont {M.~H.}\ \bibnamefont {{Devoret}}},\ }\bibfield  {title}
  {\enquote {\bibinfo {title} {{The Kerr-Cat Qubit: Stabilization, Readout, and
  Gates}},}\ }\href@noop {} {\bibfield  {journal} {\bibinfo  {journal} {arXiv
  e-prints}\ ,\ \bibinfo {eid} {arXiv:1907.12131}} (\bibinfo {year} {2019})},\
  \Eprint {http://arxiv.org/abs/1907.12131} {arXiv:1907.12131 [quant-ph]}
  \BibitemShut {NoStop}%
\bibitem [{\citenamefont {{Campagne-Ibarcq}}\ \emph {et~al.}(2019)\citenamefont
  {{Campagne-Ibarcq}}, \citenamefont {{Eickbusch}}, \citenamefont {{Touzard}},
  \citenamefont {{Zalys-Geller}}, \citenamefont {{Frattini}}, \citenamefont
  {{Sivak}}, \citenamefont {{Reinhold}}, \citenamefont {{Puri}}, \citenamefont
  {{Shankar}}, \citenamefont {{Schoelkopf}}, \citenamefont {{Frunzio}},
  \citenamefont {{Mirrahimi}},\ and\ \citenamefont {{Devoret}}}]{Campagne2019}%
  \BibitemOpen
  \bibfield  {author} {\bibinfo {author} {\bibfnamefont {P.}~\bibnamefont
  {{Campagne-Ibarcq}}}, \bibinfo {author} {\bibfnamefont {A.}~\bibnamefont
  {{Eickbusch}}}, \bibinfo {author} {\bibfnamefont {S.}~\bibnamefont
  {{Touzard}}}, \bibinfo {author} {\bibfnamefont {E.}~\bibnamefont
  {{Zalys-Geller}}}, \bibinfo {author} {\bibfnamefont {N.~E.}\ \bibnamefont
  {{Frattini}}}, \bibinfo {author} {\bibfnamefont {V.~V.}\ \bibnamefont
  {{Sivak}}}, \bibinfo {author} {\bibfnamefont {P.}~\bibnamefont {{Reinhold}}},
  \bibinfo {author} {\bibfnamefont {S.}~\bibnamefont {{Puri}}}, \bibinfo
  {author} {\bibfnamefont {S.}~\bibnamefont {{Shankar}}}, \bibinfo {author}
  {\bibfnamefont {R.~J.}\ \bibnamefont {{Schoelkopf}}}, \bibinfo {author}
  {\bibfnamefont {L.}~\bibnamefont {{Frunzio}}}, \bibinfo {author}
  {\bibfnamefont {M.}~\bibnamefont {{Mirrahimi}}}, \ and\ \bibinfo {author}
  {\bibfnamefont {M.~H.}\ \bibnamefont {{Devoret}}},\ }\bibfield  {title}
  {\enquote {\bibinfo {title} {{A stabilized logical quantum bit encoded in
  grid states of a superconducting cavity}},}\ }\href@noop {} {\bibfield
  {journal} {\bibinfo  {journal} {arXiv e-prints}\ ,\ \bibinfo {eid}
  {arXiv:1907.12487}} (\bibinfo {year} {2019})},\ \Eprint
  {http://arxiv.org/abs/1907.12487} {arXiv:1907.12487 [quant-ph]} \BibitemShut
  {NoStop}%
\bibitem [{\citenamefont {Lloyd}\ and\ \citenamefont
  {Slotine}(1998)}]{Lloyd1998}%
  \BibitemOpen
  \bibfield  {author} {\bibinfo {author} {\bibfnamefont {S.}~\bibnamefont
  {Lloyd}}\ and\ \bibinfo {author} {\bibfnamefont {J.-J.~E.}\ \bibnamefont
  {Slotine}},\ }\bibfield  {title} {\enquote {\bibinfo {title} {{Analog Quantum
  Error Correction}},}\ }\href {\doibase 10.1103/PhysRevLett.80.4088}
  {\bibfield  {journal} {\bibinfo  {journal} {Phys. Rev. Lett.}\ }\textbf
  {\bibinfo {volume} {80}},\ \bibinfo {pages} {4088--4091} (\bibinfo {year}
  {1998})}\BibitemShut {NoStop}%
\bibitem [{\citenamefont {Braunstein}(1998{\natexlab{a}})}]{Braunstein1998}%
  \BibitemOpen
  \bibfield  {author} {\bibinfo {author} {\bibfnamefont {S.~L.}\ \bibnamefont
  {Braunstein}},\ }\bibfield  {title} {\enquote {\bibinfo {title} {Error
  correction for continuous quantum variables},}\ }\href {\doibase
  10.1103/PhysRevLett.80.4084} {\bibfield  {journal} {\bibinfo  {journal}
  {Phys. Rev. Lett.}\ }\textbf {\bibinfo {volume} {80}},\ \bibinfo {pages}
  {4084--4087} (\bibinfo {year} {1998}{\natexlab{a}})}\BibitemShut {NoStop}%
\bibitem [{\citenamefont {Braunstein}(1998{\natexlab{b}})}]{Braunstein1998b}%
  \BibitemOpen
  \bibfield  {author} {\bibinfo {author} {\bibfnamefont {Samuel~L.}\
  \bibnamefont {Braunstein}},\ }\bibfield  {title} {\enquote {\bibinfo {title}
  {Quantum error correction for communication with linear optics},}\ }\href
  {https://doi.org/10.1038/27850} {\bibfield  {journal} {\bibinfo  {journal}
  {Nature}\ }\textbf {\bibinfo {volume} {394}},\ \bibinfo {pages} {47--49}
  (\bibinfo {year} {1998}{\natexlab{b}})}\BibitemShut {NoStop}%
\bibitem [{\citenamefont {Aoki}\ \emph {et~al.}(2009)\citenamefont {Aoki},
  \citenamefont {Takahashi}, \citenamefont {Kajiya}, \citenamefont {Yoshikawa},
  \citenamefont {Braunstein}, \citenamefont {van Loock},\ and\ \citenamefont
  {Furusawa}}]{Aoki2009}%
  \BibitemOpen
  \bibfield  {author} {\bibinfo {author} {\bibfnamefont {T.}~\bibnamefont
  {Aoki}}, \bibinfo {author} {\bibfnamefont {G.}~\bibnamefont {Takahashi}},
  \bibinfo {author} {\bibfnamefont {T.}~\bibnamefont {Kajiya}}, \bibinfo
  {author} {\bibfnamefont {J.}~\bibnamefont {Yoshikawa}}, \bibinfo {author}
  {\bibfnamefont {S.~L.}\ \bibnamefont {Braunstein}}, \bibinfo {author}
  {\bibfnamefont {P.}~\bibnamefont {van Loock}}, \ and\ \bibinfo {author}
  {\bibfnamefont {A.}~\bibnamefont {Furusawa}},\ }\bibfield  {title} {\enquote
  {\bibinfo {title} {Quantum error correction beyond qubits},}\ }\href
  {https://doi.org/10.1038/nphys1309} {\bibfield  {journal} {\bibinfo
  {journal} {Nature Physics}\ }\textbf {\bibinfo {volume} {5}},\ \bibinfo
  {pages} {541–546} (\bibinfo {year} {2009})}\BibitemShut {NoStop}%
\bibitem [{\citenamefont {Hayden}\ \emph {et~al.}(2016)\citenamefont {Hayden},
  \citenamefont {Nezami}, \citenamefont {Salton},\ and\ \citenamefont
  {Sanders}}]{Hayden2016}%
  \BibitemOpen
  \bibfield  {author} {\bibinfo {author} {\bibfnamefont {P.}~\bibnamefont
  {Hayden}}, \bibinfo {author} {\bibfnamefont {S.}~\bibnamefont {Nezami}},
  \bibinfo {author} {\bibfnamefont {G.}~\bibnamefont {Salton}}, \ and\ \bibinfo
  {author} {\bibfnamefont {B.~C.}\ \bibnamefont {Sanders}},\ }\bibfield
  {title} {\enquote {\bibinfo {title} {{Spacetime replication of continuous
  variable quantum information}},}\ }\href {\doibase
  10.1088/1367-2630/18/8/083043} {\bibfield  {journal} {\bibinfo  {journal}
  {New J. Phys.}\ }\textbf {\bibinfo {volume} {18}},\ \bibinfo {pages} {083043}
  (\bibinfo {year} {2016})}\BibitemShut {NoStop}%
\bibitem [{\citenamefont {{Hayden}}\ \emph {et~al.}(2017)\citenamefont
  {{Hayden}}, \citenamefont {{Nezami}}, \citenamefont {{Popescu}},\ and\
  \citenamefont {{Salton}}}]{Hayden2017}%
  \BibitemOpen
  \bibfield  {author} {\bibinfo {author} {\bibfnamefont {P.}~\bibnamefont
  {{Hayden}}}, \bibinfo {author} {\bibfnamefont {S.}~\bibnamefont {{Nezami}}},
  \bibinfo {author} {\bibfnamefont {S.}~\bibnamefont {{Popescu}}}, \ and\
  \bibinfo {author} {\bibfnamefont {G.}~\bibnamefont {{Salton}}},\ }\bibfield
  {title} {\enquote {\bibinfo {title} {{Error Correction of Quantum Reference
  Frame Information}},}\ }\href@noop {} {\bibfield  {journal} {\bibinfo
  {journal} {arXiv e-prints}\ ,\ \bibinfo {eid} {arXiv:1709.04471}} (\bibinfo
  {year} {2017})},\ \Eprint {http://arxiv.org/abs/1709.04471} {arXiv:1709.04471
  [quant-ph]} \BibitemShut {NoStop}%
\bibitem [{\citenamefont {{Faist}}\ \emph {et~al.}(2019)\citenamefont
  {{Faist}}, \citenamefont {{Nezami}}, \citenamefont {{Albert}}, \citenamefont
  {{Salton}}, \citenamefont {{Pastawski}}, \citenamefont {{Hayden}},\ and\
  \citenamefont {{Preskill}}}]{Faist2019}%
  \BibitemOpen
  \bibfield  {author} {\bibinfo {author} {\bibfnamefont {P.}~\bibnamefont
  {{Faist}}}, \bibinfo {author} {\bibfnamefont {S.}~\bibnamefont {{Nezami}}},
  \bibinfo {author} {\bibfnamefont {V.~V.}\ \bibnamefont {{Albert}}}, \bibinfo
  {author} {\bibfnamefont {G.}~\bibnamefont {{Salton}}}, \bibinfo {author}
  {\bibfnamefont {F.}~\bibnamefont {{Pastawski}}}, \bibinfo {author}
  {\bibfnamefont {P.}~\bibnamefont {{Hayden}}}, \ and\ \bibinfo {author}
  {\bibfnamefont {J.}~\bibnamefont {{Preskill}}},\ }\bibfield  {title}
  {\enquote {\bibinfo {title} {{Continuous symmetries and approximate quantum
  error correction}},}\ }\href@noop {} {\bibfield  {journal} {\bibinfo
  {journal} {arXiv e-prints}\ ,\ \bibinfo {eid} {arXiv:1902.07714}} (\bibinfo
  {year} {2019})},\ \Eprint {http://arxiv.org/abs/1902.07714} {arXiv:1902.07714
  [quant-ph]} \BibitemShut {NoStop}%
\bibitem [{\citenamefont {{Woods}}\ and\ \citenamefont
  {{Alhambra}}(2019)}]{Woods2019}%
  \BibitemOpen
  \bibfield  {author} {\bibinfo {author} {\bibfnamefont {M.~P.}\ \bibnamefont
  {{Woods}}}\ and\ \bibinfo {author} {\bibfnamefont {{\'A}.~M.}\ \bibnamefont
  {{Alhambra}}},\ }\bibfield  {title} {\enquote {\bibinfo {title} {{Continuous
  groups of transversal gates for quantum error correcting codes from finite
  clock reference frames}},}\ }\href@noop {} {\bibfield  {journal} {\bibinfo
  {journal} {arXiv e-prints}\ ,\ \bibinfo {eid} {arXiv:1902.07725}} (\bibinfo
  {year} {2019})},\ \Eprint {http://arxiv.org/abs/1902.07725} {arXiv:1902.07725
  [quant-ph]} \BibitemShut {NoStop}%
\bibitem [{\citenamefont {Eisert}\ \emph {et~al.}(2002)\citenamefont {Eisert},
  \citenamefont {Scheel},\ and\ \citenamefont {Plenio}}]{Eisert2002}%
  \BibitemOpen
  \bibfield  {author} {\bibinfo {author} {\bibfnamefont {J.}~\bibnamefont
  {Eisert}}, \bibinfo {author} {\bibfnamefont {S.}~\bibnamefont {Scheel}}, \
  and\ \bibinfo {author} {\bibfnamefont {M.~B.}\ \bibnamefont {Plenio}},\
  }\bibfield  {title} {\enquote {\bibinfo {title} {Distilling {G}aussian states
  with {G}aussian operations is impossible},}\ }\href {\doibase
  10.1103/PhysRevLett.89.137903} {\bibfield  {journal} {\bibinfo  {journal}
  {Phys. Rev. Lett.}\ }\textbf {\bibinfo {volume} {89}},\ \bibinfo {pages}
  {137903} (\bibinfo {year} {2002})}\BibitemShut {NoStop}%
\bibitem [{\citenamefont {Niset}\ \emph {et~al.}(2009)\citenamefont {Niset},
  \citenamefont {Fiur\'a\ifmmode~\check{s}\else \v{s}\fi{}ek},\ and\
  \citenamefont {Cerf}}]{Niset2009}%
  \BibitemOpen
  \bibfield  {author} {\bibinfo {author} {\bibfnamefont {J.}~\bibnamefont
  {Niset}}, \bibinfo {author} {\bibfnamefont {J.}~\bibnamefont
  {Fiur\'a\ifmmode~\check{s}\else \v{s}\fi{}ek}}, \ and\ \bibinfo {author}
  {\bibfnamefont {N.~J.}\ \bibnamefont {Cerf}},\ }\bibfield  {title} {\enquote
  {\bibinfo {title} {No-go theorem for {G}aussian quantum error correction},}\
  }\href {\doibase 10.1103/PhysRevLett.102.120501} {\bibfield  {journal}
  {\bibinfo  {journal} {Phys. Rev. Lett.}\ }\textbf {\bibinfo {volume} {102}},\
  \bibinfo {pages} {120501} (\bibinfo {year} {2009})}\BibitemShut {NoStop}%
\bibitem [{\citenamefont {Zhuang}\ \emph {et~al.}(2018)\citenamefont {Zhuang},
  \citenamefont {Shor},\ and\ \citenamefont {Shapiro}}]{Zhuang2018}%
  \BibitemOpen
  \bibfield  {author} {\bibinfo {author} {\bibfnamefont {Q.}~\bibnamefont
  {Zhuang}}, \bibinfo {author} {\bibfnamefont {P.~W.}\ \bibnamefont {Shor}}, \
  and\ \bibinfo {author} {\bibfnamefont {J.~H.}\ \bibnamefont {Shapiro}},\
  }\bibfield  {title} {\enquote {\bibinfo {title} {Resource theory of
  non-gaussian operations},}\ }\href {\doibase 10.1103/PhysRevA.97.052317}
  {\bibfield  {journal} {\bibinfo  {journal} {Phys. Rev. A}\ }\textbf {\bibinfo
  {volume} {97}},\ \bibinfo {pages} {052317} (\bibinfo {year}
  {2018})}\BibitemShut {NoStop}%
\bibitem [{\citenamefont {Takagi}\ and\ \citenamefont
  {Zhuang}(2018)}]{Takagi2018}%
  \BibitemOpen
  \bibfield  {author} {\bibinfo {author} {\bibfnamefont {R.}~\bibnamefont
  {Takagi}}\ and\ \bibinfo {author} {\bibfnamefont {Q.}~\bibnamefont
  {Zhuang}},\ }\bibfield  {title} {\enquote {\bibinfo {title} {Convex resource
  theory of non-gaussianity},}\ }\href {\doibase 10.1103/PhysRevA.97.062337}
  {\bibfield  {journal} {\bibinfo  {journal} {Phys. Rev. A}\ }\textbf {\bibinfo
  {volume} {97}},\ \bibinfo {pages} {062337} (\bibinfo {year}
  {2018})}\BibitemShut {NoStop}%
\bibitem [{\citenamefont {Knill}\ \emph {et~al.}(2001)\citenamefont {Knill},
  \citenamefont {Laflamme},\ and\ \citenamefont {Milburn}}]{Knill2001}%
  \BibitemOpen
  \bibfield  {author} {\bibinfo {author} {\bibfnamefont {E.}~\bibnamefont
  {Knill}}, \bibinfo {author} {\bibfnamefont {R.}~\bibnamefont {Laflamme}}, \
  and\ \bibinfo {author} {\bibfnamefont {G.~J.}\ \bibnamefont {Milburn}},\
  }\bibfield  {title} {\enquote {\bibinfo {title} {A scheme for efficient
  quantum computation with linear optics},}\ }\href
  {http://dx.doi.org/10.1038/35051009} {\bibfield  {journal} {\bibinfo
  {journal} {Nature}\ }\textbf {\bibinfo {volume} {409}},\ \bibinfo {pages}
  {46--52} (\bibinfo {year} {2001})}\BibitemShut {NoStop}%
\bibitem [{\citenamefont {Lloyd}\ and\ \citenamefont
  {Braunstein}(1999)}]{Lloyd1999}%
  \BibitemOpen
  \bibfield  {author} {\bibinfo {author} {\bibfnamefont {S.}~\bibnamefont
  {Lloyd}}\ and\ \bibinfo {author} {\bibfnamefont {S.~L.}\ \bibnamefont
  {Braunstein}},\ }\bibfield  {title} {\enquote {\bibinfo {title} {Quantum
  computation over continuous variables},}\ }\href {\doibase
  10.1103/PhysRevLett.82.1784} {\bibfield  {journal} {\bibinfo  {journal}
  {Phys. Rev. Lett.}\ }\textbf {\bibinfo {volume} {82}},\ \bibinfo {pages}
  {1784--1787} (\bibinfo {year} {1999})}\BibitemShut {NoStop}%
\bibitem [{\citenamefont {Krastanov}\ \emph {et~al.}(2015)\citenamefont
  {Krastanov}, \citenamefont {Albert}, \citenamefont {Shen}, \citenamefont
  {Zou}, \citenamefont {Heeres}, \citenamefont {Vlastakis}, \citenamefont
  {Schoelkopf},\ and\ \citenamefont {Jiang}}]{Krastanov2015}%
  \BibitemOpen
  \bibfield  {author} {\bibinfo {author} {\bibfnamefont {S.}~\bibnamefont
  {Krastanov}}, \bibinfo {author} {\bibfnamefont {V.~V.}\ \bibnamefont
  {Albert}}, \bibinfo {author} {\bibfnamefont {C.}~\bibnamefont {Shen}},
  \bibinfo {author} {\bibfnamefont {C.-L.}\ \bibnamefont {Zou}}, \bibinfo
  {author} {\bibfnamefont {R.~W.}\ \bibnamefont {Heeres}}, \bibinfo {author}
  {\bibfnamefont {B.}~\bibnamefont {Vlastakis}}, \bibinfo {author}
  {\bibfnamefont {R.~J.}\ \bibnamefont {Schoelkopf}}, \ and\ \bibinfo {author}
  {\bibfnamefont {L.}~\bibnamefont {Jiang}},\ }\bibfield  {title} {\enquote
  {\bibinfo {title} {Universal control of an oscillator with dispersive
  coupling to a qubit},}\ }\href {\doibase 10.1103/PhysRevA.92.040303}
  {\bibfield  {journal} {\bibinfo  {journal} {Phys. Rev. A}\ }\textbf {\bibinfo
  {volume} {92}},\ \bibinfo {pages} {040303} (\bibinfo {year}
  {2015})}\BibitemShut {NoStop}%
\bibitem [{\citenamefont {Baragiola}\ \emph {et~al.}(2019)\citenamefont
  {Baragiola}, \citenamefont {Pantaleoni}, \citenamefont {Alexander},
  \citenamefont {Karanjai},\ and\ \citenamefont {Menicucci}}]{Baragiola2019}%
  \BibitemOpen
  \bibfield  {author} {\bibinfo {author} {\bibfnamefont {B.~Q.}\ \bibnamefont
  {Baragiola}}, \bibinfo {author} {\bibfnamefont {G.}~\bibnamefont
  {Pantaleoni}}, \bibinfo {author} {\bibfnamefont {R.~N.}\ \bibnamefont
  {Alexander}}, \bibinfo {author} {\bibfnamefont {A.}~\bibnamefont {Karanjai}},
  \ and\ \bibinfo {author} {\bibfnamefont {N.~C.}\ \bibnamefont {Menicucci}},\
  }\bibfield  {title} {\enquote {\bibinfo {title} {All-gaussian universality
  and fault tolerance with the gottesman-kitaev-preskill code},}\ }\href
  {\doibase 10.1103/PhysRevLett.123.200502} {\bibfield  {journal} {\bibinfo
  {journal} {Phys. Rev. Lett.}\ }\textbf {\bibinfo {volume} {123}},\ \bibinfo
  {pages} {200502} (\bibinfo {year} {2019})}\BibitemShut {NoStop}%
\bibitem [{\citenamefont {Duivenvoorden}\ \emph {et~al.}(2017)\citenamefont
  {Duivenvoorden}, \citenamefont {Terhal},\ and\ \citenamefont
  {Weigand}}]{Duivenvoorden2017}%
  \BibitemOpen
  \bibfield  {author} {\bibinfo {author} {\bibfnamefont {K.}~\bibnamefont
  {Duivenvoorden}}, \bibinfo {author} {\bibfnamefont {B.~M.}\ \bibnamefont
  {Terhal}}, \ and\ \bibinfo {author} {\bibfnamefont {D.}~\bibnamefont
  {Weigand}},\ }\bibfield  {title} {\enquote {\bibinfo {title} {Single-mode
  displacement sensor},}\ }\href {\doibase 10.1103/PhysRevA.95.012305}
  {\bibfield  {journal} {\bibinfo  {journal} {Phys. Rev. A}\ }\textbf {\bibinfo
  {volume} {95}},\ \bibinfo {pages} {012305} (\bibinfo {year}
  {2017})}\BibitemShut {NoStop}%
\bibitem [{\citenamefont {Travaglione}\ and\ \citenamefont
  {Milburn}(2002)}]{Travaglione2002}%
  \BibitemOpen
  \bibfield  {author} {\bibinfo {author} {\bibfnamefont {B.~C.}\ \bibnamefont
  {Travaglione}}\ and\ \bibinfo {author} {\bibfnamefont {G.~J.}\ \bibnamefont
  {Milburn}},\ }\bibfield  {title} {\enquote {\bibinfo {title} {Preparing
  encoded states in an oscillator},}\ }\href {\doibase
  10.1103/PhysRevA.66.052322} {\bibfield  {journal} {\bibinfo  {journal} {Phys.
  Rev. A}\ }\textbf {\bibinfo {volume} {66}},\ \bibinfo {pages} {052322}
  (\bibinfo {year} {2002})}\BibitemShut {NoStop}%
\bibitem [{\citenamefont {Pirandola}\ \emph {et~al.}(2004)\citenamefont
  {Pirandola}, \citenamefont {Mancini}, \citenamefont {Vitali},\ and\
  \citenamefont {Tombesi}}]{Pirandola2004}%
  \BibitemOpen
  \bibfield  {author} {\bibinfo {author} {\bibfnamefont {S.}~\bibnamefont
  {Pirandola}}, \bibinfo {author} {\bibfnamefont {S.}~\bibnamefont {Mancini}},
  \bibinfo {author} {\bibfnamefont {D.}~\bibnamefont {Vitali}}, \ and\ \bibinfo
  {author} {\bibfnamefont {P.}~\bibnamefont {Tombesi}},\ }\bibfield  {title}
  {\enquote {\bibinfo {title} {Constructing finite-dimensional codes with
  optical continuous variables},}\ }\href
  {http://stacks.iop.org/0295-5075/68/i=3/a=323} {\bibfield  {journal}
  {\bibinfo  {journal} {EPL (Europhysics Letters)}\ }\textbf {\bibinfo {volume}
  {68}},\ \bibinfo {pages} {323} (\bibinfo {year} {2004})}\BibitemShut
  {NoStop}%
\bibitem [{\citenamefont {Pirandola}\ \emph {et~al.}(2006)\citenamefont
  {Pirandola}, \citenamefont {Mancini}, \citenamefont {Vitali},\ and\
  \citenamefont {Tombesi}}]{Pirandola2006}%
  \BibitemOpen
  \bibfield  {author} {\bibinfo {author} {\bibfnamefont {S.}~\bibnamefont
  {Pirandola}}, \bibinfo {author} {\bibfnamefont {S.}~\bibnamefont {Mancini}},
  \bibinfo {author} {\bibfnamefont {D.}~\bibnamefont {Vitali}}, \ and\ \bibinfo
  {author} {\bibfnamefont {P.}~\bibnamefont {Tombesi}},\ }\bibfield  {title}
  {\enquote {\bibinfo {title} {Generating continuous variable quantum codewords
  in the near-field atomic lithography},}\ }\href
  {http://stacks.iop.org/0953-4075/39/i=4/a=023} {\bibfield  {journal}
  {\bibinfo  {journal} {Journal of Physics B: Atomic, Molecular and Optical
  Physics}\ }\textbf {\bibinfo {volume} {39}},\ \bibinfo {pages} {997}
  (\bibinfo {year} {2006})}\BibitemShut {NoStop}%
\bibitem [{\citenamefont {Vasconcelos}\ \emph {et~al.}(2010)\citenamefont
  {Vasconcelos}, \citenamefont {Sanz},\ and\ \citenamefont
  {Glancy}}]{Vasconcelos2010}%
  \BibitemOpen
  \bibfield  {author} {\bibinfo {author} {\bibfnamefont {H.~M.}\ \bibnamefont
  {Vasconcelos}}, \bibinfo {author} {\bibfnamefont {L.}~\bibnamefont {Sanz}}, \
  and\ \bibinfo {author} {\bibfnamefont {S.}~\bibnamefont {Glancy}},\
  }\bibfield  {title} {\enquote {\bibinfo {title} {All-optical generation of
  states for ``encoding a qubit in an oscillator''},}\ }\href {\doibase
  10.1364/OL.35.003261} {\bibfield  {journal} {\bibinfo  {journal} {Opt.
  Lett.}\ }\textbf {\bibinfo {volume} {35}},\ \bibinfo {pages} {3261--3263}
  (\bibinfo {year} {2010})}\BibitemShut {NoStop}%
\bibitem [{\citenamefont {Terhal}\ and\ \citenamefont
  {Weigand}(2016)}]{Terhal2016}%
  \BibitemOpen
  \bibfield  {author} {\bibinfo {author} {\bibfnamefont {B.~M.}\ \bibnamefont
  {Terhal}}\ and\ \bibinfo {author} {\bibfnamefont {D.}~\bibnamefont
  {Weigand}},\ }\bibfield  {title} {\enquote {\bibinfo {title} {Encoding a
  qubit into a cavity mode in circuit {QED} using phase estimation},}\ }\href
  {\doibase 10.1103/PhysRevA.93.012315} {\bibfield  {journal} {\bibinfo
  {journal} {Phys. Rev. A}\ }\textbf {\bibinfo {volume} {93}},\ \bibinfo
  {pages} {012315} (\bibinfo {year} {2016})}\BibitemShut {NoStop}%
\bibitem [{\citenamefont {Motes}\ \emph {et~al.}(2017)\citenamefont {Motes},
  \citenamefont {Baragiola}, \citenamefont {Gilchrist},\ and\ \citenamefont
  {Menicucci}}]{Motes2017}%
  \BibitemOpen
  \bibfield  {author} {\bibinfo {author} {\bibfnamefont {K.~R.}\ \bibnamefont
  {Motes}}, \bibinfo {author} {\bibfnamefont {B.~Q.}\ \bibnamefont
  {Baragiola}}, \bibinfo {author} {\bibfnamefont {A.}~\bibnamefont
  {Gilchrist}}, \ and\ \bibinfo {author} {\bibfnamefont {N.~C.}\ \bibnamefont
  {Menicucci}},\ }\bibfield  {title} {\enquote {\bibinfo {title} {Encoding
  qubits into oscillators with atomic ensembles and squeezed light},}\ }\href
  {\doibase 10.1103/PhysRevA.95.053819} {\bibfield  {journal} {\bibinfo
  {journal} {Phys. Rev. A}\ }\textbf {\bibinfo {volume} {95}},\ \bibinfo
  {pages} {053819} (\bibinfo {year} {2017})}\BibitemShut {NoStop}%
\bibitem [{\citenamefont {Weigand}\ and\ \citenamefont
  {Terhal}(2018)}]{Weigand2017}%
  \BibitemOpen
  \bibfield  {author} {\bibinfo {author} {\bibfnamefont {D.~J.}\ \bibnamefont
  {Weigand}}\ and\ \bibinfo {author} {\bibfnamefont {B.~M.}\ \bibnamefont
  {Terhal}},\ }\bibfield  {title} {\enquote {\bibinfo {title} {Generating grid
  states from {S}chr\"odinger-cat states without postselection},}\ }\href
  {\doibase 10.1103/PhysRevA.97.022341} {\bibfield  {journal} {\bibinfo
  {journal} {Phys. Rev. A}\ }\textbf {\bibinfo {volume} {97}},\ \bibinfo
  {pages} {022341} (\bibinfo {year} {2018})}\BibitemShut {NoStop}%
\bibitem [{\citenamefont {Arrazola}\ \emph {et~al.}(2019)\citenamefont
  {Arrazola}, \citenamefont {Bromley}, \citenamefont {Izaac}, \citenamefont
  {Myers}, \citenamefont {Br{\'{a}}dler},\ and\ \citenamefont
  {Killoran}}]{Arrazola2019}%
  \BibitemOpen
  \bibfield  {author} {\bibinfo {author} {\bibfnamefont {J.~M.}\ \bibnamefont
  {Arrazola}}, \bibinfo {author} {\bibfnamefont {T.~R.}\ \bibnamefont
  {Bromley}}, \bibinfo {author} {\bibfnamefont {J.}~\bibnamefont {Izaac}},
  \bibinfo {author} {\bibfnamefont {C.~R.}\ \bibnamefont {Myers}}, \bibinfo
  {author} {\bibfnamefont {K.}~\bibnamefont {Br{\'{a}}dler}}, \ and\ \bibinfo
  {author} {\bibfnamefont {N.}~\bibnamefont {Killoran}},\ }\bibfield  {title}
  {\enquote {\bibinfo {title} {Machine learning method for state preparation
  and gate synthesis on photonic quantum computers},}\ }\href {\doibase
  10.1088/2058-9565/aaf59e} {\bibfield  {journal} {\bibinfo  {journal} {Quantum
  Science and Technology}\ }\textbf {\bibinfo {volume} {4}},\ \bibinfo {pages}
  {024004} (\bibinfo {year} {2019})}\BibitemShut {NoStop}%
\bibitem [{\citenamefont {Su}\ \emph {et~al.}(2019)\citenamefont {Su},
  \citenamefont {Myers},\ and\ \citenamefont {Sabapathy}}]{Su2019}%
  \BibitemOpen
  \bibfield  {author} {\bibinfo {author} {\bibfnamefont {D.}~\bibnamefont
  {Su}}, \bibinfo {author} {\bibfnamefont {C.~R.}\ \bibnamefont {Myers}}, \
  and\ \bibinfo {author} {\bibfnamefont {K.~K.}\ \bibnamefont {Sabapathy}},\
  }\bibfield  {title} {\enquote {\bibinfo {title} {Conversion of gaussian
  states to non-gaussian states using photon-number-resolving detectors},}\
  }\href {\doibase 10.1103/PhysRevA.100.052301} {\bibfield  {journal} {\bibinfo
   {journal} {Phys. Rev. A}\ }\textbf {\bibinfo {volume} {100}},\ \bibinfo
  {pages} {052301} (\bibinfo {year} {2019})}\BibitemShut {NoStop}%
\bibitem [{\citenamefont {Eaton}\ \emph {et~al.}(2019)\citenamefont {Eaton},
  \citenamefont {Nehra},\ and\ \citenamefont {Pfister}}]{Eaton2019}%
  \BibitemOpen
  \bibfield  {author} {\bibinfo {author} {\bibfnamefont {M.}~\bibnamefont
  {Eaton}}, \bibinfo {author} {\bibfnamefont {R.}~\bibnamefont {Nehra}}, \ and\
  \bibinfo {author} {\bibfnamefont {O.}~\bibnamefont {Pfister}},\ }\bibfield
  {title} {\enquote {\bibinfo {title} {Non-gaussian and
  gottesman{\textendash}kitaev{\textendash}preskill state preparation by photon
  catalysis},}\ }\href {\doibase 10.1088/1367-2630/ab5330} {\bibfield
  {journal} {\bibinfo  {journal} {New Journal of Physics}\ }\textbf {\bibinfo
  {volume} {21}},\ \bibinfo {pages} {113034} (\bibinfo {year}
  {2019})}\BibitemShut {NoStop}%
\bibitem [{\citenamefont {Shi}\ \emph {et~al.}(2019)\citenamefont {Shi},
  \citenamefont {Chamberland},\ and\ \citenamefont {Cross}}]{Shi2019}%
  \BibitemOpen
  \bibfield  {author} {\bibinfo {author} {\bibfnamefont {Y.}~\bibnamefont
  {Shi}}, \bibinfo {author} {\bibfnamefont {C.}~\bibnamefont {Chamberland}}, \
  and\ \bibinfo {author} {\bibfnamefont {A.}~\bibnamefont {Cross}},\ }\bibfield
   {title} {\enquote {\bibinfo {title} {Fault-tolerant preparation of
  approximate {GKP} states},}\ }\href {\doibase 10.1088/1367-2630/ab3a62}
  {\bibfield  {journal} {\bibinfo  {journal} {New Journal of Physics}\ }\textbf
  {\bibinfo {volume} {21}},\ \bibinfo {pages} {093007} (\bibinfo {year}
  {2019})}\BibitemShut {NoStop}%
\bibitem [{\citenamefont {{Weigand}}\ and\ \citenamefont
  {{Terhal}}(2019)}]{Weigand2019}%
  \BibitemOpen
  \bibfield  {author} {\bibinfo {author} {\bibfnamefont {D.~J.}\ \bibnamefont
  {{Weigand}}}\ and\ \bibinfo {author} {\bibfnamefont {B.~M.}\ \bibnamefont
  {{Terhal}}},\ }\bibfield  {title} {\enquote {\bibinfo {title} {{Realizing
  modular quadrature measurements via a tunable photon-pressure coupling in
  circuit-QED}},}\ }\href@noop {} {\bibfield  {journal} {\bibinfo  {journal}
  {arXiv e-prints}\ ,\ \bibinfo {eid} {arXiv:1909.10075}} (\bibinfo {year}
  {2019})},\ \Eprint {http://arxiv.org/abs/1909.10075} {arXiv:1909.10075
  [quant-ph]} \BibitemShut {NoStop}%
\bibitem [{\citenamefont {{Hastrup}}\ \emph {et~al.}(2019)\citenamefont
  {{Hastrup}}, \citenamefont {{Park}}, \citenamefont {{Bohr Brask}},
  \citenamefont {{Filip}},\ and\ \citenamefont {{Andersen}}}]{Hastrup2019}%
  \BibitemOpen
  \bibfield  {author} {\bibinfo {author} {\bibfnamefont {J.}~\bibnamefont
  {{Hastrup}}}, \bibinfo {author} {\bibfnamefont {K.}~\bibnamefont {{Park}}},
  \bibinfo {author} {\bibfnamefont {J.}~\bibnamefont {{Bohr Brask}}}, \bibinfo
  {author} {\bibfnamefont {R.}~\bibnamefont {{Filip}}}, \ and\ \bibinfo
  {author} {\bibfnamefont {U.~L.}\ \bibnamefont {{Andersen}}},\ }\bibfield
  {title} {\enquote {\bibinfo {title} {{Measurement-free preparation of grid
  states}},}\ }\href@noop {} {\bibfield  {journal} {\bibinfo  {journal} {arXiv
  e-prints}\ ,\ \bibinfo {eid} {arXiv:1912.12645}} (\bibinfo {year} {2019})},\
  \Eprint {http://arxiv.org/abs/1912.12645} {arXiv:1912.12645 [quant-ph]}
  \BibitemShut {NoStop}%
\bibitem [{\citenamefont {{Noh}}\ \emph {et~al.}(2019)\citenamefont {{Noh}},
  \citenamefont {{Albert}},\ and\ \citenamefont {{Jiang}}}]{Noh2019}%
  \BibitemOpen
  \bibfield  {author} {\bibinfo {author} {\bibfnamefont {K.}~\bibnamefont
  {{Noh}}}, \bibinfo {author} {\bibfnamefont {V.~V.}\ \bibnamefont {{Albert}}},
  \ and\ \bibinfo {author} {\bibfnamefont {L.}~\bibnamefont {{Jiang}}},\
  }\bibfield  {title} {\enquote {\bibinfo {title} {Quantum capacity bounds of
  {G}aussian thermal loss channels and achievable rates with
  {G}ottesman-{K}itaev-{P}reskill codes},}\ }\href {\doibase
  10.1109/TIT.2018.2873764} {\bibfield  {journal} {\bibinfo  {journal} {IEEE
  Transactions on Information Theory}\ }\textbf {\bibinfo {volume} {65}},\
  \bibinfo {pages} {2563--2582} (\bibinfo {year} {2019})}\BibitemShut {NoStop}%
\bibitem [{\citenamefont {{Shapiro}}\ \emph {et~al.}(1979)\citenamefont
  {{Shapiro}}, \citenamefont {{Yuen}},\ and\ \citenamefont
  {{Mata}}}]{Shapiro1979}%
  \BibitemOpen
  \bibfield  {author} {\bibinfo {author} {\bibfnamefont {J.}~\bibnamefont
  {{Shapiro}}}, \bibinfo {author} {\bibfnamefont {H.}~\bibnamefont {{Yuen}}}, \
  and\ \bibinfo {author} {\bibfnamefont {A.}~\bibnamefont {{Mata}}},\
  }\bibfield  {title} {\enquote {\bibinfo {title} {Optical communication with
  two-photon coherent states--part ii: Photoemissive detection and structured
  receiver performance},}\ }\href {\doibase 10.1109/TIT.1979.1056033}
  {\bibfield  {journal} {\bibinfo  {journal} {IEEE Transactions on Information
  Theory}\ }\textbf {\bibinfo {volume} {25}},\ \bibinfo {pages} {179--192}
  (\bibinfo {year} {1979})}\BibitemShut {NoStop}%
\bibitem [{\citenamefont {Nielsen}\ and\ \citenamefont
  {Chuang}(2000)}]{Nielsen2000}%
  \BibitemOpen
  \bibfield  {author} {\bibinfo {author} {\bibfnamefont {M.~A.}\ \bibnamefont
  {Nielsen}}\ and\ \bibinfo {author} {\bibfnamefont {I.~L.}\ \bibnamefont
  {Chuang}},\ }\href {https://books.google.co.kr/books?id=65FqEKQOfP8C} {\emph
  {\bibinfo {title} {{Quantum Computation and Quantum Information}}}},\
  Cambridge Series on Information and the Natural Sciences\ (\bibinfo
  {publisher} {Cambridge University Press},\ \bibinfo {year}
  {2000})\BibitemShut {NoStop}%
\bibitem [{\citenamefont {Laflamme}\ \emph {et~al.}(1996)\citenamefont
  {Laflamme}, \citenamefont {Miquel}, \citenamefont {Paz},\ and\ \citenamefont
  {Zurek}}]{Laflamme1996}%
  \BibitemOpen
  \bibfield  {author} {\bibinfo {author} {\bibfnamefont {R.}~\bibnamefont
  {Laflamme}}, \bibinfo {author} {\bibfnamefont {C.}~\bibnamefont {Miquel}},
  \bibinfo {author} {\bibfnamefont {J.~P.}\ \bibnamefont {Paz}}, \ and\
  \bibinfo {author} {\bibfnamefont {W.~H.}\ \bibnamefont {Zurek}},\ }\bibfield
  {title} {\enquote {\bibinfo {title} {Perfect quantum error correcting
  code},}\ }\href {\doibase 10.1103/PhysRevLett.77.198} {\bibfield  {journal}
  {\bibinfo  {journal} {Phys. Rev. Lett.}\ }\textbf {\bibinfo {volume} {77}},\
  \bibinfo {pages} {198--201} (\bibinfo {year} {1996})}\BibitemShut {NoStop}%
\bibitem [{\citenamefont {Bennett}\ \emph {et~al.}(1996)\citenamefont
  {Bennett}, \citenamefont {DiVincenzo}, \citenamefont {Smolin},\ and\
  \citenamefont {Wootters}}]{Bennett1996}%
  \BibitemOpen
  \bibfield  {author} {\bibinfo {author} {\bibfnamefont {C.~H.}\ \bibnamefont
  {Bennett}}, \bibinfo {author} {\bibfnamefont {D.~P.}\ \bibnamefont
  {DiVincenzo}}, \bibinfo {author} {\bibfnamefont {J.~A.}\ \bibnamefont
  {Smolin}}, \ and\ \bibinfo {author} {\bibfnamefont {W.~K.}\ \bibnamefont
  {Wootters}},\ }\bibfield  {title} {\enquote {\bibinfo {title} {Mixed-state
  entanglement and quantum error correction},}\ }\href {\doibase
  10.1103/PhysRevA.54.3824} {\bibfield  {journal} {\bibinfo  {journal} {Phys.
  Rev. A}\ }\textbf {\bibinfo {volume} {54}},\ \bibinfo {pages} {3824--3851}
  (\bibinfo {year} {1996})}\BibitemShut {NoStop}%
\bibitem [{\citenamefont {{Gottesman}}(1997)}]{Gottesman1997}%
  \BibitemOpen
  \bibfield  {author} {\bibinfo {author} {\bibfnamefont {D.}~\bibnamefont
  {{Gottesman}}},\ }\emph {\bibinfo {title} {{Stabilizer codes and quantum
  error correction}}},\ \href@noop {} {Ph.D. thesis},\ \bibinfo  {school}
  {California Institute of Technology} (\bibinfo {year} {1997})\BibitemShut
  {NoStop}%
\bibitem [{\citenamefont {Kitaev}(1997)}]{Kitaev1997}%
  \BibitemOpen
  \bibfield  {author} {\bibinfo {author} {\bibfnamefont {A.}~\bibnamefont
  {Kitaev}},\ }\enquote {\bibinfo {title} {Quantum error correction with
  imperfect gates},}\ in\ \href {\doibase 10.1007/978-1-4615-5923-8_19} {\emph
  {\bibinfo {booktitle} {Quantum Communication, Computing, and Measurement}}},\
  \bibinfo {editor} {edited by\ \bibinfo {editor} {\bibfnamefont
  {O.}~\bibnamefont {Hirota}}, \bibinfo {editor} {\bibfnamefont {A.~S.}\
  \bibnamefont {Holevo}}, \ and\ \bibinfo {editor} {\bibfnamefont {C.~M.}\
  \bibnamefont {Caves}}}\ (\bibinfo  {publisher} {Springer US},\ \bibinfo
  {address} {Boston, MA},\ \bibinfo {year} {1997})\ pp.\ \bibinfo {pages}
  {181--188}\BibitemShut {NoStop}%
\bibitem [{\citenamefont {Fowler}\ \emph {et~al.}(2012)\citenamefont {Fowler},
  \citenamefont {Mariantoni}, \citenamefont {Martinis},\ and\ \citenamefont
  {Cleland}}]{Fowler2012}%
  \BibitemOpen
  \bibfield  {author} {\bibinfo {author} {\bibfnamefont {A.~G.}\ \bibnamefont
  {Fowler}}, \bibinfo {author} {\bibfnamefont {M.}~\bibnamefont {Mariantoni}},
  \bibinfo {author} {\bibfnamefont {J.~M.}\ \bibnamefont {Martinis}}, \ and\
  \bibinfo {author} {\bibfnamefont {A.~N.}\ \bibnamefont {Cleland}},\
  }\bibfield  {title} {\enquote {\bibinfo {title} {Surface codes: Towards
  practical large-scale quantum computation},}\ }\href {\doibase
  10.1103/PhysRevA.86.032324} {\bibfield  {journal} {\bibinfo  {journal} {Phys.
  Rev. A}\ }\textbf {\bibinfo {volume} {86}},\ \bibinfo {pages} {032324}
  (\bibinfo {year} {2012})}\BibitemShut {NoStop}%
\bibitem [{\citenamefont {Bravyi}\ and\ \citenamefont
  {Kitaev}(2005)}]{Bravyi2005}%
  \BibitemOpen
  \bibfield  {author} {\bibinfo {author} {\bibfnamefont {S.}~\bibnamefont
  {Bravyi}}\ and\ \bibinfo {author} {\bibfnamefont {A.}~\bibnamefont
  {Kitaev}},\ }\bibfield  {title} {\enquote {\bibinfo {title} {Universal
  quantum computation with ideal clifford gates and noisy ancillas},}\ }\href
  {\doibase 10.1103/PhysRevA.71.022316} {\bibfield  {journal} {\bibinfo
  {journal} {Phys. Rev. A}\ }\textbf {\bibinfo {volume} {71}},\ \bibinfo
  {pages} {022316} (\bibinfo {year} {2005})}\BibitemShut {NoStop}%
\bibitem [{\citenamefont {{Zhuang}}\ \emph {et~al.}(2019)\citenamefont
  {{Zhuang}}, \citenamefont {{Preskill}},\ and\ \citenamefont
  {{Jiang}}}]{Zhuang2019}%
  \BibitemOpen
  \bibfield  {author} {\bibinfo {author} {\bibfnamefont {Q.}~\bibnamefont
  {{Zhuang}}}, \bibinfo {author} {\bibfnamefont {J.}~\bibnamefont
  {{Preskill}}}, \ and\ \bibinfo {author} {\bibfnamefont {L.}~\bibnamefont
  {{Jiang}}},\ }\bibfield  {title} {\enquote {\bibinfo {title} {{Distributed
  quantum sensing enhanced by continuous-variable error correction}},}\
  }\href@noop {} {\bibfield  {journal} {\bibinfo  {journal} {arXiv e-prints}\
  ,\ \bibinfo {eid} {arXiv:1910.14156}} (\bibinfo {year} {2019})},\ \Eprint
  {http://arxiv.org/abs/1910.14156} {arXiv:1910.14156 [quant-ph]} \BibitemShut
  {NoStop}%
\bibitem [{\citenamefont {Zhang}\ \emph {et~al.}(2018)\citenamefont {Zhang},
  \citenamefont {Zou},\ and\ \citenamefont {Jiang}}]{Zhang2018}%
  \BibitemOpen
  \bibfield  {author} {\bibinfo {author} {\bibfnamefont {M.}~\bibnamefont
  {Zhang}}, \bibinfo {author} {\bibfnamefont {C.-L.}\ \bibnamefont {Zou}}, \
  and\ \bibinfo {author} {\bibfnamefont {L.}~\bibnamefont {Jiang}},\ }\bibfield
   {title} {\enquote {\bibinfo {title} {Quantum transduction with adaptive
  control},}\ }\href {\doibase 10.1103/PhysRevLett.120.020502} {\bibfield
  {journal} {\bibinfo  {journal} {Phys. Rev. Lett.}\ }\textbf {\bibinfo
  {volume} {120}},\ \bibinfo {pages} {020502} (\bibinfo {year}
  {2018})}\BibitemShut {NoStop}%
\bibitem [{\citenamefont {Lau}\ and\ \citenamefont {Clerk}(2019)}]{Lau2019}%
  \BibitemOpen
  \bibfield  {author} {\bibinfo {author} {\bibfnamefont {H.-K.}\ \bibnamefont
  {Lau}}\ and\ \bibinfo {author} {\bibfnamefont {A.~A.}\ \bibnamefont
  {Clerk}},\ }\bibfield  {title} {\enquote {\bibinfo {title} {High-fidelity
  bosonic quantum state transfer using imperfect transducers and
  interference},}\ }\href {\doibase 10.1038/s41534-019-0143-1} {\bibfield
  {journal} {\bibinfo  {journal} {npj Quantum Information}\ }\textbf {\bibinfo
  {volume} {5}},\ \bibinfo {pages} {31} (\bibinfo {year} {2019})}\BibitemShut
  {NoStop}%
\end{thebibliography}%

%\vspace*{6em} 

%\newpage

%%%%%%%%%%%%%%%%%%%%%%%%%%%%%%%%%%%%%%%%%%%%%%%%%%%%%%%%%%%%%%%%%%%%%%%%%\newpage
%\begin{widetext}
%\setcounter{equation}{0}
%\setcounter{figure}{0}
%\setcounter{table}{0}
%\setcounter{page}{1}
%\renewcommand{\theequation}{S\arabic{equation}}
%\renewcommand{\thefigure}{S\arabic{figure}}
%\renewcommand{\thetheorem}{S\arabic{theorem}}

%%\widetext

%\end{widetext}

\end{document}